\documentclass[]{aa}
\usepackage{psfig,latexsym,longtable,lscape}

\def\ergsec{\hbox{erg s$^{-1}$ }}
\def\ergcm{\hbox{erg cm$^{-2}$ s$^{-1}$ }}

\def\et{et al. }

\begin{document}
 
   \thesaurus{11; % Galaxies
              (13.25.2; %X-rays: galaxies,
               11.01.2; %Galaxies: active,
               11.02.1; %BL Lacertae objects: general
               4.19.1; %Surveys               
             }
   \title{New active galactic nuclei detected in ROSAT All Sky 
          Survey galaxies}

   \subtitle{Part I: Verification of selection strategy
             \thanks{based partially on observations collected at
             the European Southern Observatory, La Silla, Chile}}

   \author{W.~Pietsch\inst{1} \and
           K. Bischoff\inst{2} \and
           Th. Boller\inst{1} \and
           S. D\"obereiner\inst{1} \and
           W. Kollatschny\inst{2} \and
           H.-U. Zimmermann\inst{1}
          }
 
   \offprints{W.~Pietsch}
   \mail{wnp@mpe.mpg.de} 
   \institute{Max-Planck-Institut f\"ur extraterrestrische Physik, 
              Giessenbachstra\ss e, D-85740 Garching, Germany
   \and Universit\"ats-Sternwarte, Geismarlandstr. 11,
              D-37083 G\"ottingen, Germany
                     }
 
   \date{Received 24 November 1997 / accepted 20 January 1998}
   \titlerunning{New AGN detected in RASS galaxies}
   \maketitle
 
   \begin{abstract}
We present the first results of a program to identify so far unknown active 
nuclei (AGN) in galaxies. Candidate galactic nuclei have been selected for optical
spectroscopy from a cross-correlation of the ROSAT All Sky Survey (RASS) bright 
source catalog with optical galaxy catalogs. A high X-ray flux has been
used as pointer to galaxies with a high probability to 
contain active nuclei. Only galaxies have been accepted for the program for
which no activity was noted in NED. For many of the galaxies no radial
velocity was reported before. The optical spectra demonstrate that the galaxies
cover a redshift range of 0.014 to 0.13 and that most of them
host active nuclei. For 75\% of the 33 candidates the X-ray emission is 
caused by the AGN.
In addition several of the remaining candidates
host Seyfert 2/LINER nuclei that, however, most certainly cannot explain the 
X-ray emission alone. 

Three BL Lac objects have been detected serendipitously in galaxy fields 
that have been followed up by short ROSAT HRI observations to confirm the  
X-ray galaxy identification with improved position accuracy and point response 
function. The sources show
X-ray to radio flux ratios typical for X-ray selected BL Lac objects.

The results presented in the paper prove the selection strategy as very 
successful to detect previously unknown AGN of all Seyfert 1 
types in nearby galaxies encouraging the extension of this program. 
The detection of new nearby AGN will be used to initiate a
detailed investigation of their multi-wavelength properties and a comparison
with the more distant AGN population.  

      \keywords{X-rays: galaxies -- Galaxies: active --
                BL Lacertae objects: general -- Surveys
               }
   \end{abstract}
 
\section{Introduction}
Active galactic nuclei are dominated by huge amounts of energy release
from their nuclear regions. This central continuum source is responsible for 
the photoionization of a central 
emission-line region. From the optical spectra one can distinguish several 
types of AGN, e.g.\ Seyfert galaxies, LINERs, and BL Lac objects. While
Seyfert galaxies and LINERs can be classified according to their optical
emission lines (see Sect. 4 for the classification scheme used),
the main characteristic of BL Lac objects is a featureless optical spectrum; 
for some BL Lac's absorption or emission lines become visible during low stages 
of contiunuum emission in optical spectra with very high S/N ratio.

AGN are well known as strong X-ray emitters often
dominating the X-ray emission of the galaxy they are hosted in. This
has nicely been demonstrated by Fabbiano et al. (1992) in their analysis
of the galaxy content of the Einstein X-ray observatory archive. They 
systematically searched for the galaxy content of the
pointed observations of the Einstein satellite resulting in 'The X-ray  
Catalog and Atlas of Galaxies'. The catalog comprises 493 galaxies, 450 
of which were imaged well, resulting in 238 detections and 212 3$\sigma$ 
upper limits. While the X-rays of E and S0
galaxies (luminosities up to 10$^{43}$ erg s$^{-1}$) are dominated by the emission 
of a hot interstellar medium, the X-ray emission of spiral galaxies 
($< 10^{42}$ erg s$^{-1}$) is dominated by the integrated output of
evolved stellar sources, such as supernova remnants and close accreting
binaries with a compact stellar remnant. If an active nucleus is
present in E and S0 galaxies their X-ray luminosity is between 10$^{41}$ \ergsec
and 10$^{43}$ \ergsec at the high end of the non-active galaxies of this type. 
For spiral galaxies hosting an active nucleus the luminosity range extents to
above 10$^{44}$ erg s$^{-1}$, out-shining non-active spirals by up to more
than two orders of magnitude. 

The X-ray satellite ROSAT (Tr\"umper 1983) performed the first all sky survey
in the soft X-ray band (0.1--2.4 keV) using an imaging telescope. Details of
the ROSAT All Sky Survey (RASS) are described by Voges (1993). A RASS bright 
source catalog has been produced (Voges et al. 1996) by visual screening of 
the sources found by the detect algorithms in the automatic processing of the
RASS data using the standard analysis software system (SASS, Voges et al. 
1992). This catalog contains  18\,811 
sources with a count rate above 0.05 cts s$^{-1}$. The 
sources have a detection likelihood of $\ge$ 15 and contain at least 
15 source photons. The typical positional accuracy is 30\arcsec. We have
analyzed the catalog for its galaxy content by correlation
with optical galaxy catalogs  (Zimmermann et al. 1998). 

The results reported above on galaxies in the Einstein energy band 
(0.2--3.5 keV) also hold for the softer ROSAT band (0.1--2.4 keV) with some 
restrictions. If the AGN is hidden behind circumnuclear material 
(for instance due to a circumnuclear disk as proposed in the
unification models) they will be heavily absorbed (e.g. AGN in Seyfert (Sy) 2 galaxies 
and - if present - in LINERs). Therefore, their observed luminosity in 
the ROSAT band will be highly reduced often preventing the detection 
of the nucleus. In such a case only 
secondary -- less luminous -- effects of present or past activity like 
anomalous arms, jets, or extended emission
from the halo can be measured by ROSAT (see e.g. the detailed ROSAT studies 
of the nearby Sy2 galaxy NGC~4258 or the LINER and starburst galaxy NGC~3079 
(Pietsch et al. 1994, 1998)).

For typical AGN spectra (e.g. power law spectrum with photon index of 2.3) and
moderate galactic foreground absorption with N$_{\rm H}$ of a 
few $10^{20}$cm$^{-2}$
the count rate limit of the RASS bright source catalog corresponds to an
unabsorbed flux of $10^{-12}$\ergcm. This flux limit does not depend strongly
on the shape of the spectrum assumed.
Therefore at distances of 100 Mpc 
only galaxies with luminosities above $10^{42}$ \ergsec can be  
detected. Extrapolating the Einstein results for the RASS bright galaxy list 
we expect to see some E and S0 galaxies to distances of 300 Mpc while 
non-active spirals should only be visible to 100 Mpc. Galaxy groups and 
unrecognized clusters of galaxies may also be found in this X-ray luminosity 
range. 

Having these considerations in mind we used the RASS bright galaxy catalog 
to identify new AGN in galaxies. Our first spectroscopic follow up 
observations that are reported here demonstrate the success of this procedure.
Further optical observations have been granted to this program and will lead 
to an extended list
of new X-ray selected AGN in nearby galaxies. Results will be published in
forthcoming papers of this series, that also will discuss sample properties.

\section{Sample selection}
Cross-correlations of sources detected in the first processing of the 
RASS with a large galaxy 
catalog merged from the most important optical catalogs of galaxies 
(RC2, de Vaucouleurs et al. 1976; RSA, Sandage \& Tamman 1981; Tully 1988; 
UGC, Nilson 1973; ESO, Lauberts 1982) not only showed identifications 
with nearby galaxies that are 
well known for their X-ray emission from investigations 
of the Einstein Observatory observations (see Fabbiano et al. 1992), 
but in addition with  galaxies located in clusters and groups 
(in which the X-ray emission may originate from hot gas in the cluster 
or group) and with galaxies known to host an active nucleus. Besides these 
classes showing exceptionally high X-ray intensities there remained some 
galaxies where no reasons for the high X-ray emission is known. 
Several of these galaxies were proposed 
for short X-ray follow up observations using the ROSAT HRI to verify the 
origin of the emission from the galaxy. While in some cases the HRI observations
clearly demonstrated that the RASS X-rays were extended emission from an 
early type galaxy or even from a cluster of galaxies, in many cases they showed 
unresolved emission from the galactic nucleus as expected from an AGN. 
We selected only these galaxic nuclei for further spectroscopic investigations.
Table~\ref{HRI} summarizes the HRI observation dates
and exposure times for the galaxies that were included into the optical observing
run we report in this paper.

\begin{table}
      \caption{Log of ROSAT HRI follow up observations for RASS galaxies included
                into our optical program}
         \label{HRI}
         \begin{flushleft}
         \begin{tabular}{lcc}
            \hline
            \noalign{\smallskip}
Object & Date& Active \\
            &     & time \\
            \noalign{\smallskip}
            \hline
            \noalign{\smallskip}
ESO~113-~G~010&  28.12.1995&2.4~ks\\
              &  10.06.1996&2.7~ks\\
\noalign{\smallskip}
NGC~427&17.06.1996--18.06.1996 &1.9~ks\\
       &10.01.1996             &3.9~ks\\
\noalign{\smallskip}
ESO~080-~G~005&22.04.1996 &4.3~ks\\
\noalign{\smallskip}
ESO~416-~G~002&28.01.1995 &2.3~ks\\
\noalign{\smallskip}
ESO~15-~IG~011&30.07.1995--17.09.1995 &6.5~ks\\
\noalign{\smallskip}
ESO~552-~G~039&21.03.1996&1.9~ks\\
\noalign{\smallskip}
ESO~254-~G~017&29.10.1996--30.10.1996 &5.5~ks\\
\noalign{\smallskip}
PMN~J0623-6436&04.07.1995--16.07.1995 &4.4~ks\\
\noalign{\smallskip}
PMN~J0630-2406&11.04.1996--12.04.1996 &4.7~ks\\
\noalign{\smallskip}
ESO~490-~IG~026&24.10.1996--25.10.1996&3.2~ks\\
\noalign{\smallskip}
ESO~209-~G~012&21.12.1995 &1.4~ks\\
\noalign{\smallskip}
ESO~602-~G~031&25.05.1995--26.05.1995 &4.5~ks\\
\noalign{\smallskip}
\hline
         \end{tabular}
         \end{flushleft}
   \end{table}

In a follow up of this work we cross-correlated the improved X-ray 
source catalog
produced from re-processing of the RASS list (RASS bright source catalog, 
Voges et al. 1996) with the Catalogue of Principal Galaxies PGC (Paturel et al. 
1989). 
The candidate galaxies have been categorized according to their X-ray
emission, separating cluster candidates and active galaxies 
(Zimmermann et al. 1998).
Correlations with the NASA Extragalactic Database (NED)\footnote{The NASA/IPAC 
Extragalactic Database is operated by the Jet
Propulsion Laboratory, California Institute of Technology, under contract with
the National Aeronautics and Space Administration.} 
have been used to accumulate additional information on these galaxies. 
This procedure resulted in a list of galaxies as candidates for the 
X-ray sources that show unusually high X-ray fluxes and were not 
known to host an active nucleus from observations at other wavelengths. 
No redshifts had been measured for most of 
these galaxies. 
To ensure the identification we produced and investigated
overlays of the X-ray contours onto optical
images extracted from the digitized sky survey (see acknowledgments). To identify
galaxies on the optical images with the cataloged ones we over-plotted galaxy positions 
from the PGC catalog and NED extracts. 

\begin{figure*}
%  \resizebox{\hsize}{!}{\vspace{22.5cm}}  
\makebox(18.,22.5)[t]{ } 
  \vspace{22.5cm}
  \caption{ROSAT HRI contours overlaid onto optical digitized sky survey images.
          The X-ray images have been smoothed with a Gaussian filter of 5\arcsec\  
          FWHM. Contours correspond to (0.1, 0.2, 0.5, 1.5, 4.5) cts arcsec$^{-2}$}
\end{figure*}
\begin{figure*}
%  \resizebox{\hsize}{!}{  
  \makebox(18.,22.5)[b]{ } 
{  \vspace{22.5cm}
{\bf Fig. 1.} continued}
\end{figure*}
\begin{figure*}
  \resizebox{12cm}{!}{
  \makebox(10.6,12.)[t]{  } 
  \vspace{12.cm}}
  \hfill
  \parbox[b]{55mm}{
  \caption{ROSAT HRI contour overlay of the extended ESO~416-~G~002 field 
     onto an optical digitized sky survey image. The X-ray image has been 
     smoothed with a Gaussian filter of 10\arcsec\  
     FWHM. Contours correspond to (0.03, 0.1, 0.3) cts arcsec$^{-2}$}
     }
\end{figure*}
%\psfull

To further characterize these X-ray bright galaxies we obtained optical 
observations for a subsample of these galaxies in an observing 
run in November 1996 at the 2.2m ESO/MPG telescope at La Silla. In this investigation we included 
from our HRI follow up studies of the RASS galaxies 
three bright nearby field objects (one source near NGC~427  
and a source pair centered on ESO~416--~G~002) 
and one candidate optical source where the X-ray emission did not 
originate from the galaxy (ESO~490--~G~008) but
from a nearby radio source (PMN~J0630-2406).
Figure~1 shows X-ray contours for the HRI fields overlaid on 
digitized optical images. In Fig.~2 the ESO~416--~G~002 field is extended to 
include the source pair mentioned above. All X-ray sources are unresolved and
are supposed to be nuclear point sources.

Table 2 summarizes the X-ray identification information for the
observed objects. Objects are identified by ROSAT name (col. 1), prefix 1RXS~J
stands for a RASS determined position (field number of galaxy 
cross-cor\-rela\-tion
from Zimmermann et al. (1998) given in col. 2), prefix RX~J for a
HRI detected source (HRI in col. 2). ROSAT PSPC count rates 
for the RASS sources or HRI count rates (0.1--2.4 keV) are given in
col. 3, object name in col. 4. The optical positions (epoch 2000.0, col. 5 
and 6) were derived from the digitized sky survey (Irwin et al. 1994) and
should be accurate to better than 2\arcsec. Only the position of NGC~1217
was taken from NED as the galaxy information from the digitized sky list was 
confused by a nearby source. Column 7 gives
the separation of the optical position from the X-ray source location. In all
cases HRI positions are closer to the optical positions of the galaxies than
the RASS positions and the
nuclei of the galaxies are within the typical X-ray error circle confirming
the identification. For  1RXS~J063059.4--240636 the cross-correlation with the 
PGC catalog suggested ESO~490--~G~008 as identification. 
The position 
however was off by 83\arcsec. The HRI follow up observation rejected the galaxy 
identification without question and identified a star-like nearby object 
coinciding with the PSPC and HRI position, which is also identified with the 
radio source PMN~J0630--2406 (see Fig.~1). 
ESO~120--~G~023 and the RASS position have a big offset. From the optical 
overlay in Zimmermann et al. (1998) 
it is clear that there is X-ray emission from this source. 
However, the automatic source detection was confused
by X-ray emission from a nearby bright star and therefore, the X-ray count rate
and position for 1RXS J055559.4-612438 are not reliable. The RASS bright
source catalog also contains information on the extent of the X-ray
sources. If one uses a conservative extent criterium (extent likelihood
$>$ 10 and extent $>$ 30") three of the sources in table 2 are X-ray extended
(gal. no. 195, 238, and 276) indicating that the X-ray emission in these galaxies 
does not, or not exclusively originate from a nuclear source. 

\section{Optical observations and data reduction}
The optical observations were performed with the 2.2m ESO/MPG
telescope at La Silla observatory from November 2 to 7, 1996.
We used EFOSC2 spectrograph with grism \#6 and a 1\farcs 5-wide long slit
and the $2048 \times 2048$ $15\mu m$ LORAL CCD, which gave a dispersion of
2 \AA\ per pixel, a spectral coverage of 3800--8000 \AA,
and a spectral resolution of 10 \AA. The seeing was typically 
between 0.8" and 1.5".

For acquisition we made images with Johnson R filter and an
integration time of typically 1 minute. These images were also used to
determine the morphological type of the galaxies.
For the spectra we chose a position angle (PA) of the slit of 90\degr\ (E-W).
The only exception was the double nucleus galaxy ESO~120-~G~016; in this
case the PA of the slit was 60\degr\ in order to take spectra of both
nuclei simultaneously. The integration times of the spectra are given
in Table 3, col.~2.

The data were reduced with standard MIDAS procedures.
Wavelength calibration was done using He-Ar comparison lamps.
The flux calibration was performed by means of the flux standard
stars HD~49798 and NGC~7293 (Turnshek \et 1990).
To extract the nuclear spectra we used an aperture of 3\arcsec.
The spectra are shown in Fig.~3.

\begin{figure*}
  \hbox{\psfig{figure=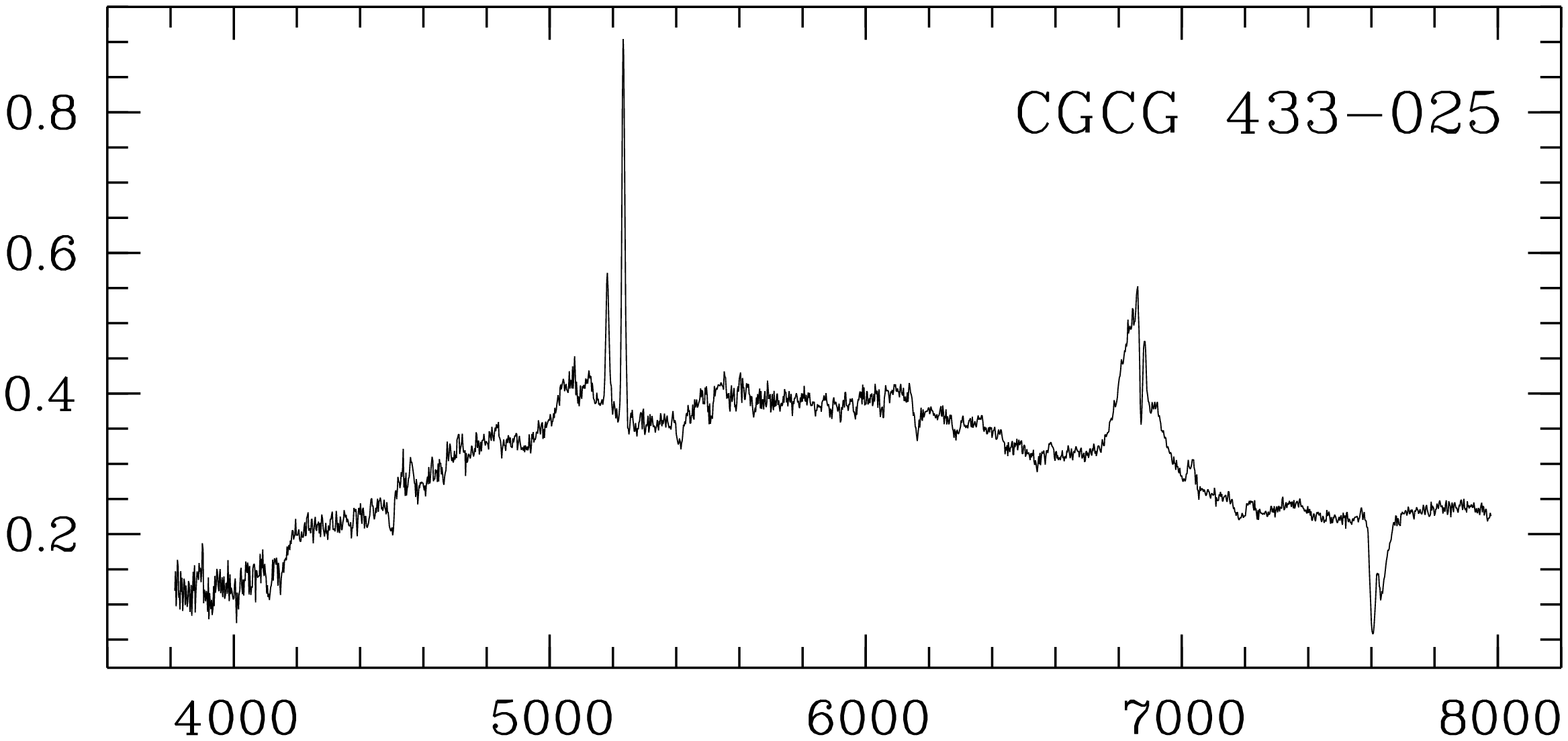,width=88mm,height=35mm,angle=270,clip=}\hfill
        \psfig{figure=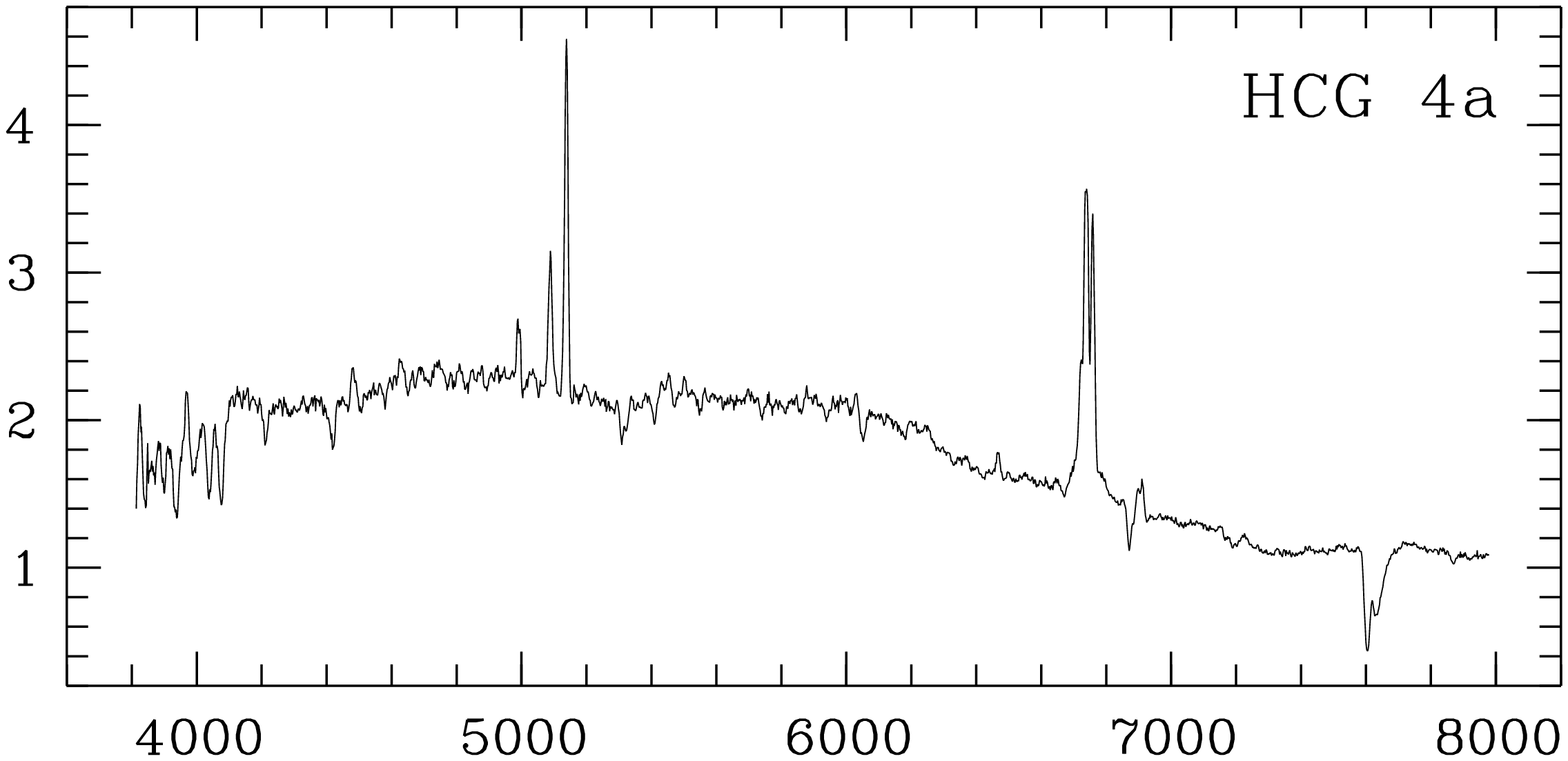,width=88mm,height=35mm,angle=270,clip=}}
  \hbox{\psfig{figure=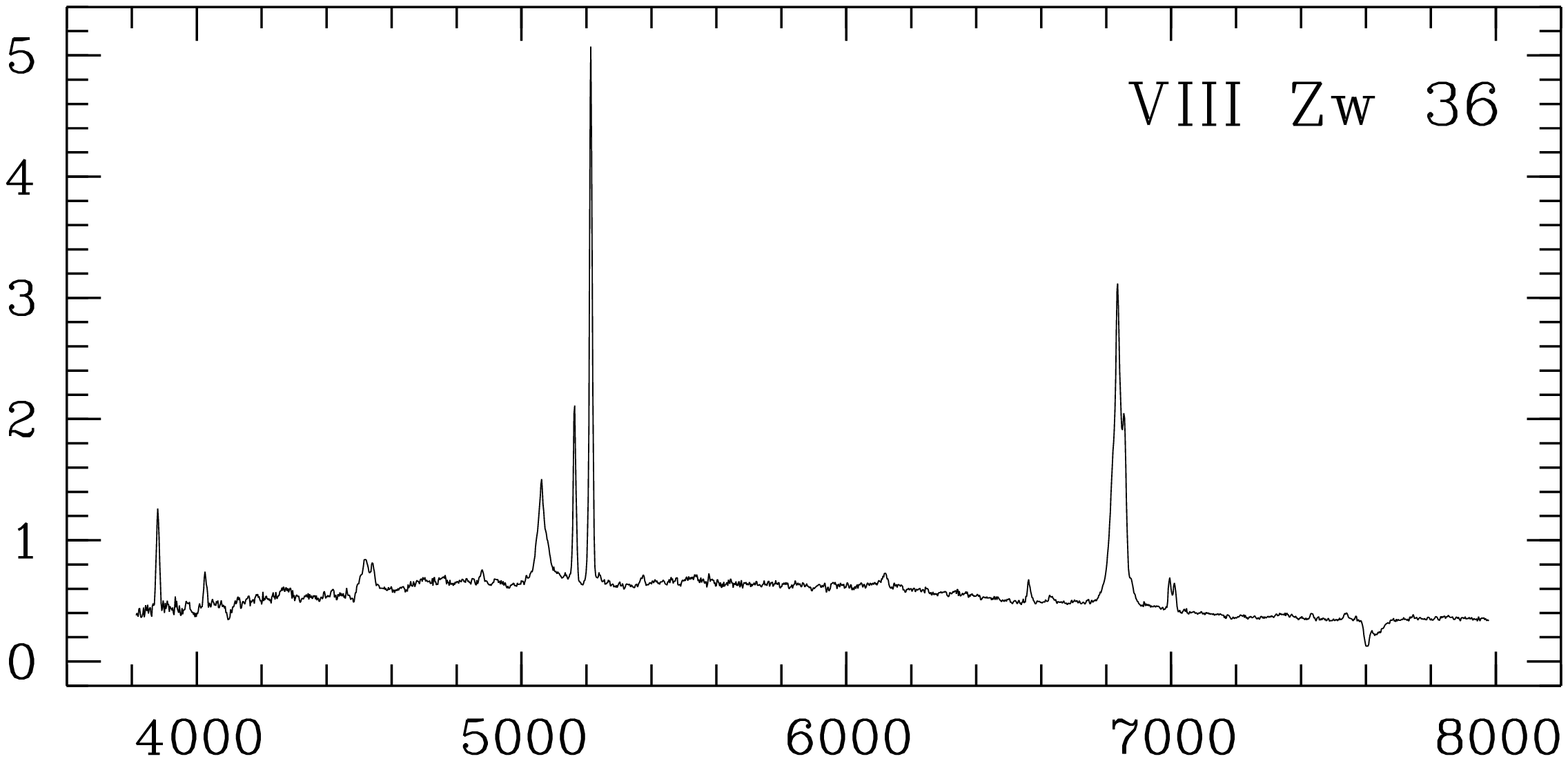,width=88mm,height=35mm,angle=270,clip=}\hfill
        \psfig{figure=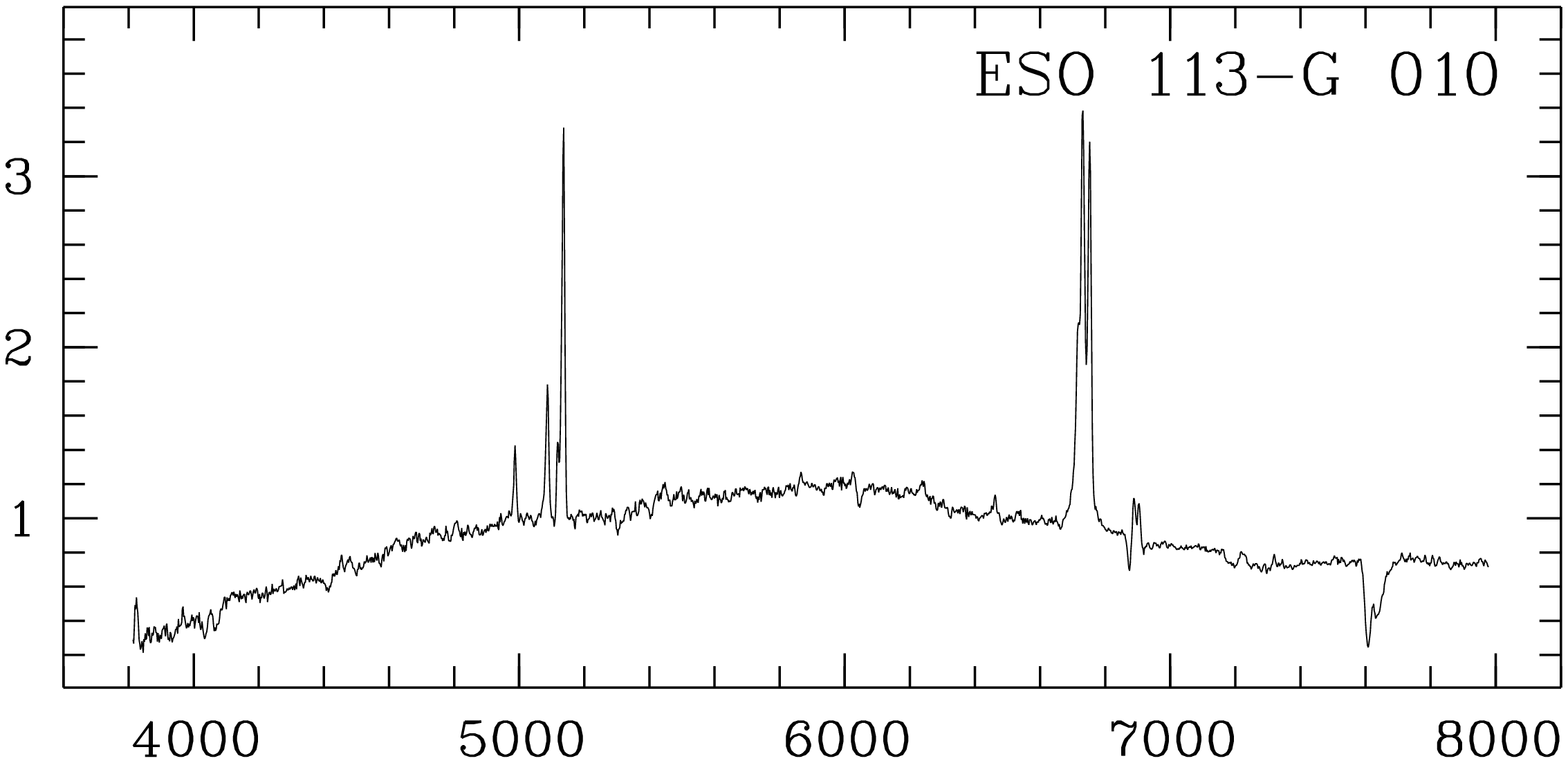,width=88mm,height=35mm,angle=270,clip=}}
  \hbox{\psfig{figure=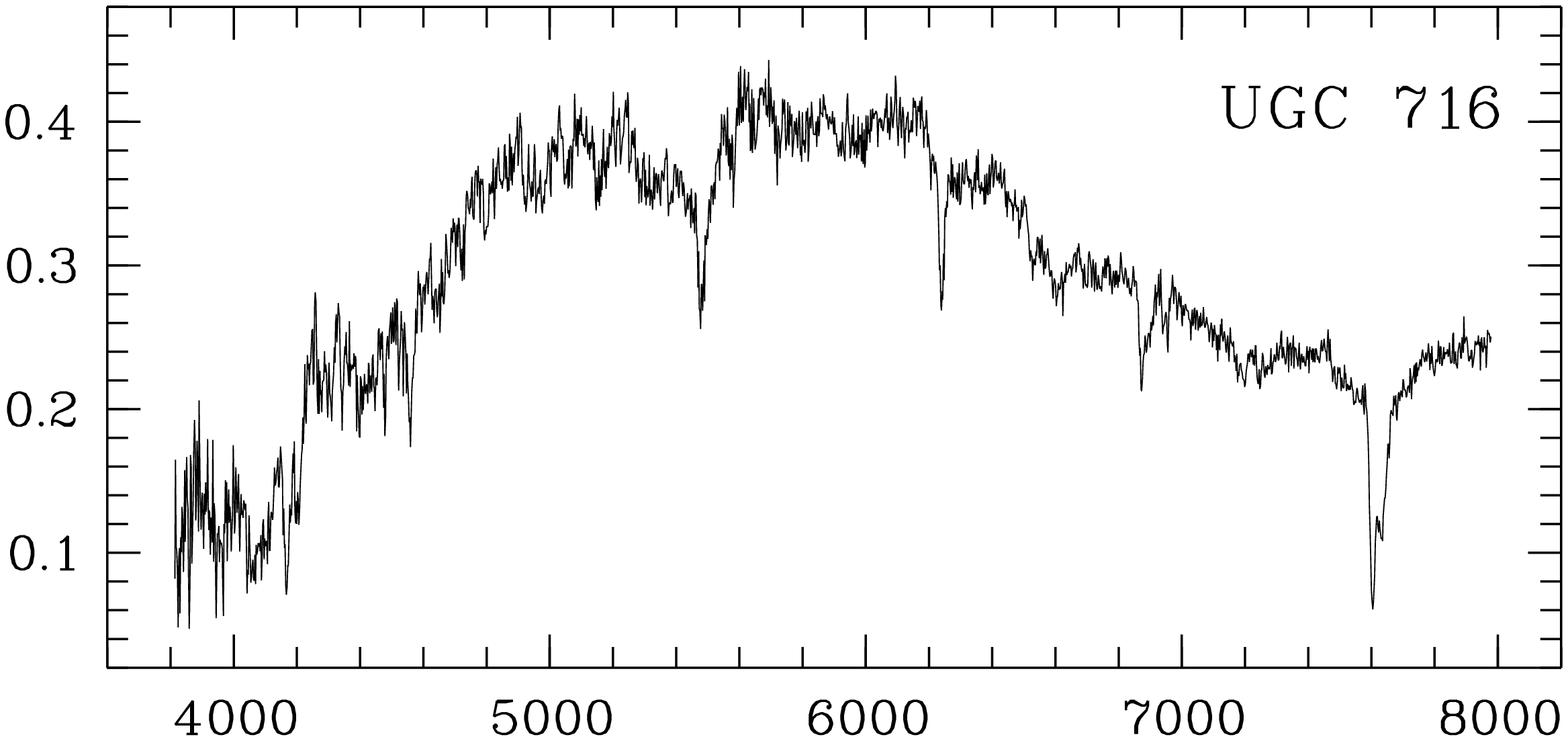,width=88mm,height=35mm,angle=270,clip=}\hfill
        \psfig{figure=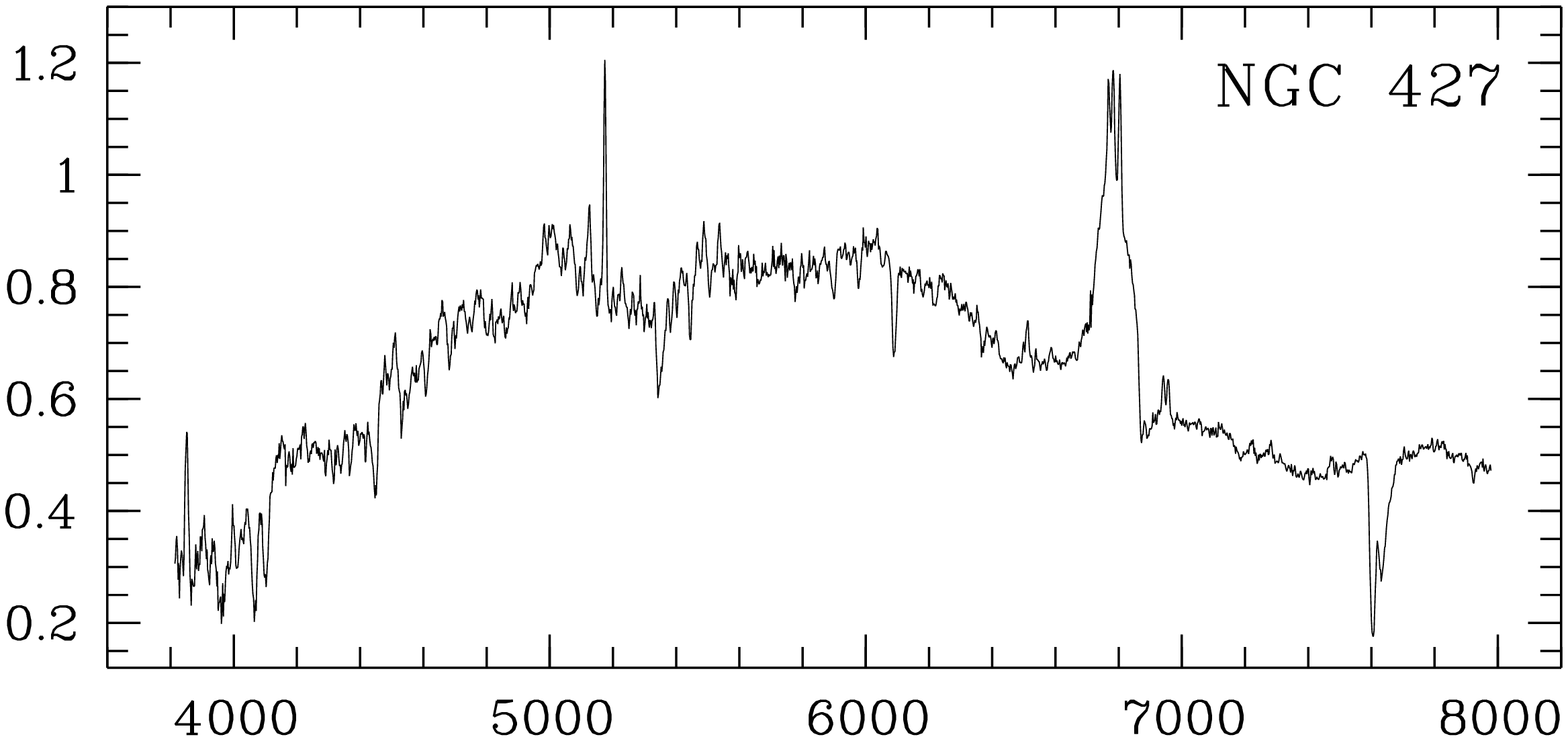,width=88mm,height=35mm,angle=270,clip=}}
  \hbox{\psfig{figure=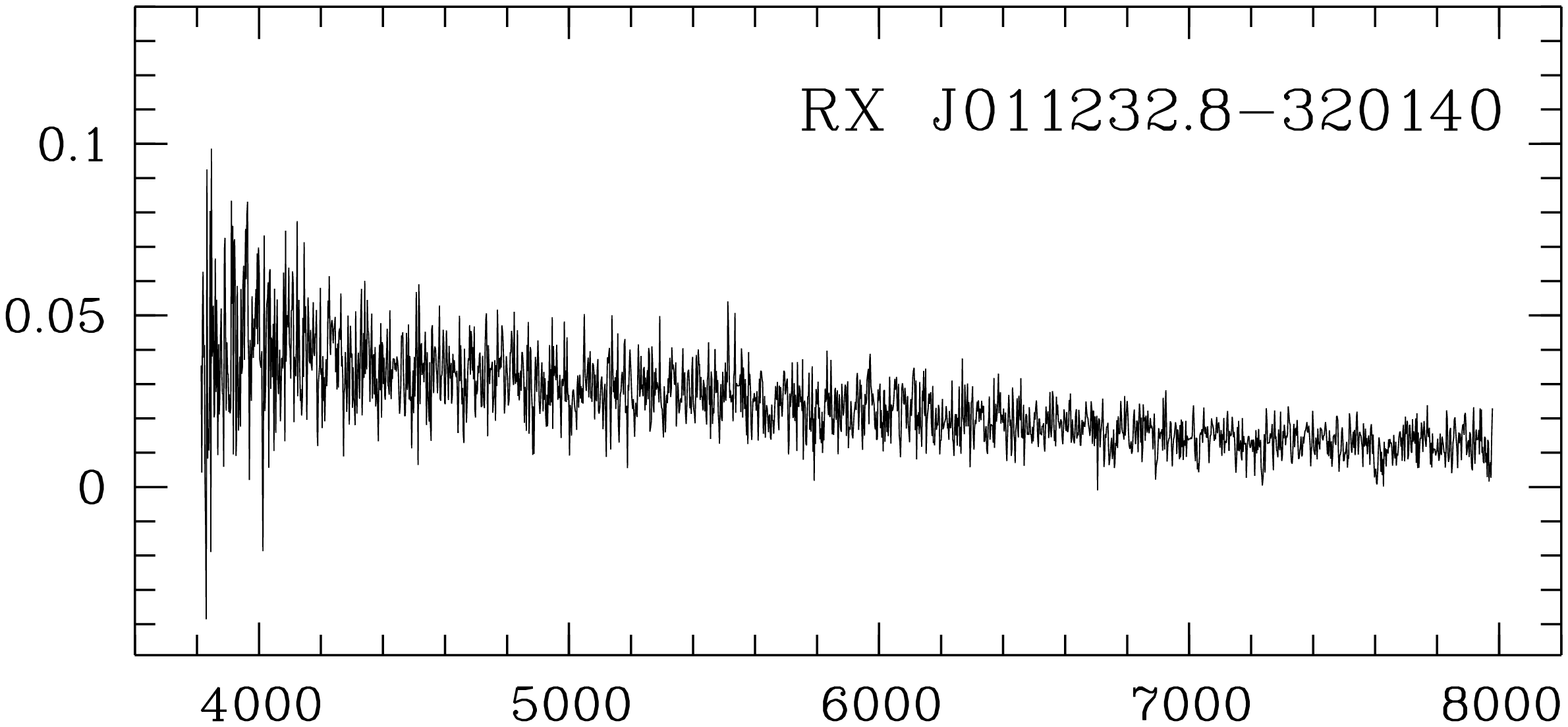,width=88mm,height=35mm,angle=270,clip=}\hfill
        \psfig{figure=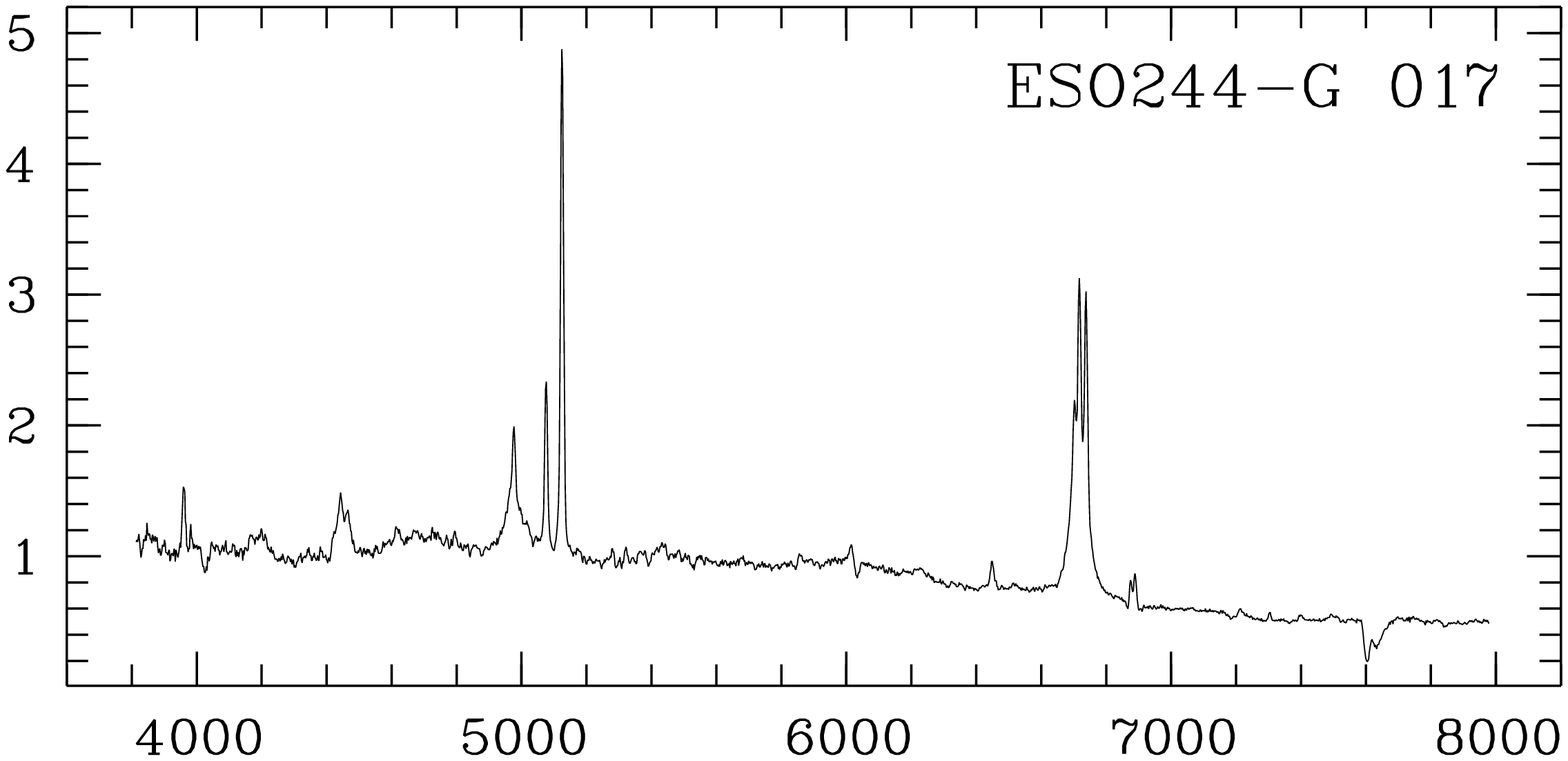,width=88mm,height=35mm,angle=270,clip=}}
  \hbox{\psfig{figure=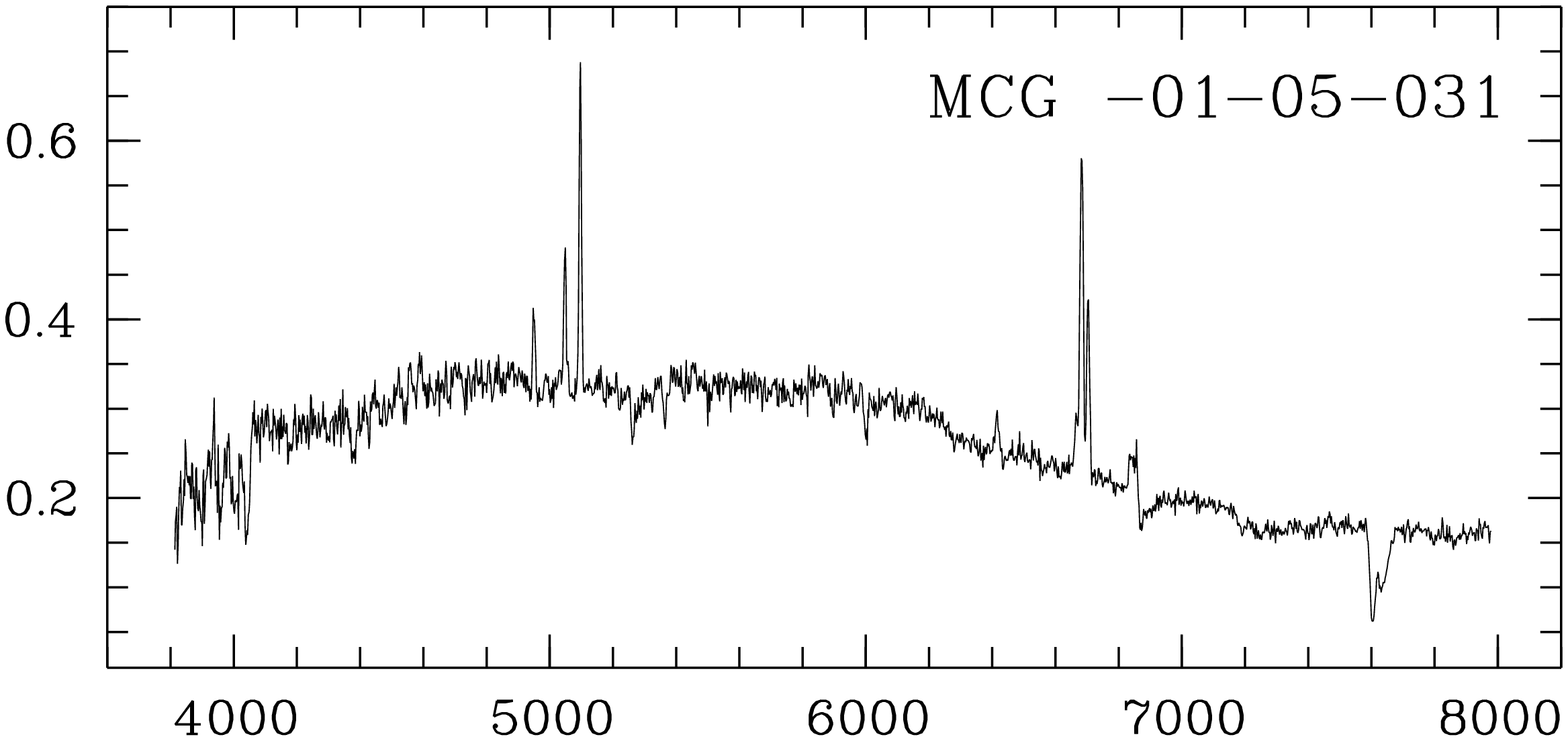,width=88mm,height=35mm,angle=270,clip=}\hfill
        \psfig{figure=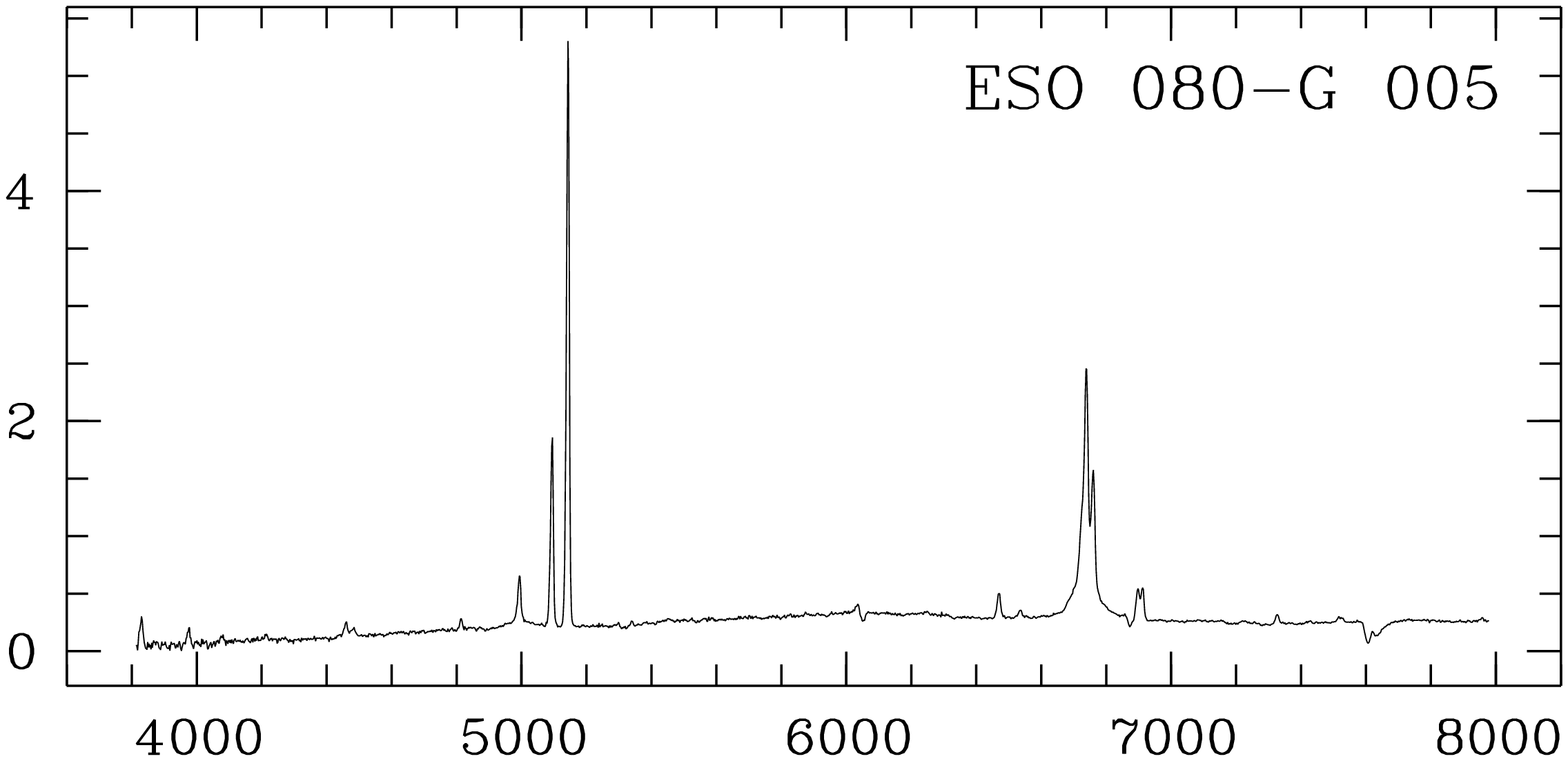,width=88mm,height=35mm,angle=270,clip=}}
  \hbox{\psfig{figure=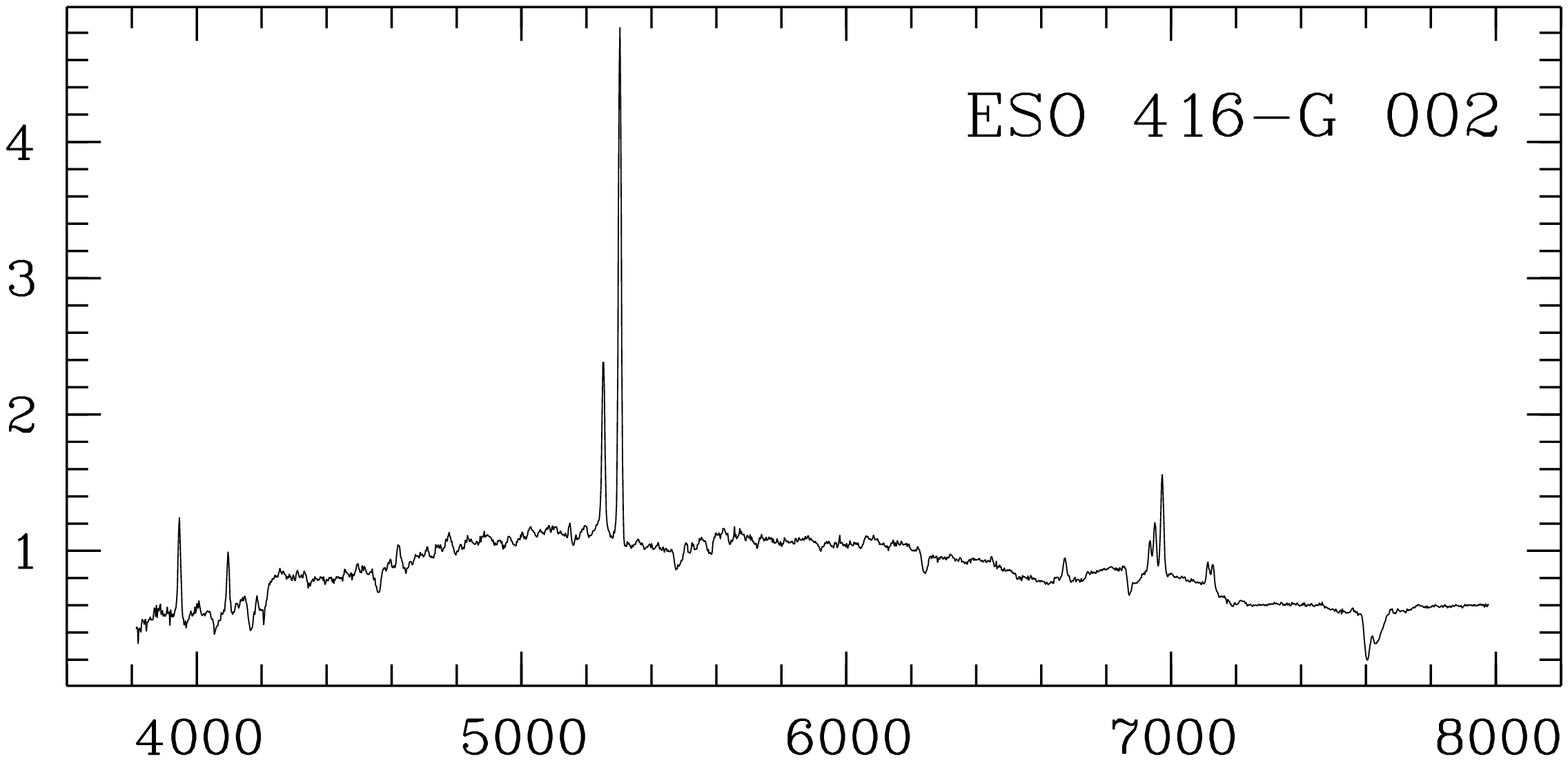,width=88mm,height=35mm,angle=270,clip=}\hfill
        \psfig{figure=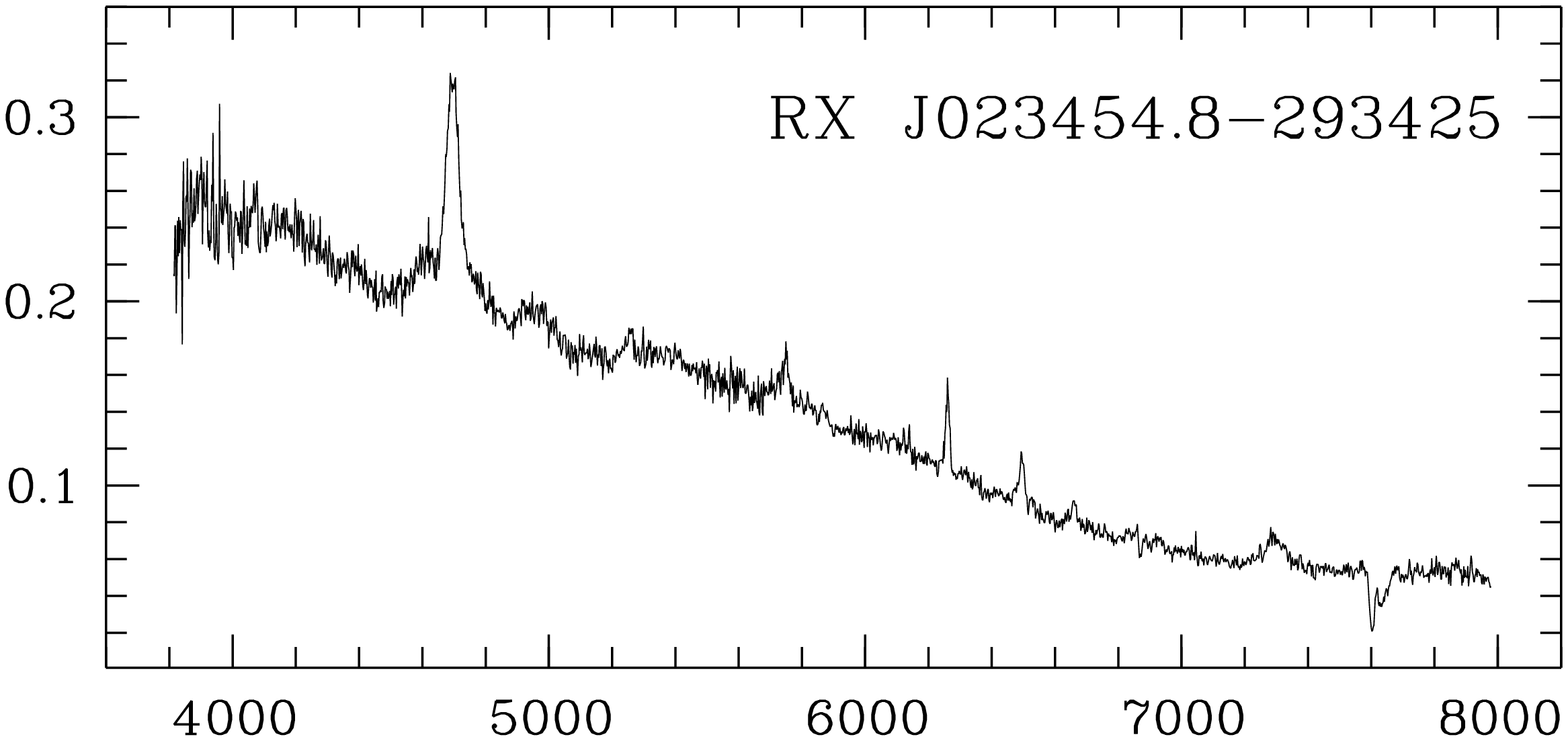,width=88mm,height=35mm,angle=270,clip=}}
  \caption{Optical spectra of galactic nuclei and AGN. f$_\lambda$
          in units of 10$^{-15}$~erg~cm$^{-2}$~s$^{-1}$~\AA$^{-1}$ is plotted against
          wavelength in \AA }
\end{figure*}
\begin{figure*}
  \hbox{\psfig{figure=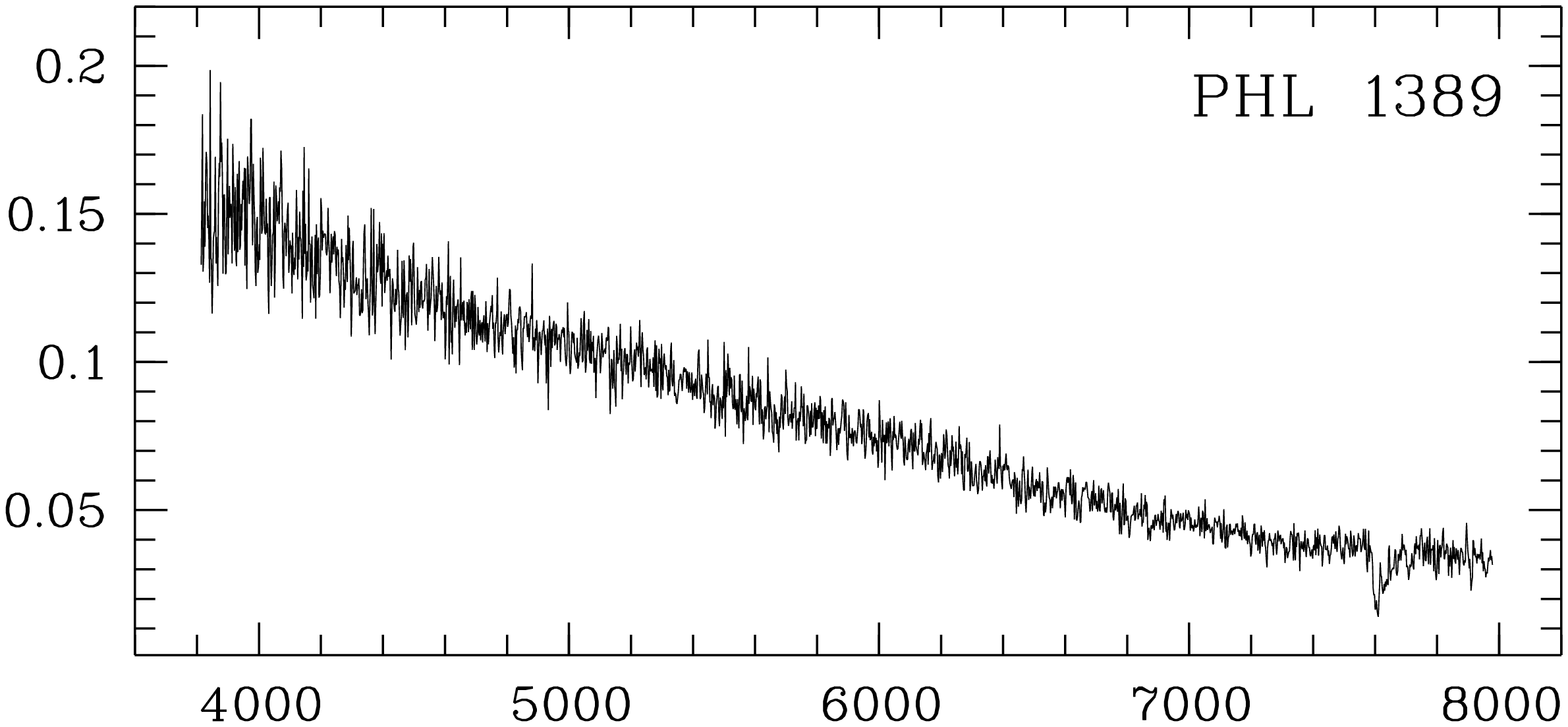,width=88mm,height=35mm,angle=270,clip=}\hfill
        \psfig{figure=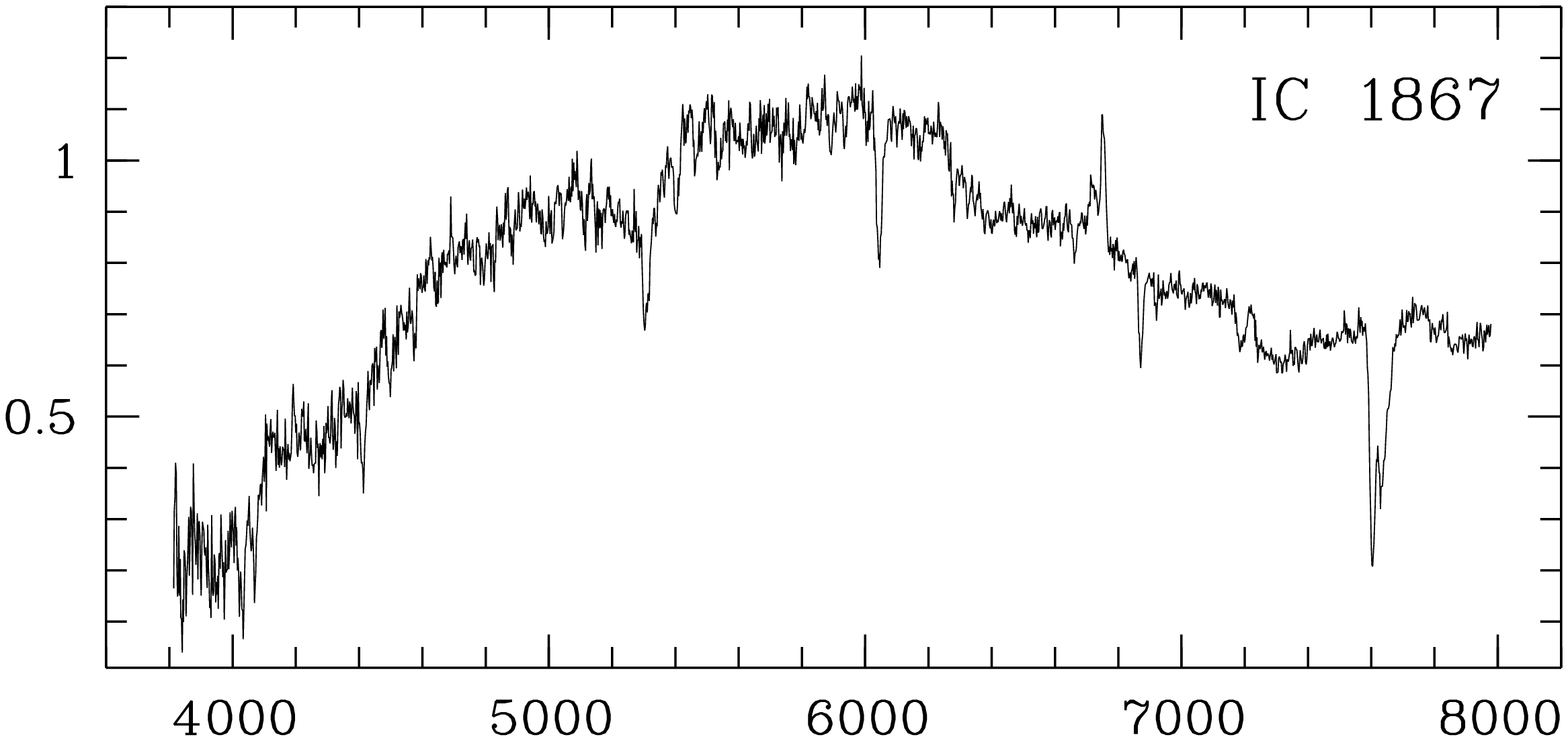,width=88mm,height=35mm,angle=270,clip=}}
  \hbox{\psfig{figure=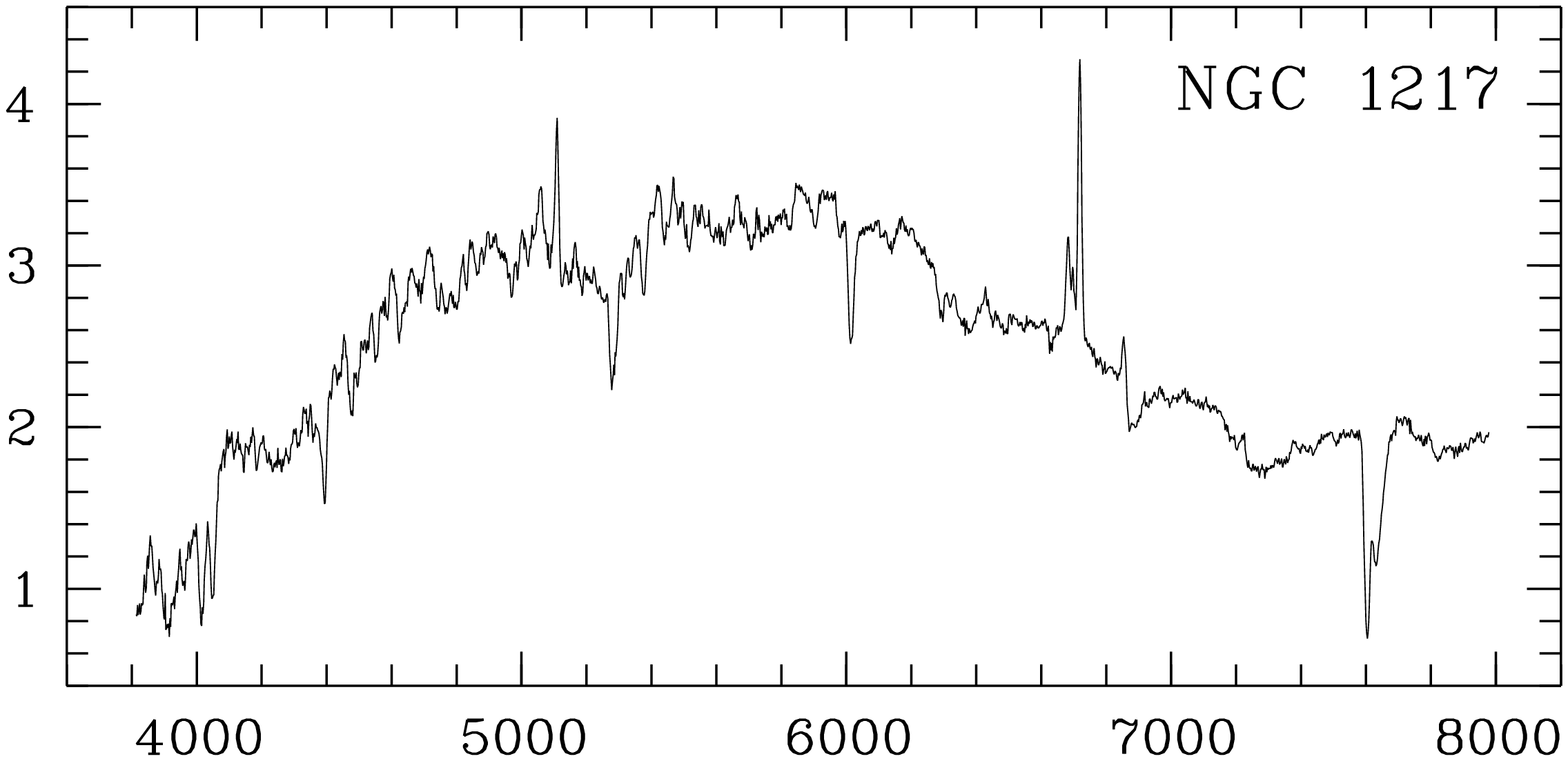,width=88mm,height=35mm,angle=270,clip=}\hfill
        \psfig{figure=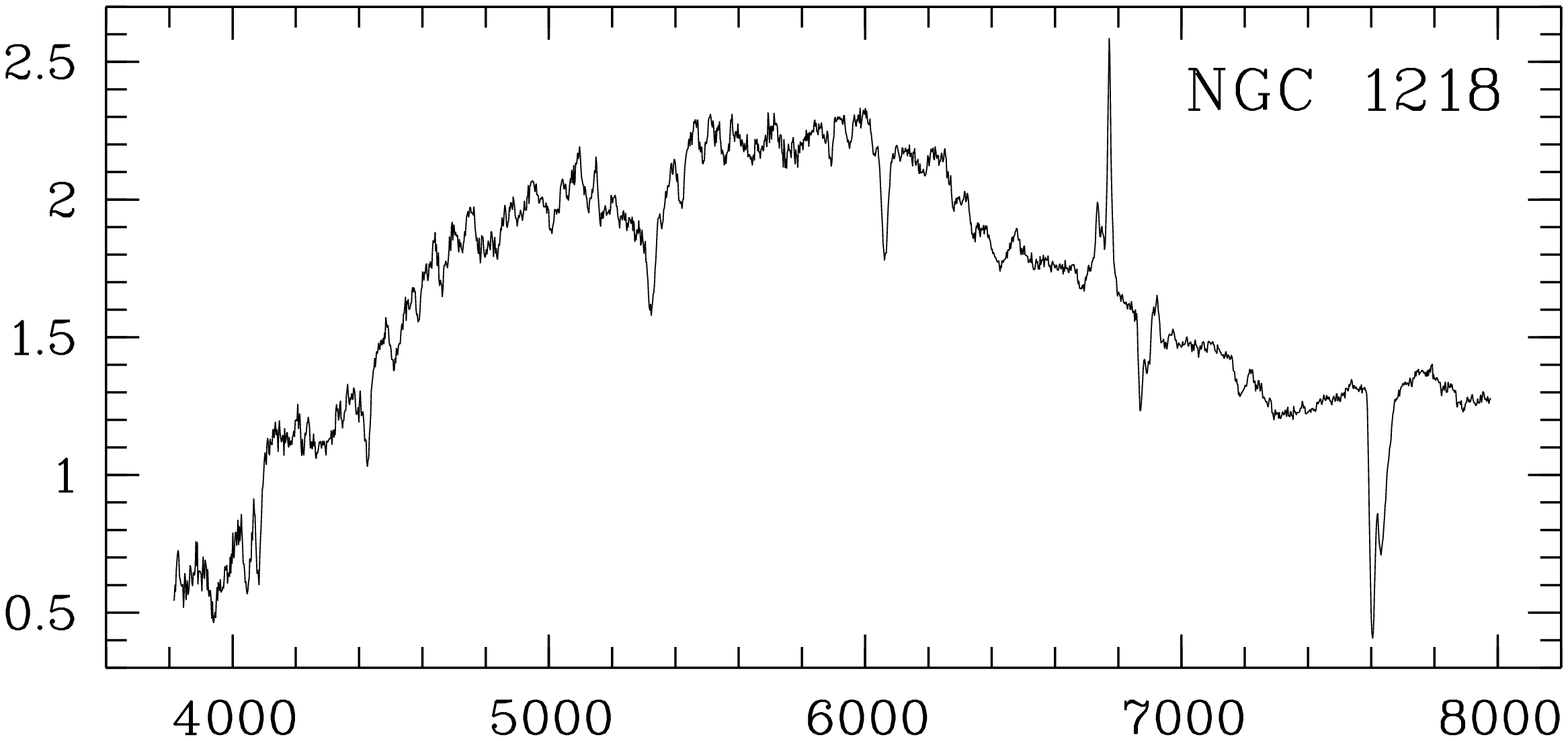,width=88mm,height=35mm,angle=270,clip=}}
  \hbox{\psfig{figure=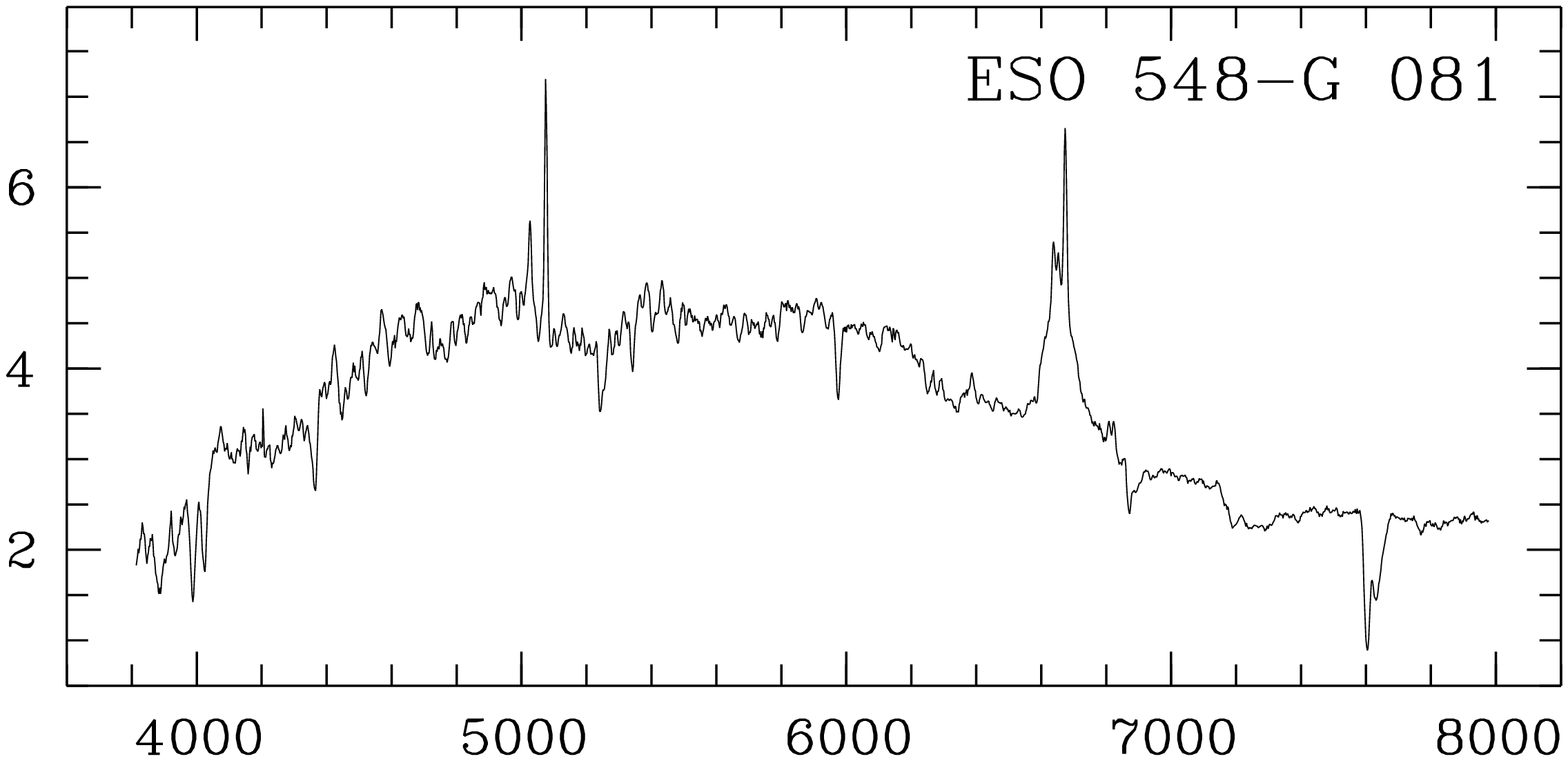,width=88mm,height=35mm,angle=270,clip=}\hfill
        \psfig{figure=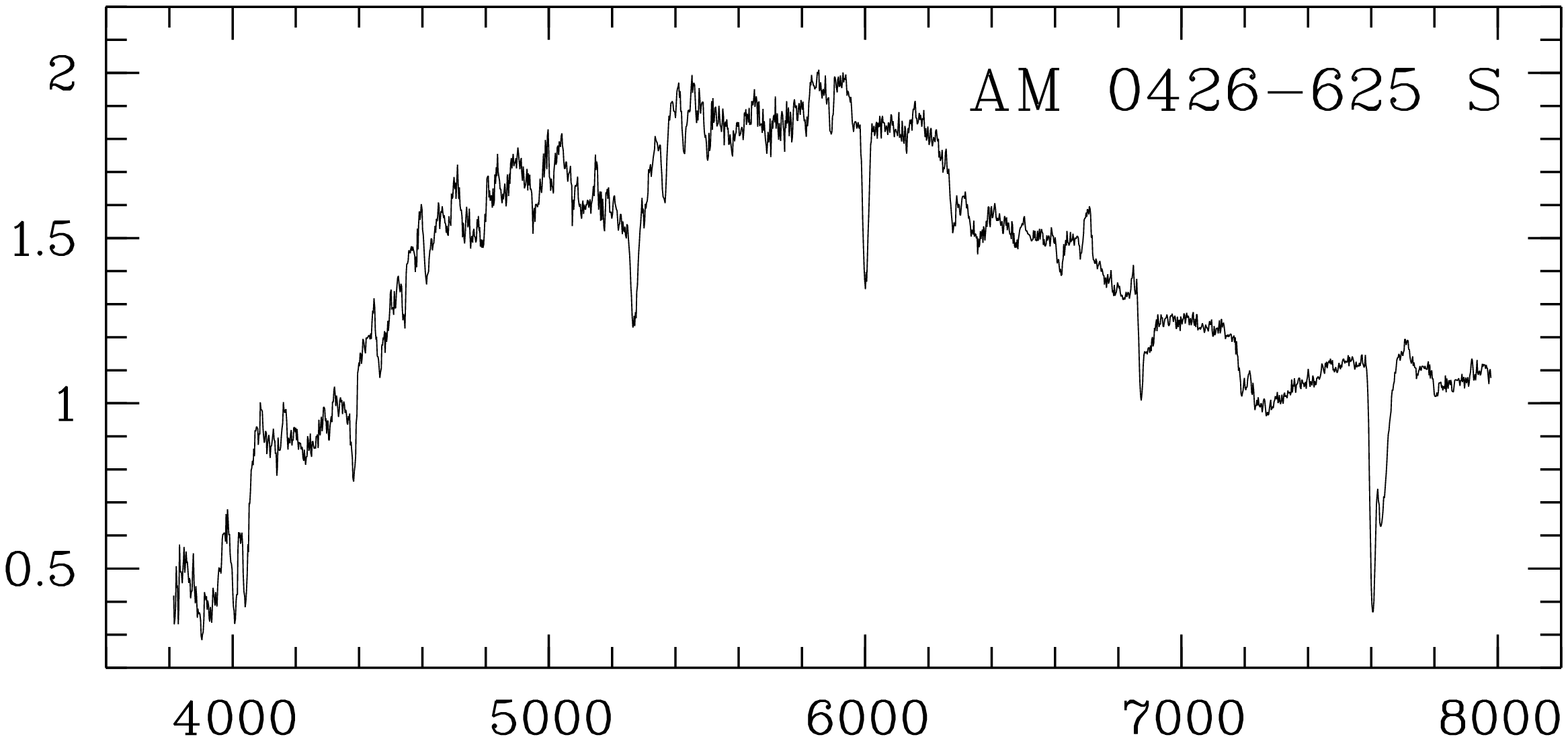,width=88mm,height=35mm,angle=270,clip=}}
  \hbox{\psfig{figure=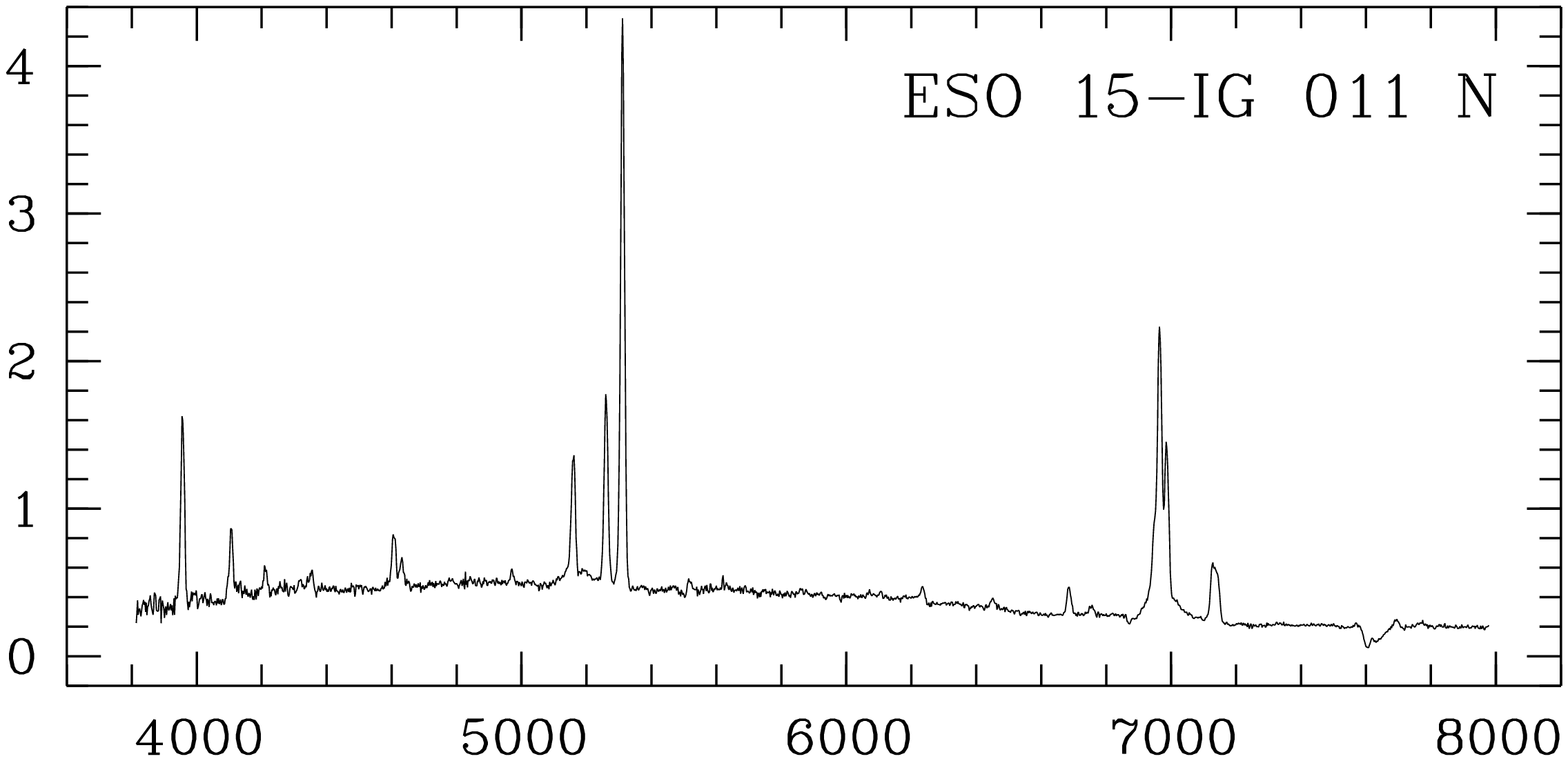,width=88mm,height=35mm,angle=270,clip=}\hfill
        \psfig{figure=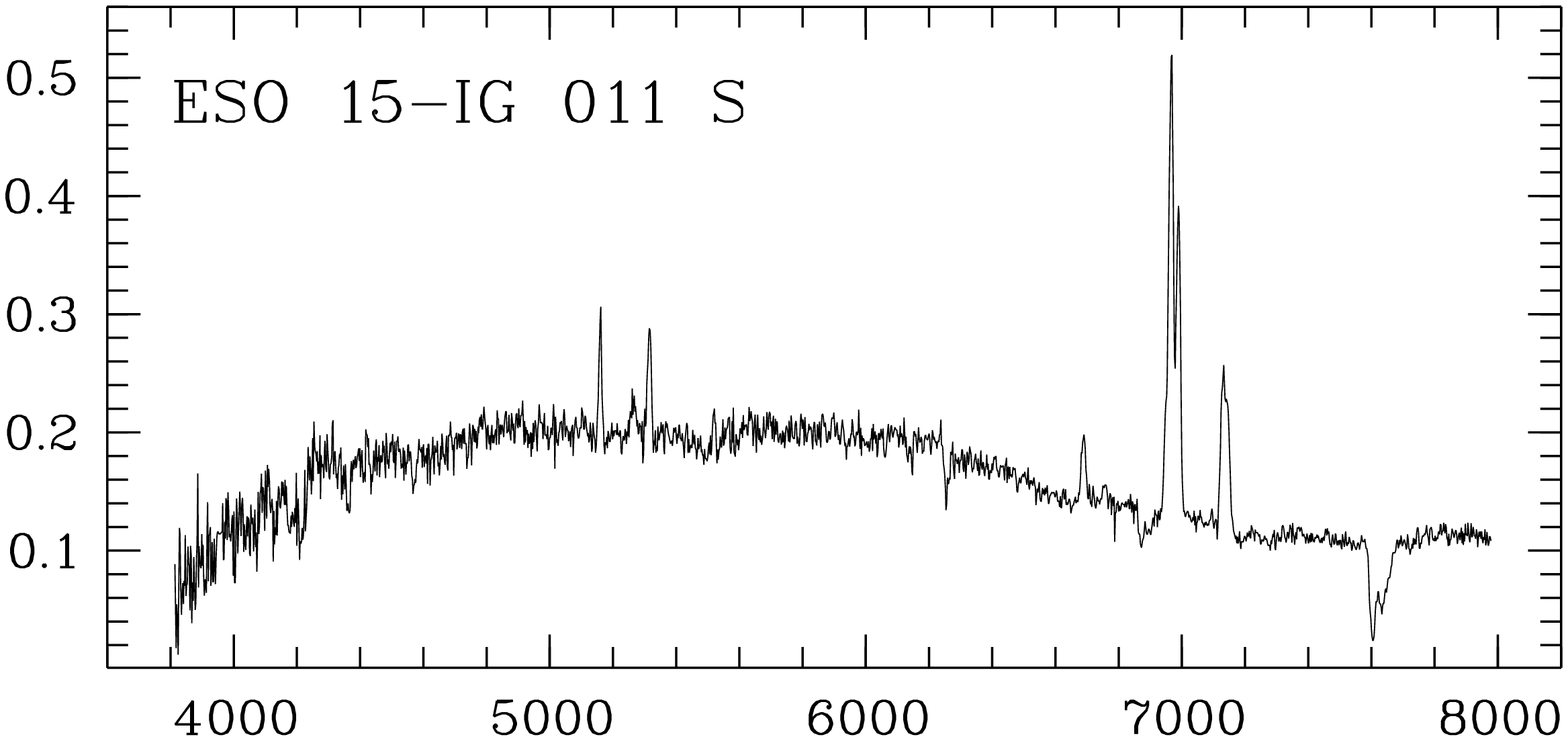,width=88mm,height=35mm,angle=270,clip=}}
  \hbox{\psfig{figure=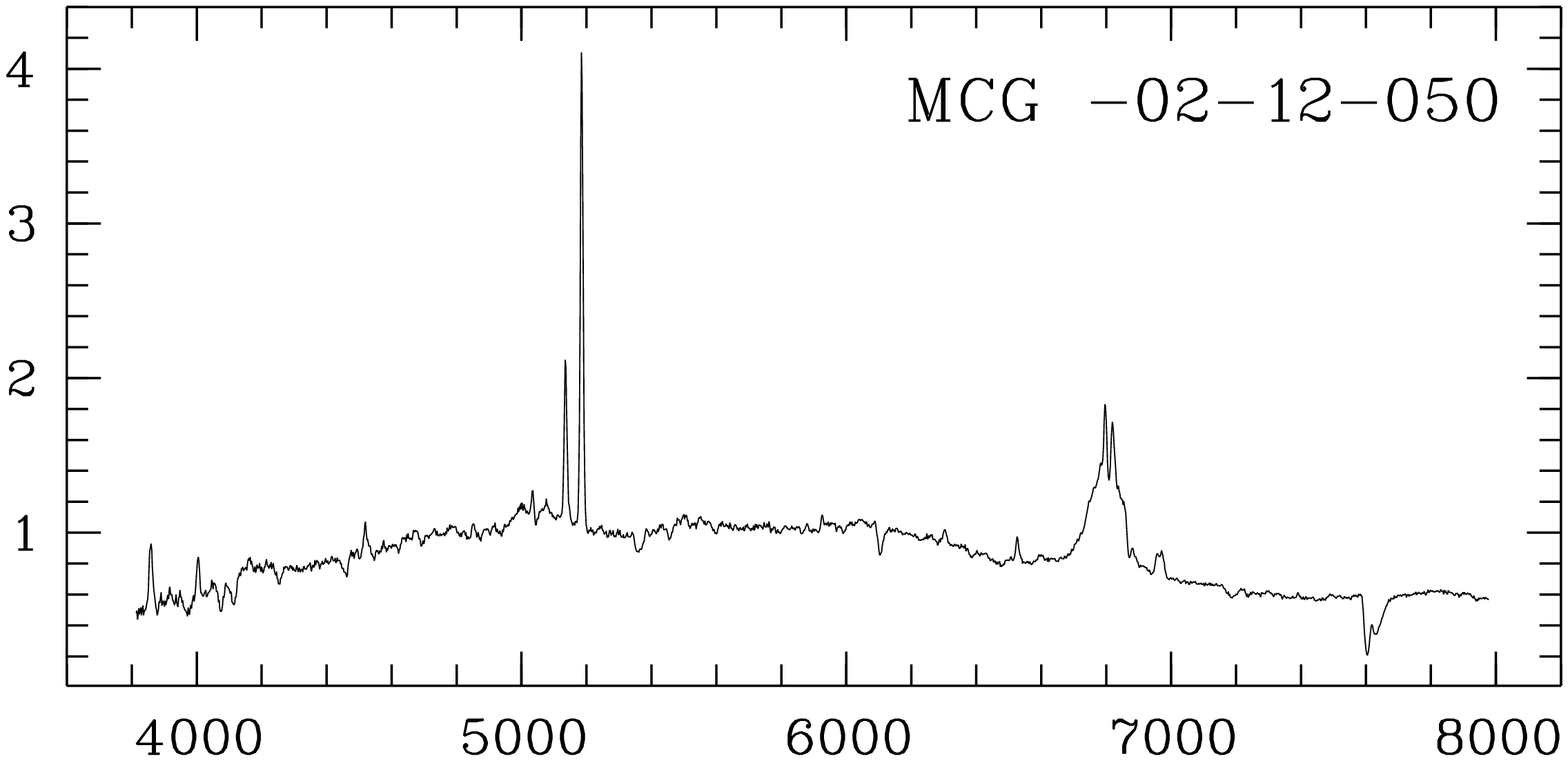,width=88mm,height=35mm,angle=270,clip=}\hfill
        \psfig{figure=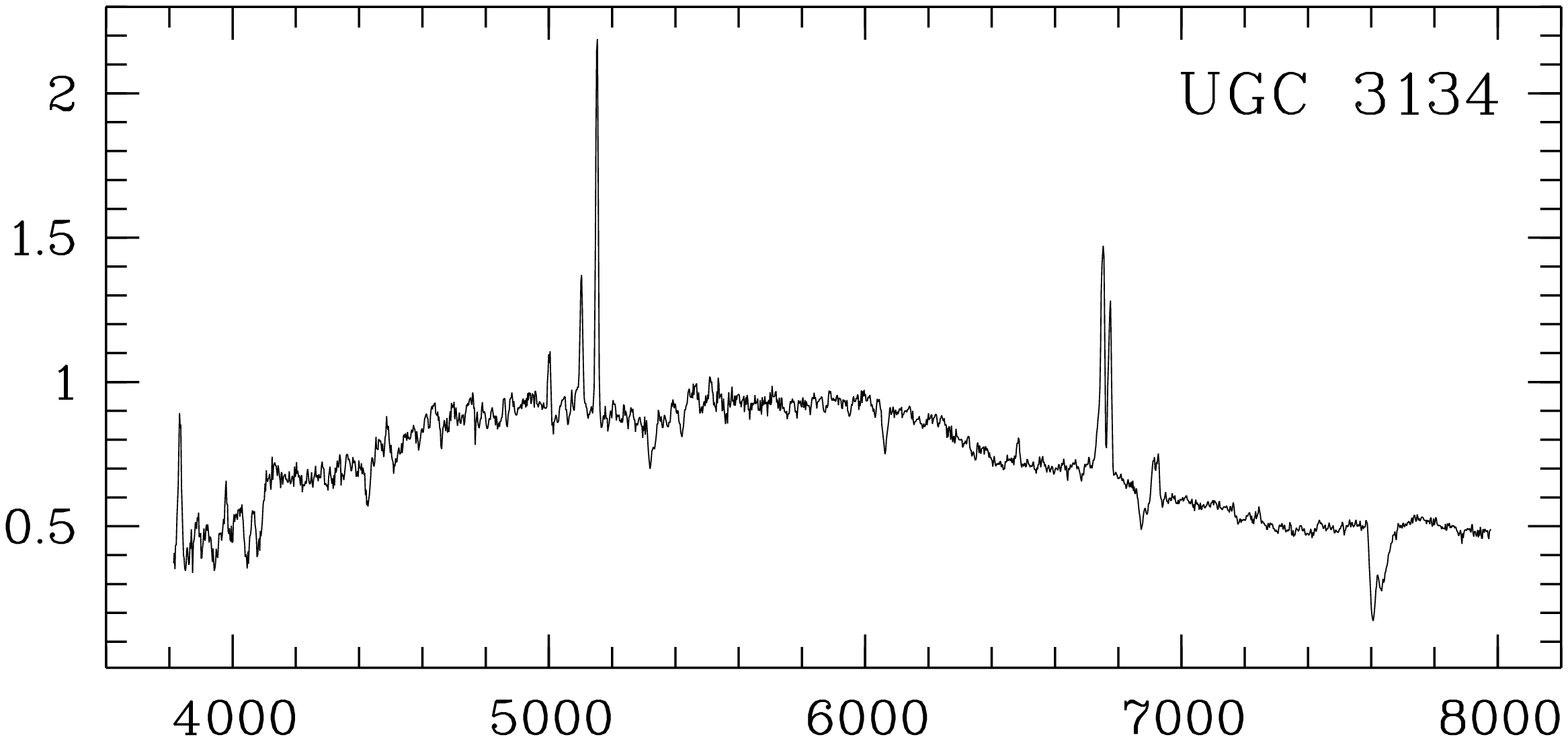,width=88mm,height=35mm,angle=270,clip=}}
  \hbox{\psfig{figure=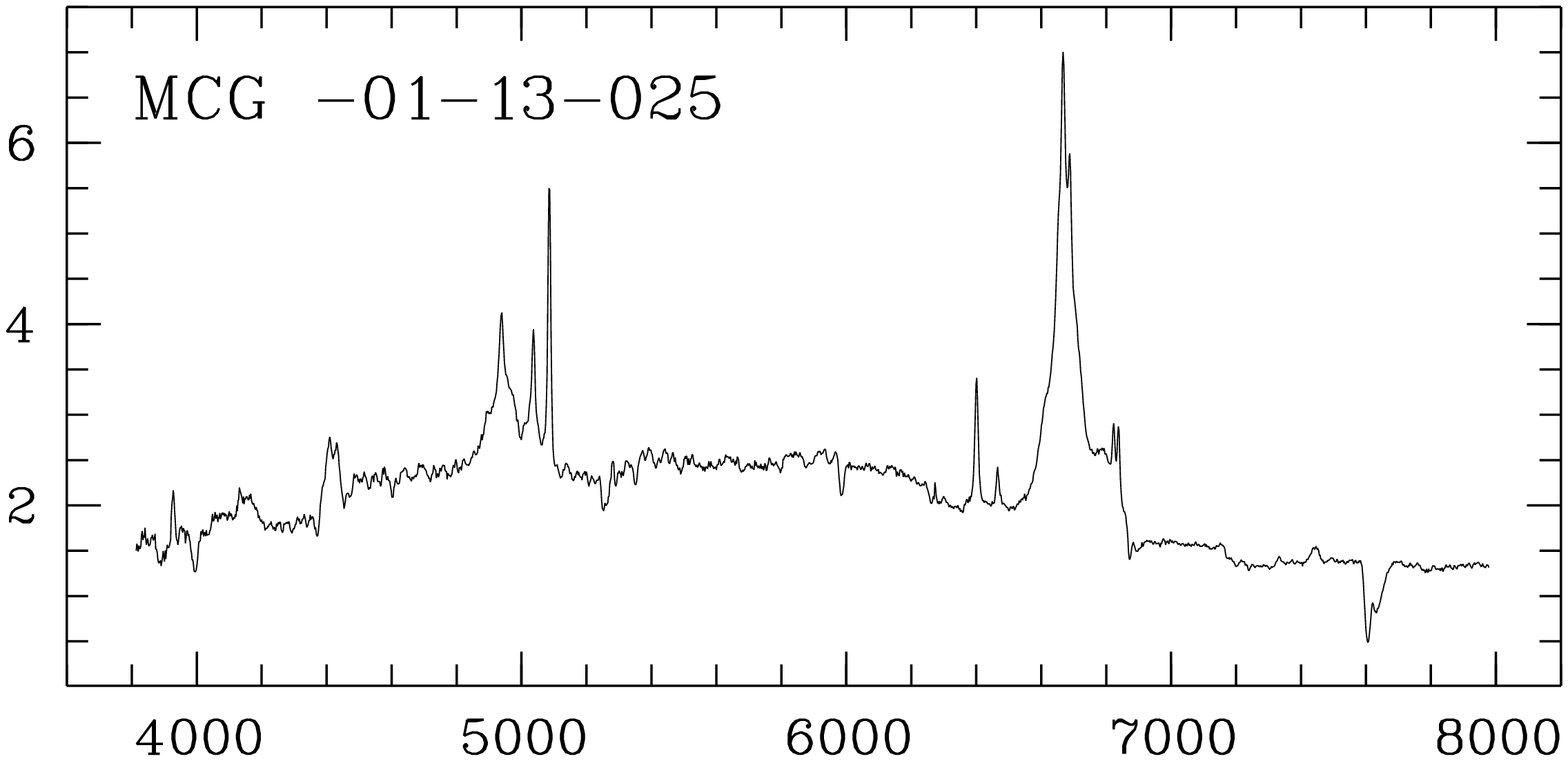,width=88mm,height=35mm,angle=270,clip=}\hfill
        \psfig{figure=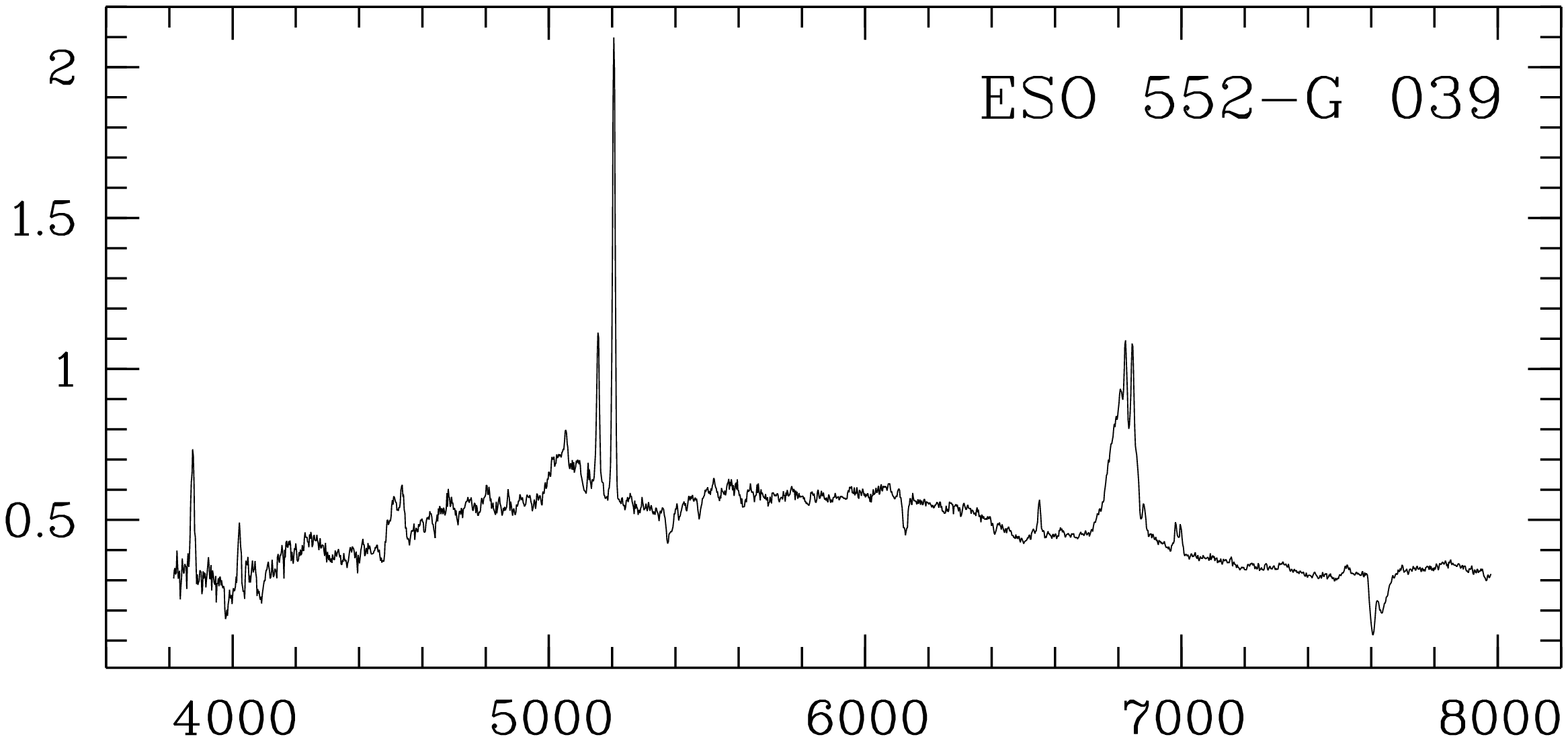,width=88mm,height=35mm,angle=270,clip=}}
{\bf Fig. 3.} continued
\end{figure*}
\begin{figure*}
  \hbox{\psfig{figure=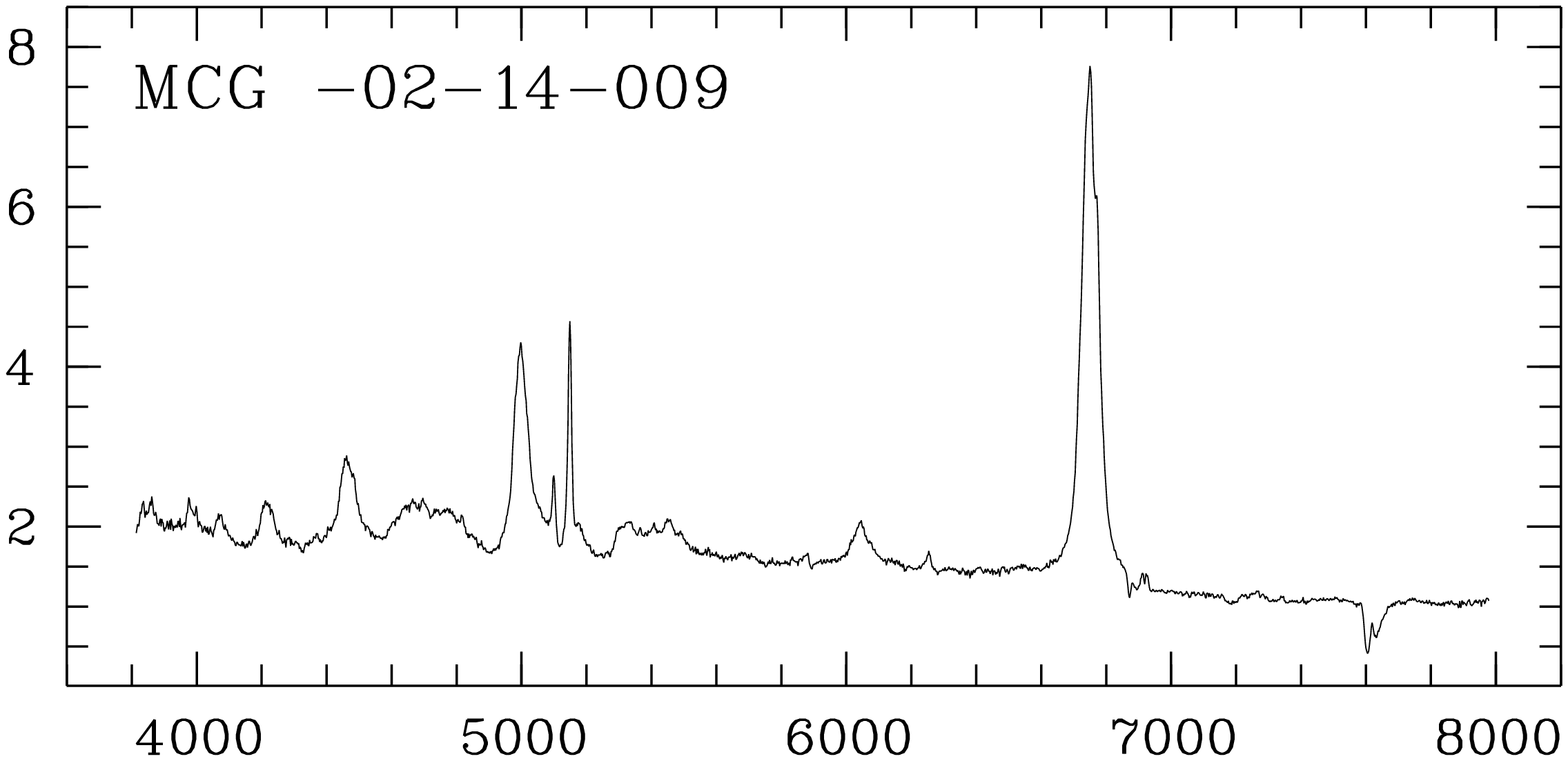,width=88mm,height=35mm,angle=270,clip=}\hfill
        \psfig{figure=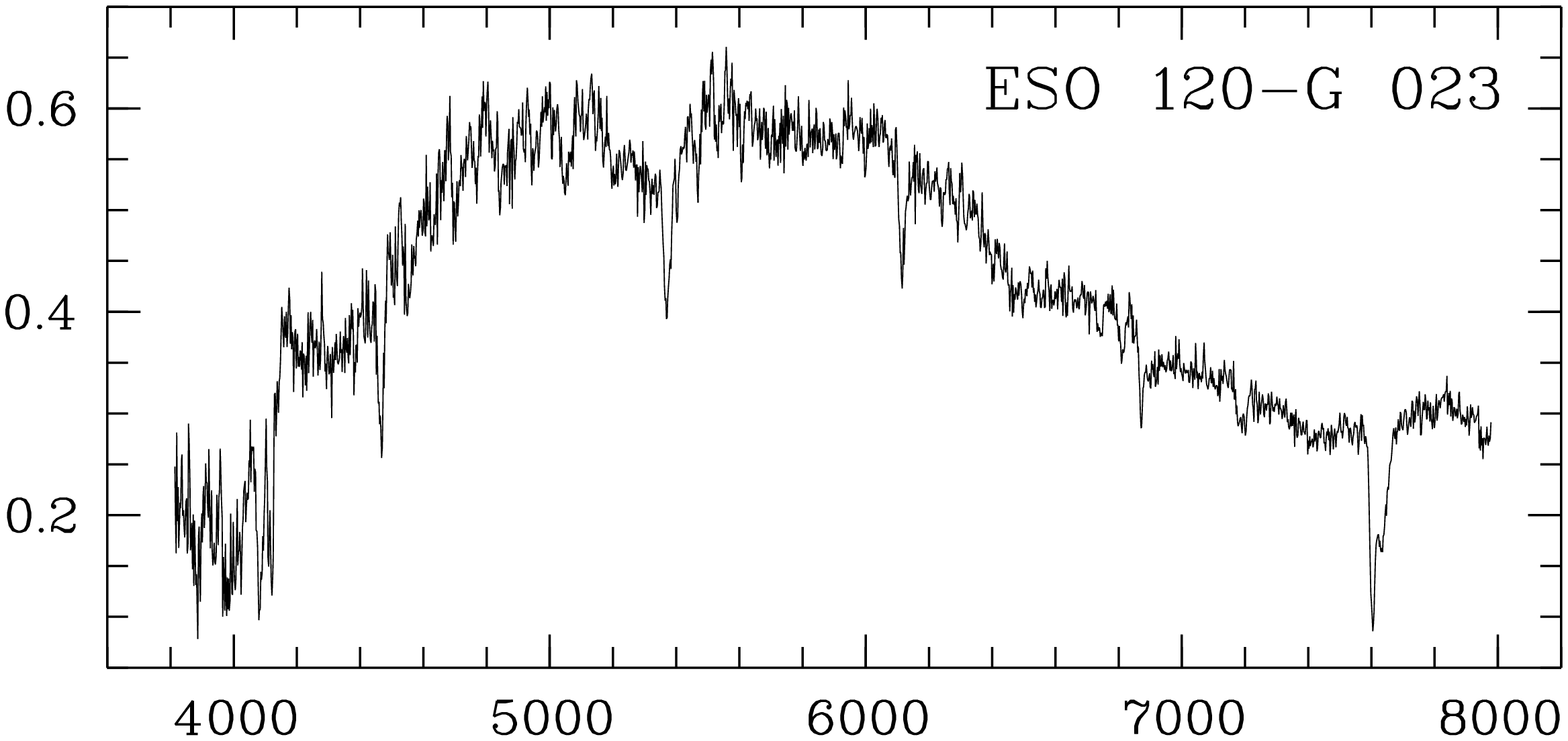,width=88mm,height=35mm,angle=270,clip=}}
  \hbox{\psfig{figure=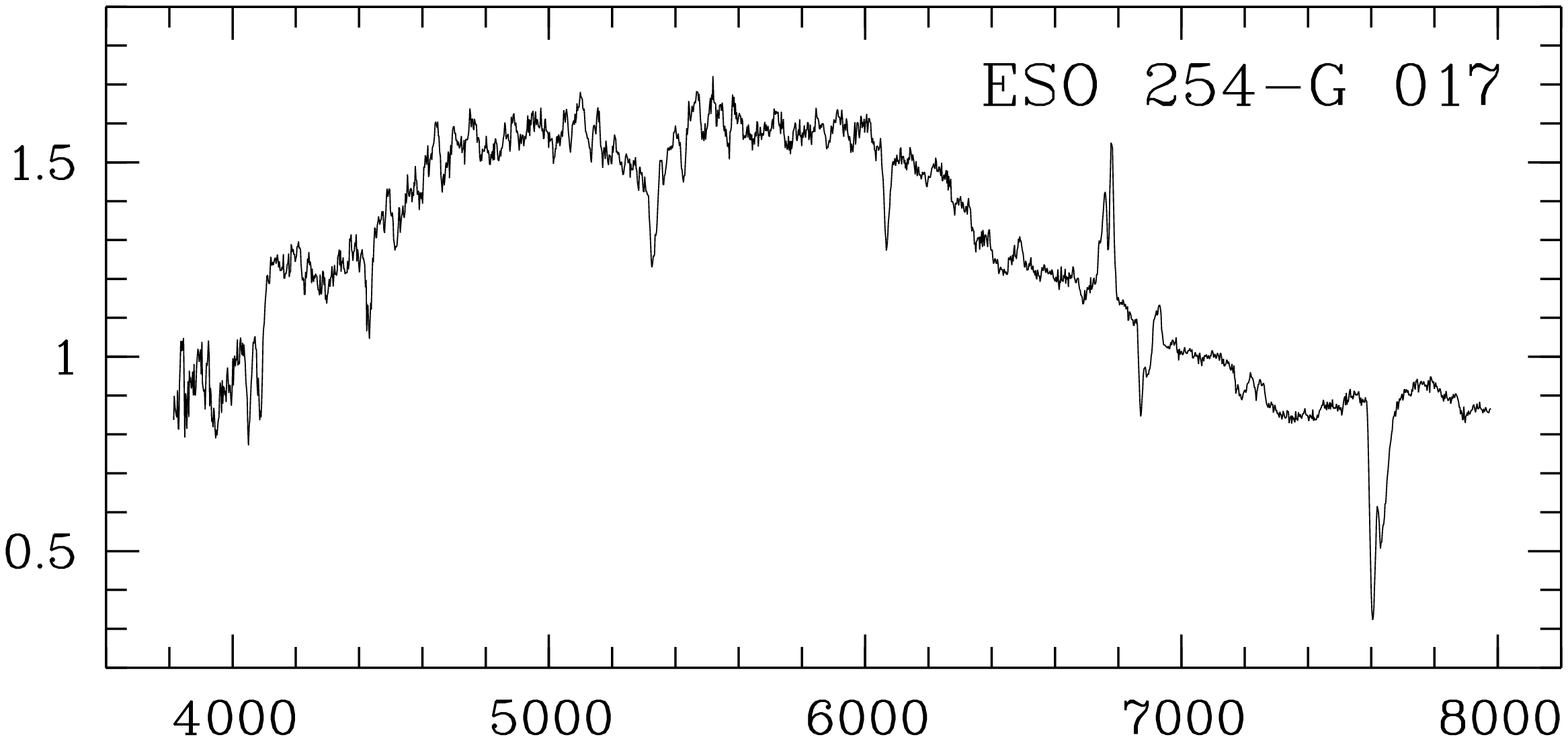,width=88mm,height=35mm,angle=270,clip=}\hfill
        \psfig{figure=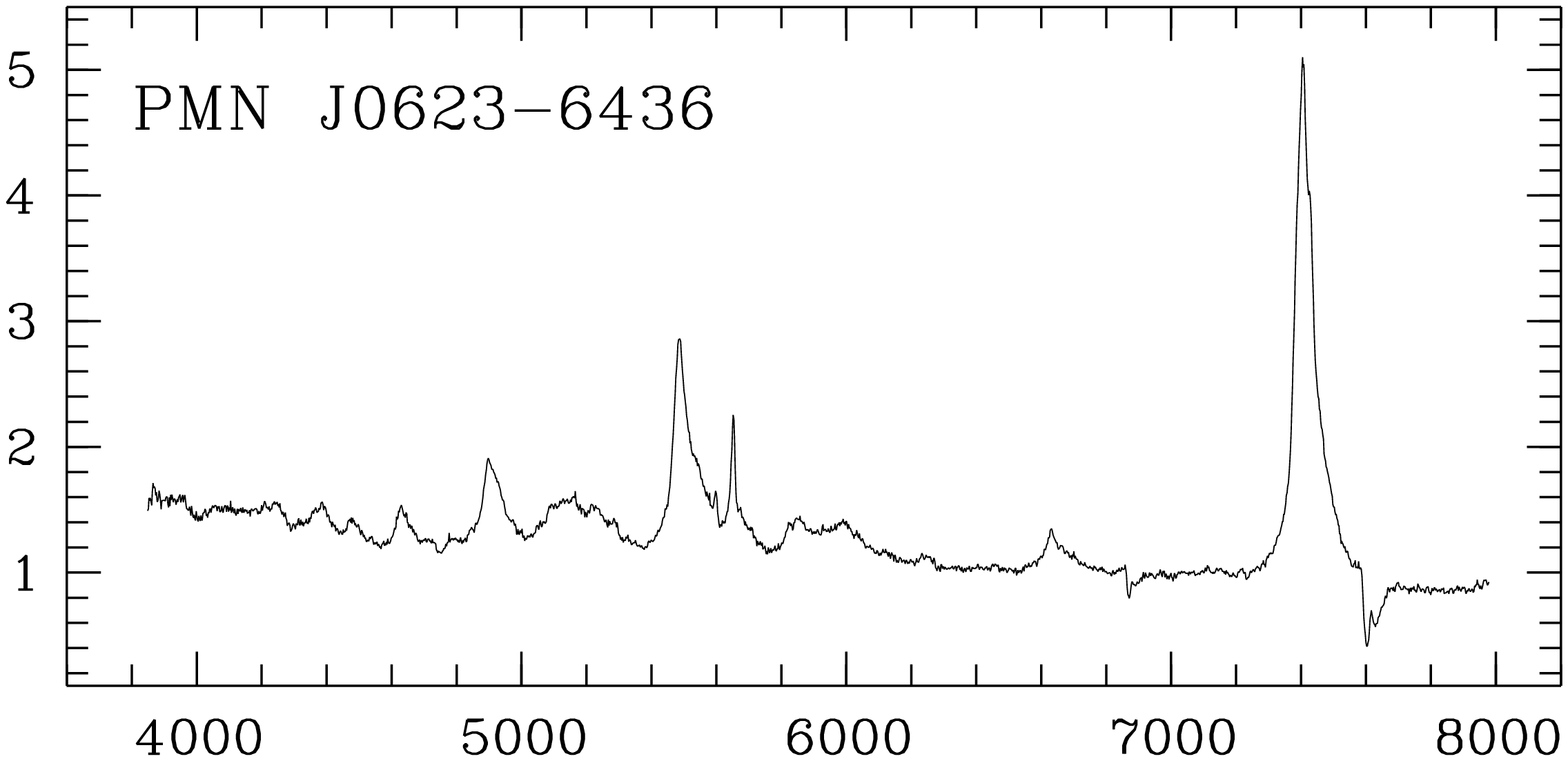,width=88mm,height=35mm,angle=270,clip=}}
  \hbox{\psfig{figure=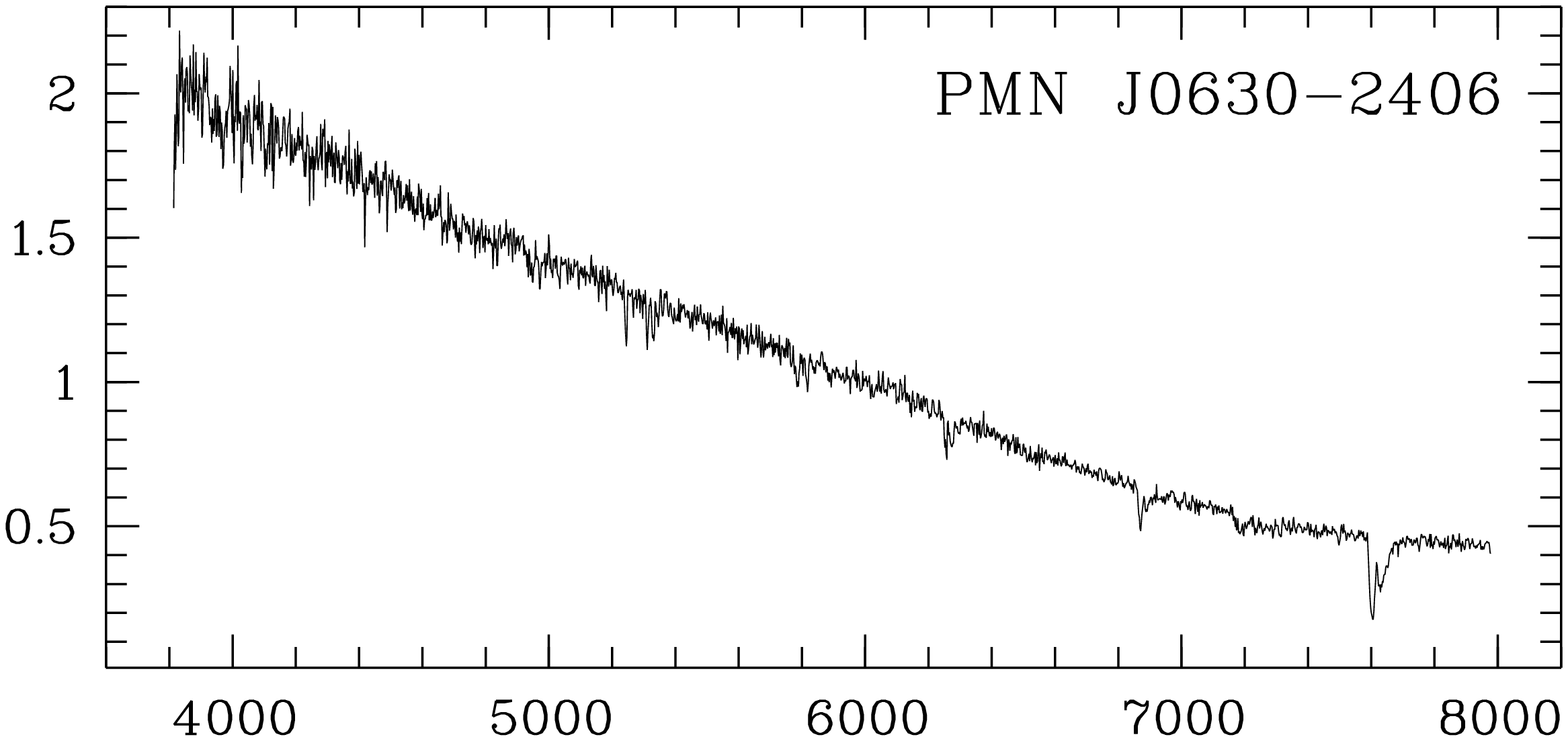,width=88mm,height=35mm,angle=270,clip=}\hfill
        \psfig{figure=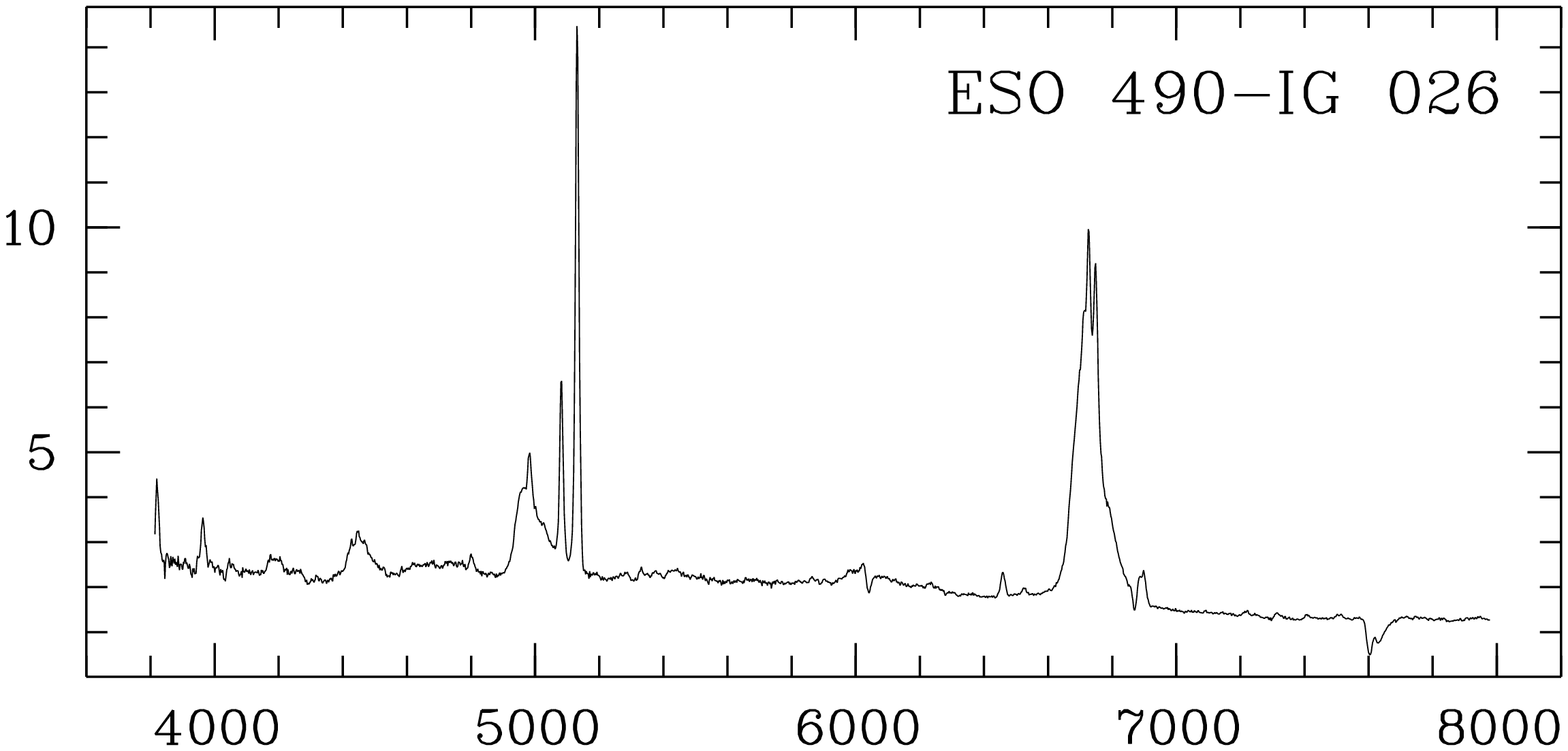,width=88mm,height=35mm,angle=270,clip=}}
  \hbox{\psfig{figure=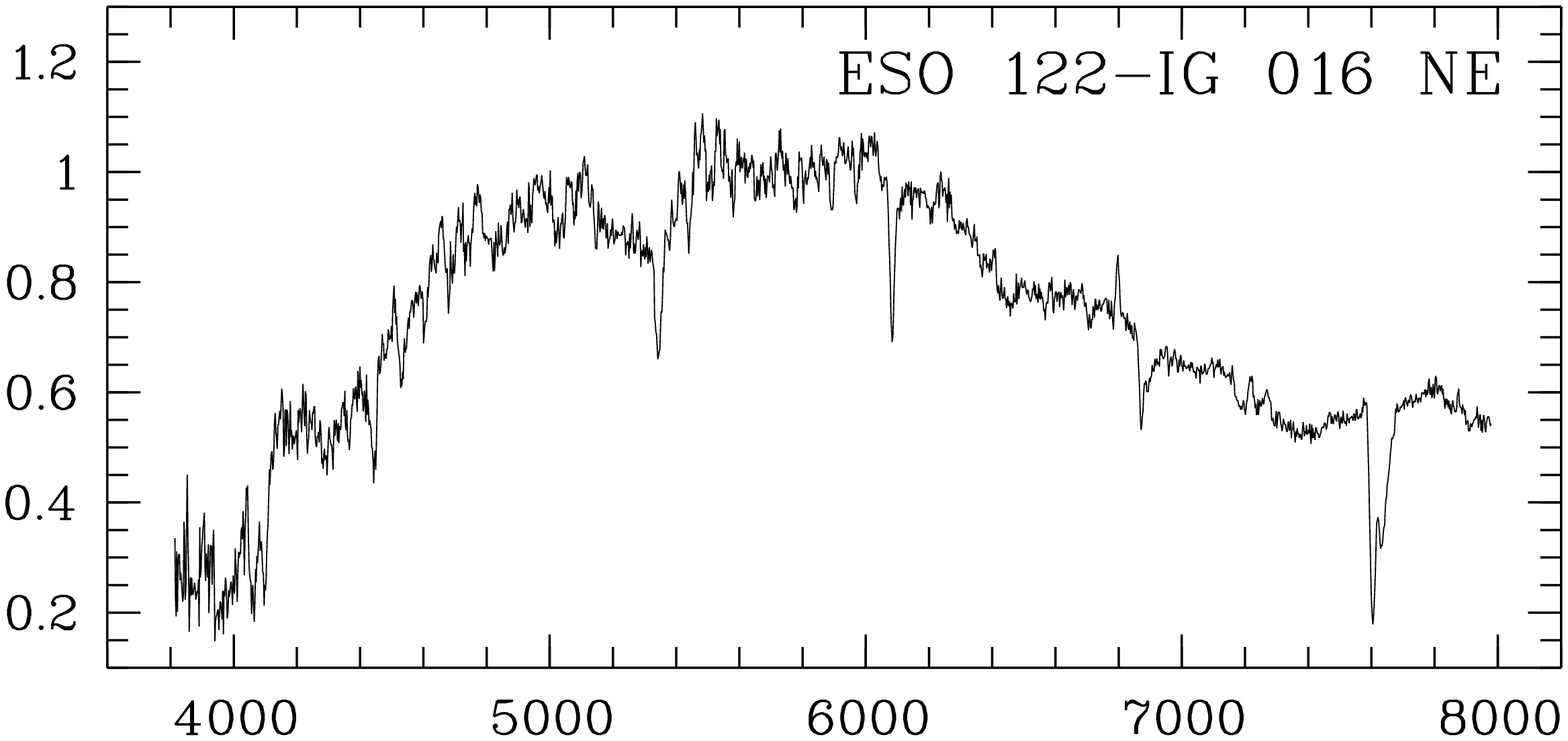,width=88mm,height=35mm,angle=270,clip=}\hfill
        \psfig{figure=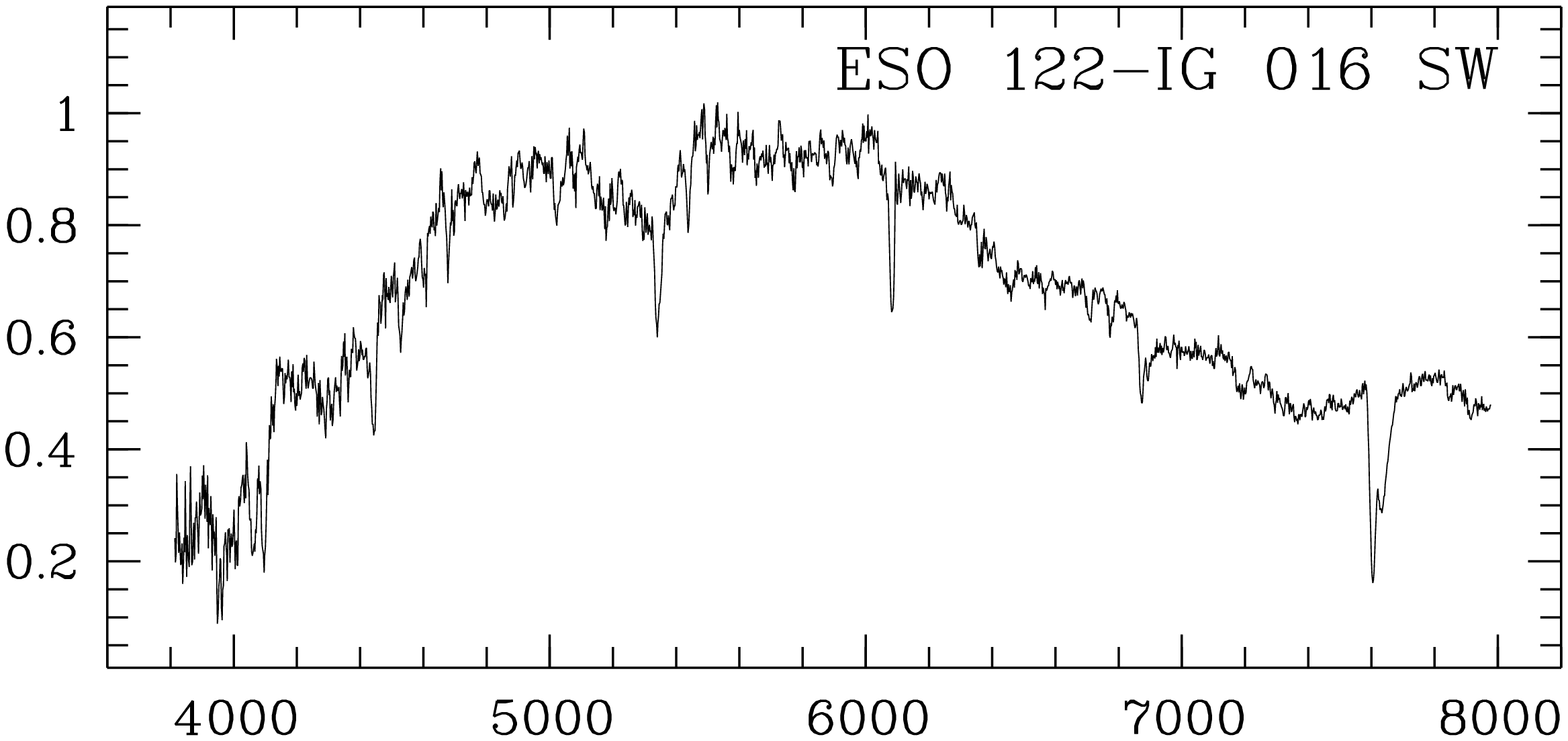,width=88mm,height=35mm,angle=270,clip=}}
  \hbox{\psfig{figure=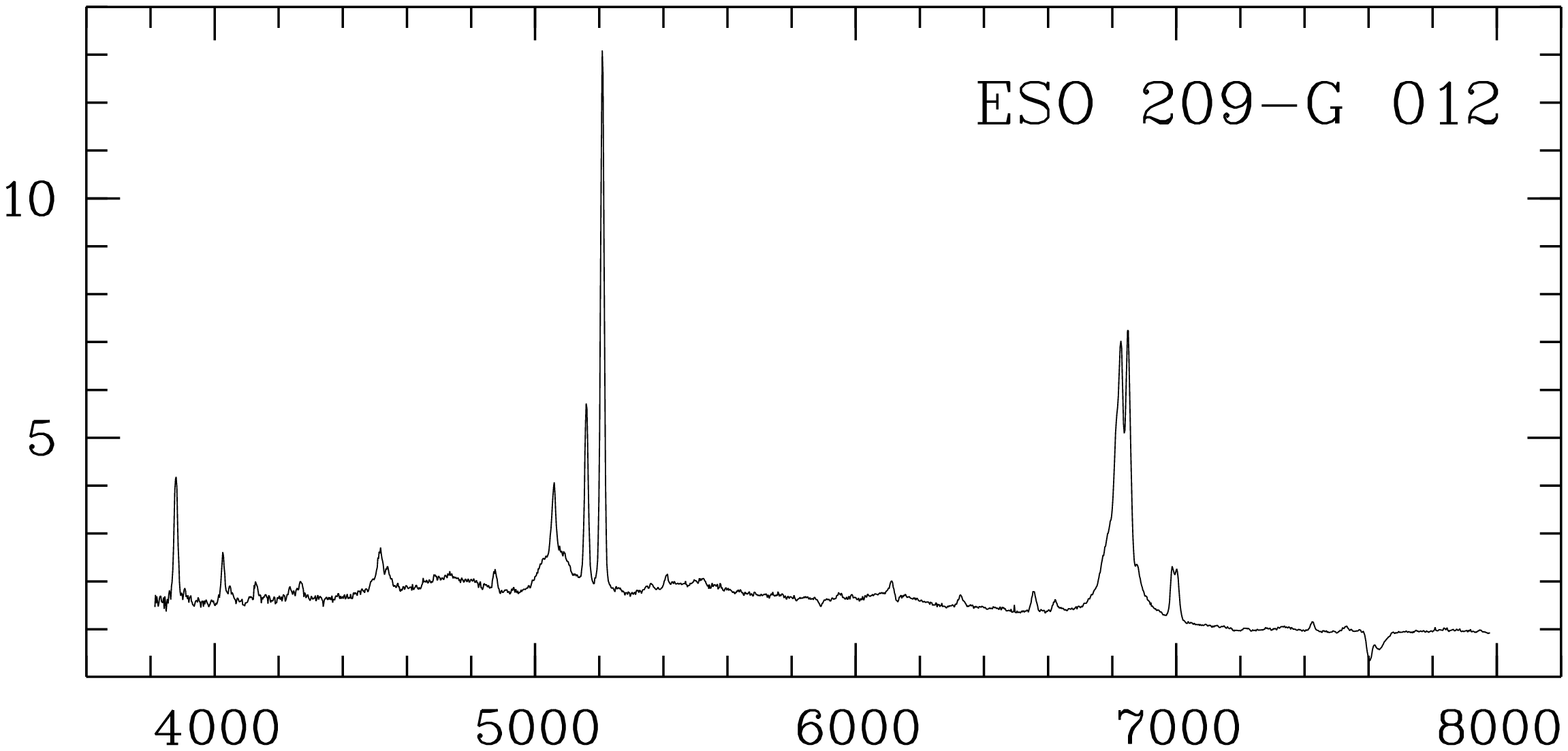,width=88mm,height=35mm,angle=270,clip=}\hfill
        \psfig{figure=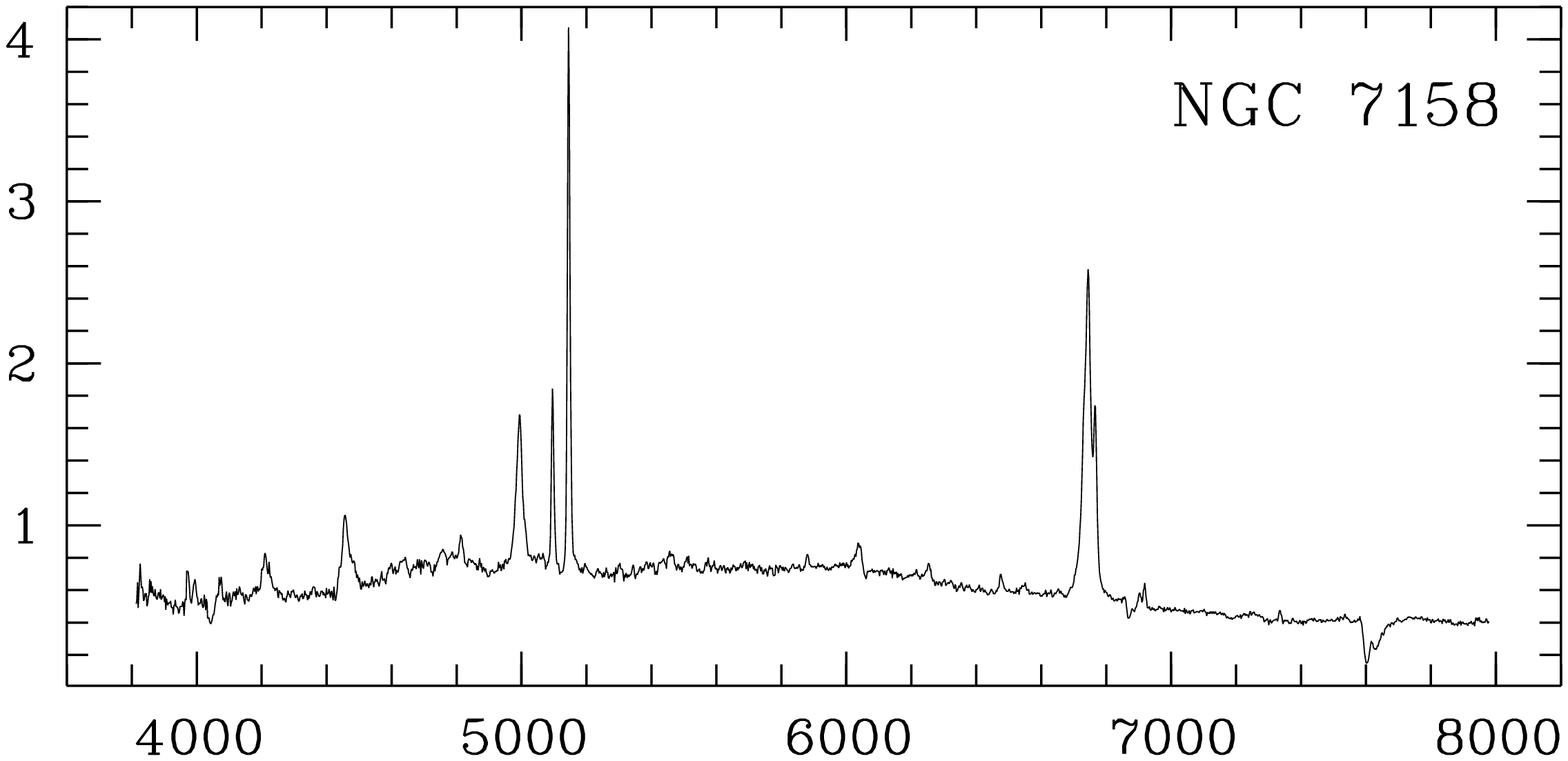,width=88mm,height=35mm,angle=270,clip=}}
  \hbox{\psfig{figure=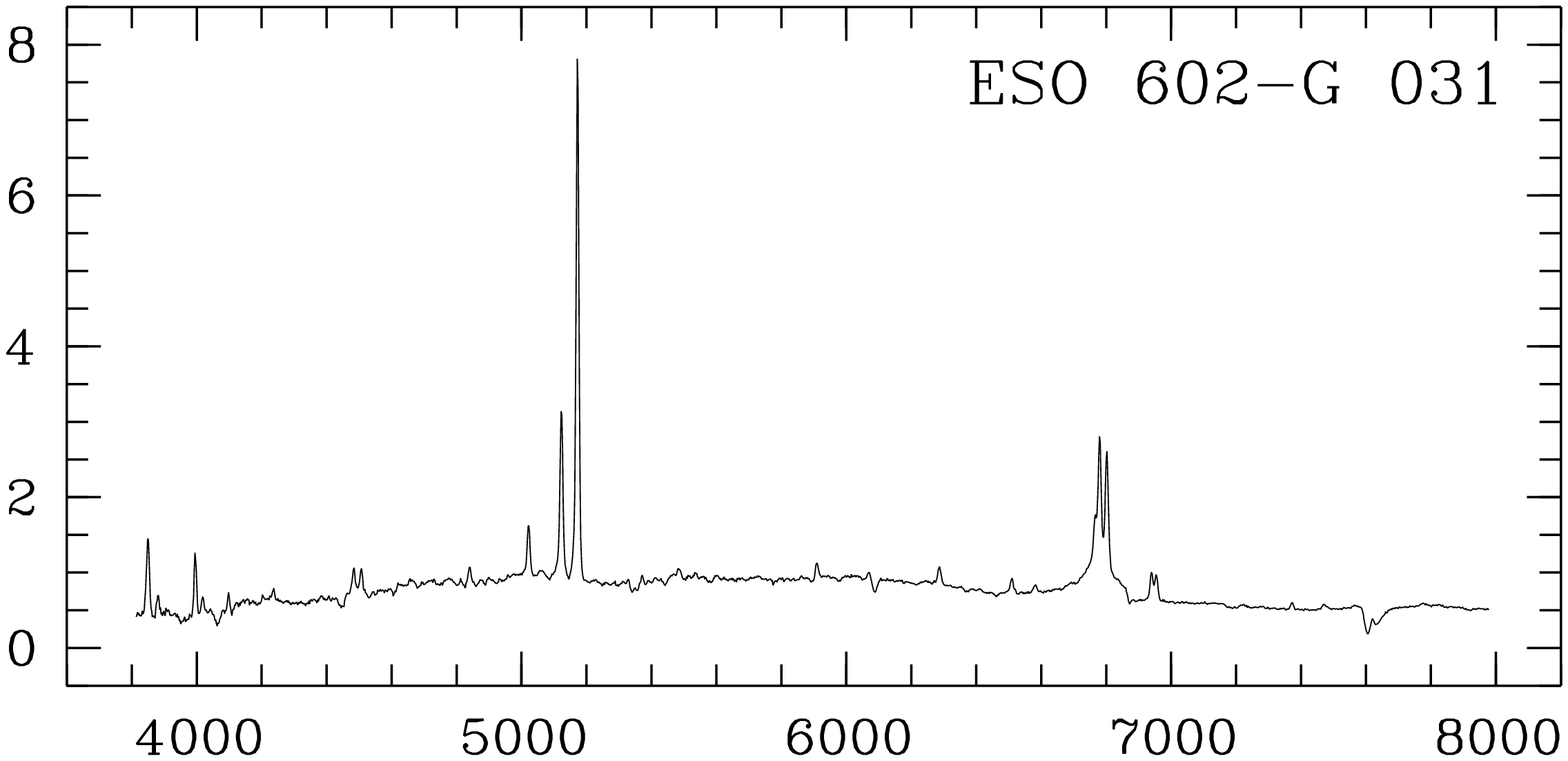,width=88mm,height=35mm,angle=270,clip=}\hfill
        \psfig{figure=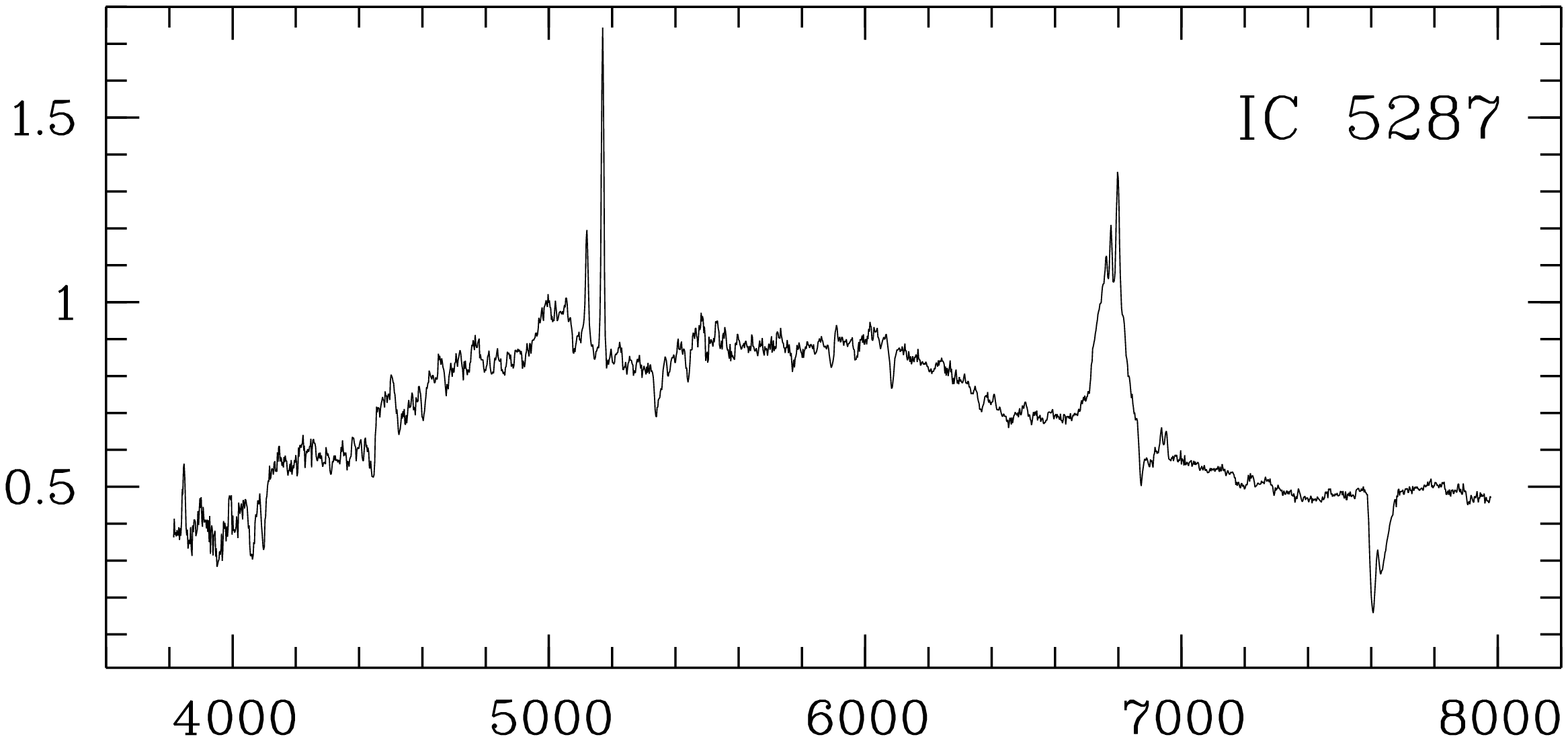,width=88mm,height=35mm,angle=270,clip=}}
{\bf Fig. 3.} continued
\end{figure*}
\begin{figure*}
  \hbox{\psfig{figure=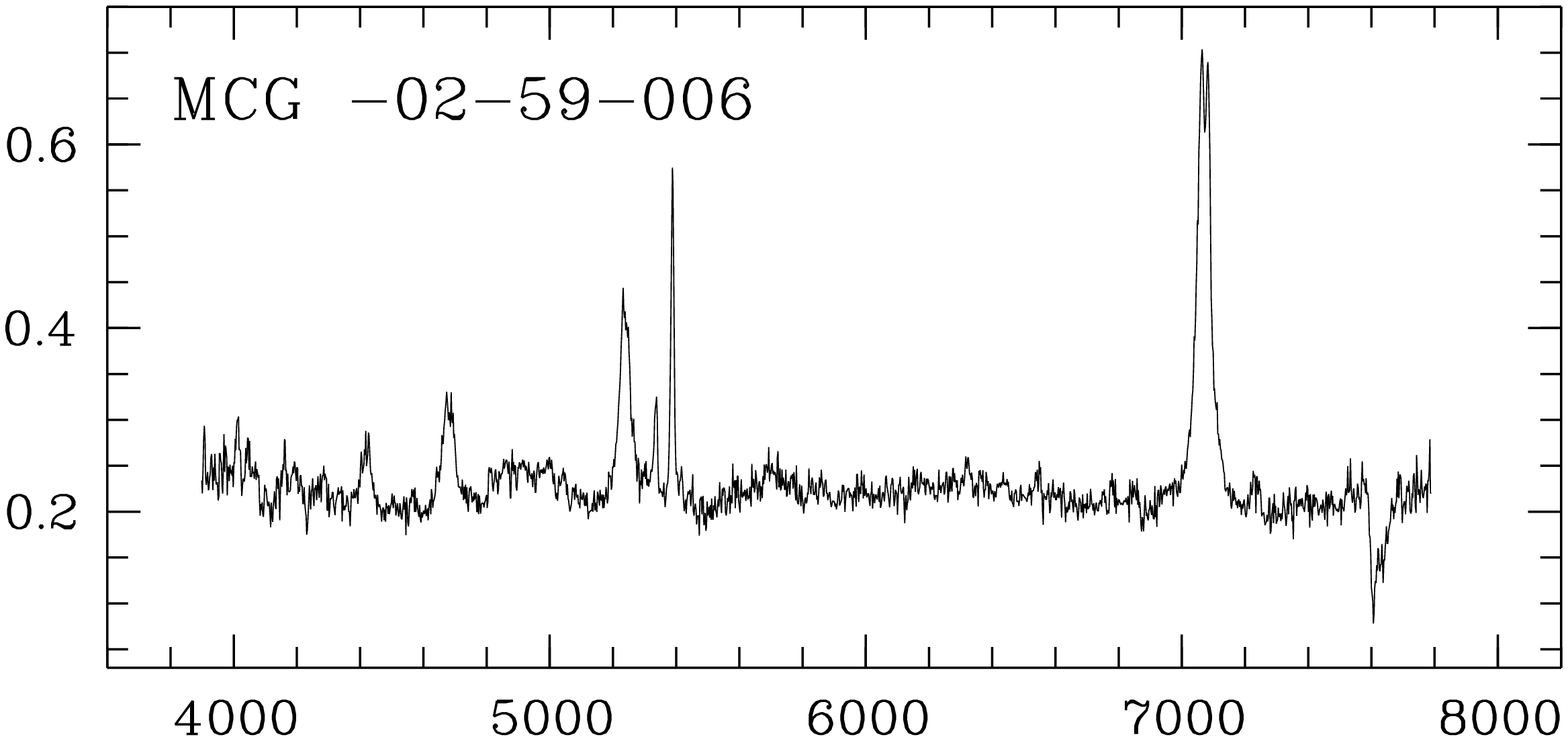,width=88mm,height=35mm,angle=270,clip=}\hfill
        \psfig{figure=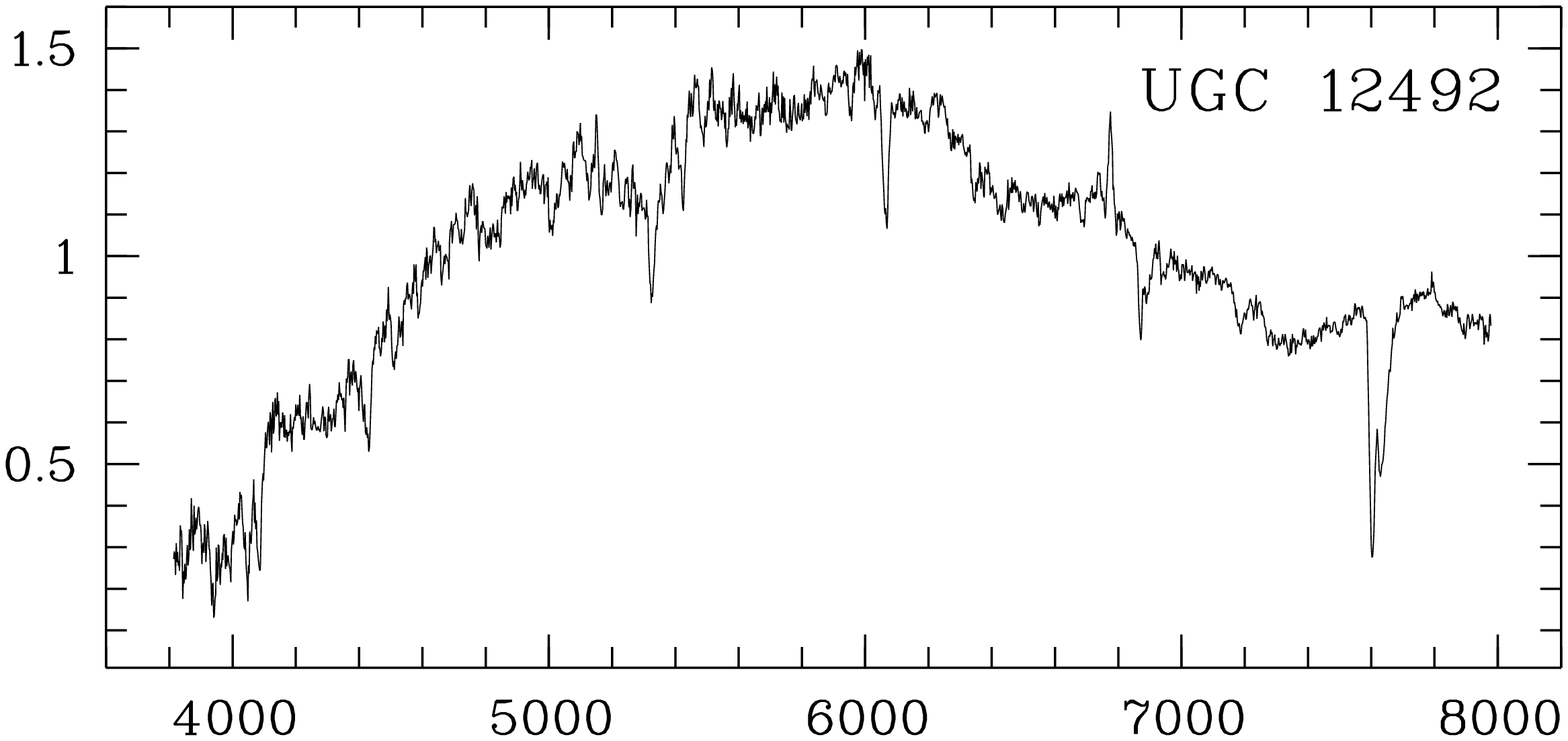,width=88mm,height=35mm,angle=270,clip=}}
{\bf Fig. 3.} continued
\end{figure*}

Redshifts and line parameters were determined by
fitting line complexes ([N\,{\sc ii}], H$_{\alpha}$ 
narrow and broad; [O\,{\sc iii}], H$_{\beta}$ narrow and broad, [S\,{\sc ii}]) and
single lines. Only the narrow Balmer line results have been used for the 
diagnostic line ratios.
For galaxies with strong underlying stellar continua we
determined redshifts from absorption lines.
With the best accuracy ($\pm 80$ km s$^{-1}$) of all spectra the redshift of 
the SW nucleus of ESO~122-~IG~016 was measured via Mg\,{\sc i $\lambda$5175} and 
Na\,{\sc d $\lambda$5890, 5896}, 
and G band {$\lambda$4304} absorption lines.
In a second step all spectra were normalized and rebinned to a logarithmic 
wavelength scale. Then the redshift was determined via cross-correlation 
with ESO~122-~IG~016 SW.
The radial velocities were converted for the Earths motion into a 
heliocentric system using a MIDAS routine with a code given by 
Stumpff (1980).

\section{Results}
The general information for our identification objects in Table 2 is
completed by redshift information (col. 8, summary from redshifts 
from Table 3 as described below) morphological and AGN type
(col. 9 and 10) and comments concerning the identification (col. 11). 
We determined a morphological type from our acquisition images and
compared them with the morphological type given in NED (available 
for most of the brighter galaxies of our sample). If our type differs from the
one given in NED we show our type marked by *. 
NED types are marked by a preceding $\Diamond$.
AGN type (col. 10) was determined from the optical spectra (see below).
Objects for which some AGN information was  
already available in NED are marked by *. In col. 11 we indicate X-ray
variability (X-var, see discussion below) and comment on group or
cluster environment, nearby stars and mark by "ID?", when the identification
of the X-ray source with the galaxy is questionable (see Section 5).
NED details and references
on sources marked by * in col. 9 to 11 are given in the notes
at the bottom of Table 2. 

In Table 3 we give results from the analysis of the optical spectra. 
If there are redshifts available in NED (col. 5, references col. 6) they are
within the errors comparable to our redshifts determined from emission 
(col. 3) and absorption lines (col. 4). If possible, 
diagnostic line ratios have been
determined from fits to the line complexes  and are given
in columns 7 to 10. If a broad H$_{\alpha}$ component was present we
determined the FWHM from a multicomponent fit (col. 11). For very asymmetric
broad line complexes the FWHM was estimated. In col. 12 we comment on
special features of the spectra.

From the spectral fits we classified the objects as AGN using classification
diagrams (Fig.~4) following Baldwin et al. (1981).
We also determined the AGN type (Table 2, col. 10).
Following the Osterbrock (1989) definition we classified the AGN in
Seyfert types according to the relative strength of narrow to broad components
of the H$_{\alpha}$ and H$_{\beta}$ emission lines; Seyfert 1 galaxies show
only broad  components; Seyfert 1.2 to 1.8 galaxies show decreasing but visible 
broad components and narrow components increasing in relative strength;
Seyfert 1.9 galaxies show only a weak broad component in H$_{\alpha}$ but no
broad H$_{\beta}$ component; Seyfert 2 galaxies show only narrow Balmer
lines. Narrow line Seyfert 1 galaxies (NLS1) are a peculiar group of Seyferts
where (1) the Balmer lines of hydrogen are only slightly broader
than the forbidden lines such as [O\,{\sc iii}], [N\,{\sc ii}]
and [S\,{\sc ii}]; (2) they often show the presence of emission lines
from Fe\,{\sc ii} (e.g. the optical multiplets centered at 4570 \AA,
5190 \AA\ and 5300 \AA) or higher ionization iron lines such as
[Fe\,{\sc vii}] $\lambda$6087 and [Fe\,{\sc x}] $\lambda$6375 (these lines are often
seen in Seyfert 1 galaxies but generally not in Seyfert 2 galaxies); and
(3) the ratio of [O\,{\sc iii}] $\lambda$5007 to H$_{\beta}$ is $<$ 3, a level which
Shuder \& Osterbrock (1981) found to discriminate well Seyfert 1s
from Seyfert 2s. The full-width at half-maximum (FWHM) of NLS1 hydrogen
Balmer lines is usually in the range $\approx$ 500--1500 km s$^{-1}$
(cf. Goodrich 1989 and Osterbrock \& Pogge 1985). 
LINERs (Low Ionization Nuclear Emission-line Region)
we define - following Ho (1997) - by [O\,{\sc i}]/H$\alpha > 0.17$, 
[N\,{\sc ii}]/H$\alpha > 0.6$, [S\,{\sc ii}]/H$\alpha > 0.4$ and
[O\,{\sc iii}]/H$_{\beta} > 3$. For some of our galaxies a separation 
between a Seyfert 2 or LINER nucleus is
not possible because [O\,{\sc iii}] and H$_{\beta}$ are not detected.
The galaxy nuclei however clearly have to be classified as active as their 
[S\,{\sc ii}]/H$\alpha$ and [N\,{\sc ii}]/H$\alpha$ ratios show.
For RX J023454.8-293425 the H$_{\alpha}$ line is shifted out of our 
spectral range. The redshift has been derived using Mg\,{\sc ii} $\lambda$2798, 
[Ne\,{\sc v}] $\lambda$3426, [O\,{\sc ii}] $\lambda$3727, 
[Ne\,{\sc iii}] $\lambda$3869, 3968
and H$\gamma$ lines. From the absolute optical B magnitude it clearly has to
be classified as a Quasar.
BL Lac candidates are 
identified from their featureless blue spectra. High f$_X$/f$_B$,
detection in the radio, and a starlike appearance in the optical 
strengthen the BL Lac nature.

We investigated the galaxies from 32 RASS correlations and 4 field sources detected
in the HRI follow up observations.
Two galaxies and one component of a galaxy with a double
nucleus show no emission lines that would indicate activity. We therefore
classify them as "non active".
The other sources are classified as Seyferts (Sy1 (5), Sy1.2 (7), Sy1.5 (3), Sy1.8 (5),
Sy1.9 (1), Sy2 (2)), NLS1 (1), LINER (3), Sy2/LINER (4), BL Lac (3), and QSO (1). 
 
The soft X-ray (0.1--2.4 keV), 
the far-infrared ($\rm 40-120 \mu m$), and
radio (1.4 GHz) 
fluxes and luminosities are listed in Table 4. A Hubble constant of
H$_0$ = 75 km s$^{-1}$ Mpc$^{-1}$ and cosmological deceleration parameter 
of q$_0$ = 0.5 were used throughout. For the objects (col. 1) and
redshifts from Table 2 we calculated the distance (col. 3). 

To compute the soft X-ray (0.1--2.4 keV) 
energy flux (col. 6) from the count rate we assumed a power-law spectrum 
\begin{equation}
f_E\ dE ~\propto~ E^{-\Gamma+1}~ dE~,
\end{equation}
where $f_{E}\ dE$ is the galaxy's energy flux between photon energies
$E$ and $E+dE$. We assume a fixed photon spectral index $\Gamma=2.3$,
which is the typical value found for extragalactic objects with ROSAT
(cf. Hasinger et al. 1991, Walter \& Fink 1993), and
an absorbing column density fixed at the individual
Galactic hydrogen value N$_{\rm H gal}$ along the 
line of sight (col. 2, Dickey \& Lockman 1990). With this procedure we 
do not correct for absorption of the X-rays within the galaxy.
Our X-ray fluxes derived with this method therefore have to be understood 
as lower limits for the intrinsic fluxes if the X-rays are emitted from 
active nuclei within the galaxies. This absorption effect could be 
corrected for if one would be able to fit spectra to the individual X-ray 
sources. While such fits are possible for RASS sources 
with more than 500 counts the few photons (typically $<100$ counts for 
our sample) do not allow such a procedure.  

For the 12 sources with follow-up ROSAT HRI observations
we used the HRI count rate to determine the X-ray flux. For 3 sources
PSPC and HRI determined fluxes correspond within 20\% (ESO~15-~IG~011, 
PMN~J0630-2406, ESO~209-~G~012), for 7 sources the HRI flux is significantly
lower (NGC~427, ESO~080-~G~005, ESO~416-~G~002, ESO~254-~G~017,
PMN~J0623-6436, ESO~490-~IG~026, and ESO~602-~G~031 by a factor of 
0.6, 0.4, 0.4, 0.7, 0.7, 0.6, and 0.2, respectively), and for 2 sources higher 
(ESO~113-~G~010 and ESO~552-~G~039 by 2.2 and 2.0). Several
explanations can be put forward for the derived flux differences.
The HRI rate could be lower, because in the RASS analysis additional unresolved 
emission is attributed to the source that is resolved by the HRI. Such additional
emission could result from an extended galaxy
halo or from other nearby X-ray sources or from 
diffuse X-ray emission due to galaxy cluster membership of our target. 
A careful check of the PSPC and HRI images most certainly rules out
this possibility for all 5 sources. On the other hand HRI and PSPC fluxes 
could differ because we did not assume the appropriate 
spectral model for the flux determination. While this effect could
explain reduced HRI fluxes by up to a factor of two it can not
explain the extreme variations and especially the strong HRI flux increase
detected in two sources. The most probable explanation for the flux 
differences therefore is true source variability. This explanation is also
consistent with the fact that we see more often a flux decrease than an increase from
RASS to HRI observations because most of the sources lie close to our detection 
threshold for the RASS. X-ray flux variability on these time scales is a
common feature for active galactic nuclei (see e.g. Ulrich et al. 1997).

For two fields the HRI observations have been split
in two observation intervals separated by half a year (see Table~\ref{HRI}).
This allowed to investigate
variability on this time scale using the same detector and therefore
avoiding the cross calibration problems mentioned above. The HRI flux of
ESO~113-~G~010 decreased by a factor of 3 between the observations. While
the average HRI flux was a factor of 2.2 above the PSPC (see above)
during December 1995 the HRI flux was higher by even a factor of 3.6 
and dropped till June 1996, when it was however still higher than the 
PSPC flux by 
a factor of 1.3. The HRI flux of NGC~427 did not vary by more than 15\%
between the observations in January and June 1996 but stayed at the  
lower intensity with respect to the PSPC observation.
The BL Lac candidate in the same field (RX~J011219.5-320140) however
showed an increase in flux between the observations by a factor of
1.5. 

The optical B magnitude and the log of the monochromatic flux at
the optical band at 4400 \AA (using the relation given by Allen (1976), p. 174)  
are listed in columns 4 and 5, respectively. The absolute B-magnitude is given in column 9.
Column 7 gives the integrated 40-120 $\mu m$ flux and column 8 the monochromatic 1.4 GHz radio flux
from the NVSS survey (Condon et al. 1996) or (if not available) the 4.85 GHz radio flux from NED,
if available.

The total far-infrared (40-120$\mu$m) fluxes, $f_{\rm FIR}$, were computed 
with the formula of Helou et al. (1985) from the IRAS 60 $\mu$m and 100 $\mu$m 
band fluxes taken from NED: 
$$
f_{\rm FIR} =1.26\times 10^{-11}  (2.58 f_{60}+f_{100})  
\rm \ erg\ cm^{-2}\ s^{-1},\eqno(2)
$$
where $f_{60}$ and $f_{100}$ are given in Jansky.
The soft X-ray, optical, far-infrared and radio fluxes were converted to luminosities using
equation (7) of Schmidt \& Green (1986) (col. 10 to 13). Column 14 to 17 give flux 
ratios between different bands.
Flux ratios have been plotted as functions of
absolute B magnitude and far infrared flux in Fig.~5.

\section{Discussion}
\begin{figure*}
  \begin{minipage}{6cm}
     \psfig{figure=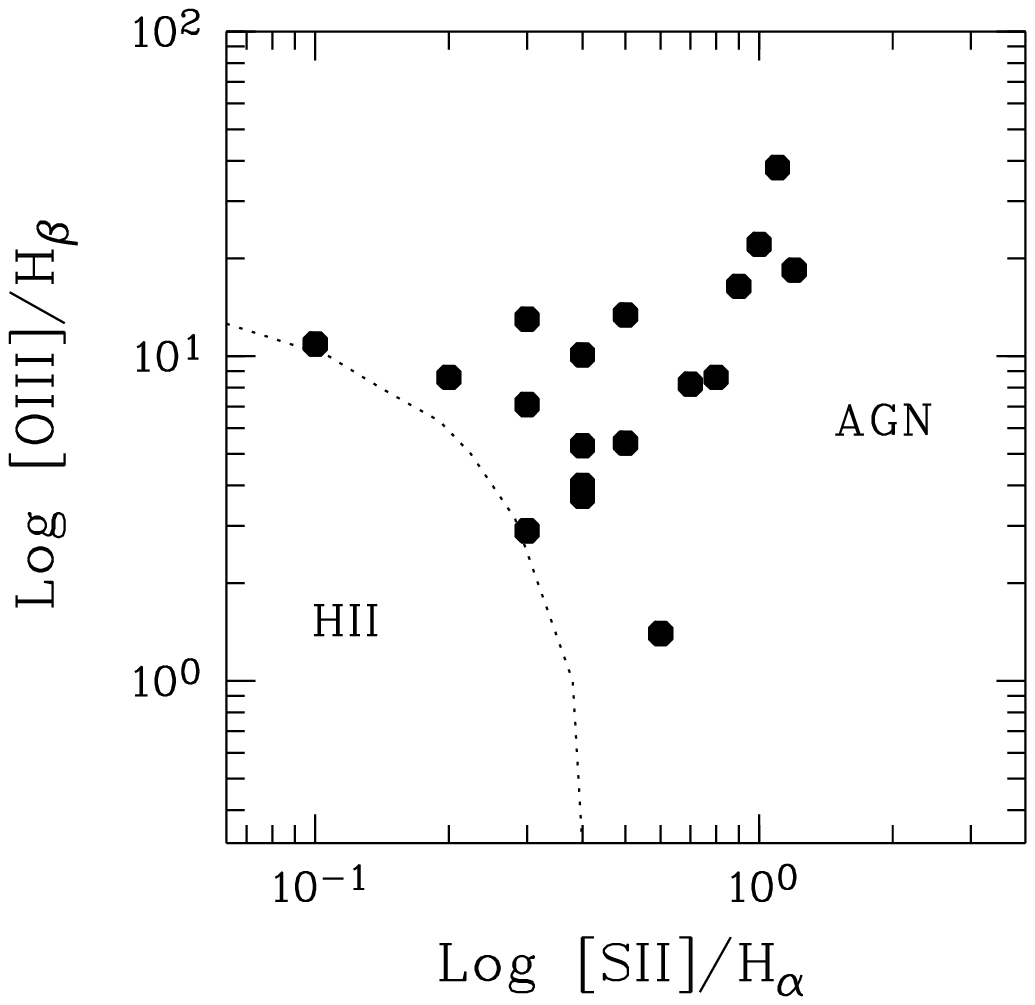,width=6cm,clip=}
  \end{minipage}
  \begin{minipage}{6cm}
     \psfig{figure=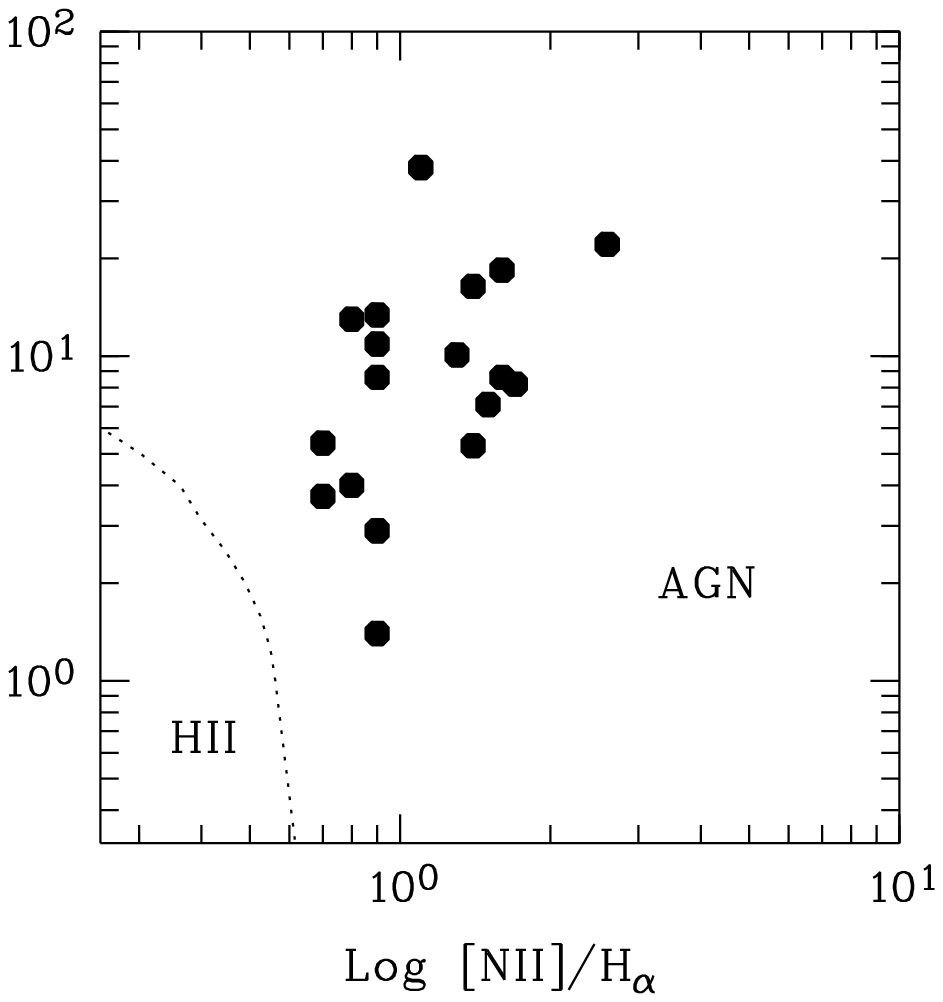,width=6cm,clip=}
  \end{minipage}
  \begin{minipage}{6cm}
     \psfig{figure=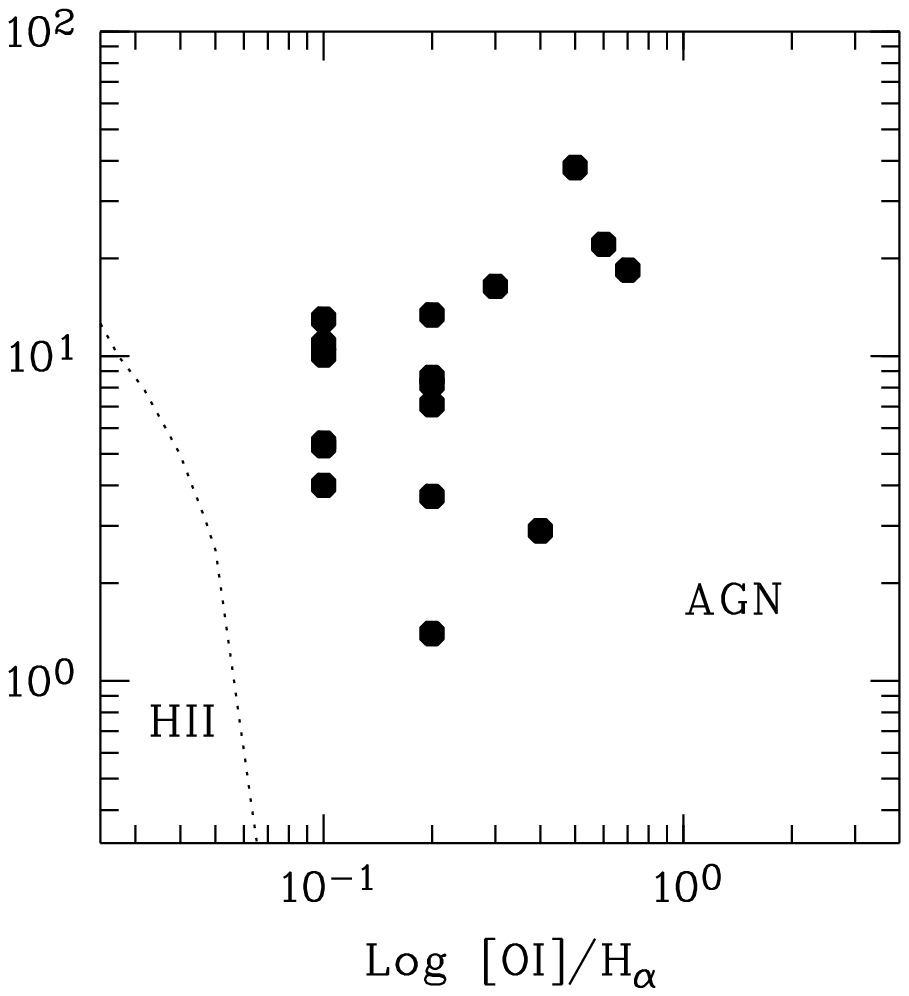,width=6cm,clip=}
  \end{minipage}
      \caption{Diagnostic diagrams for emission-line galaxies.
               The dotted curve is the best empirical dividing line between
               H~II regions and AGN (Osterbrock 1989). 
              }
\end{figure*}
\begin{figure*}
  \begin{minipage}{6cm}
     \psfig{figure=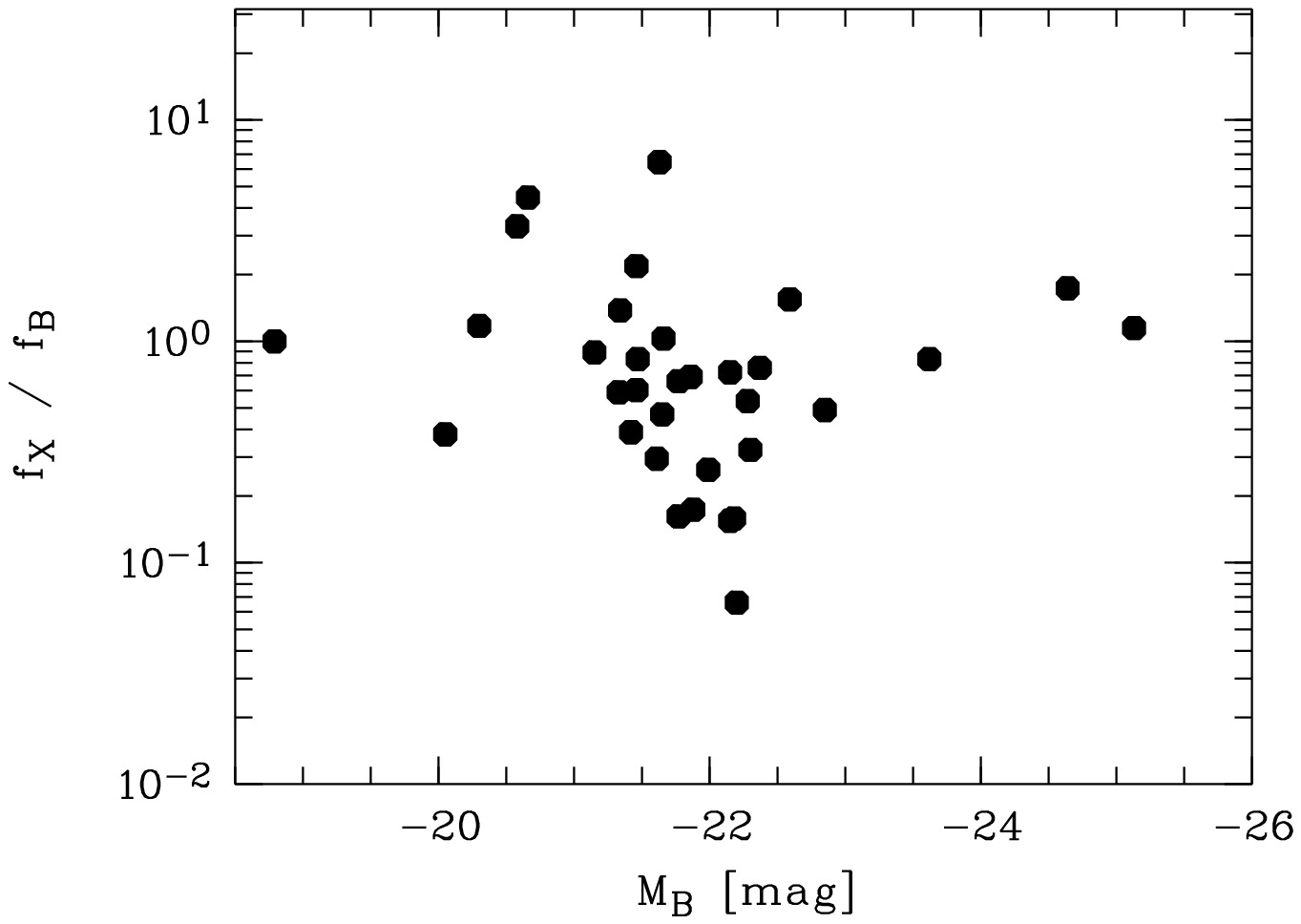,width=6cm,clip=}
  \end{minipage}
  \begin{minipage}{6cm}
     \psfig{figure=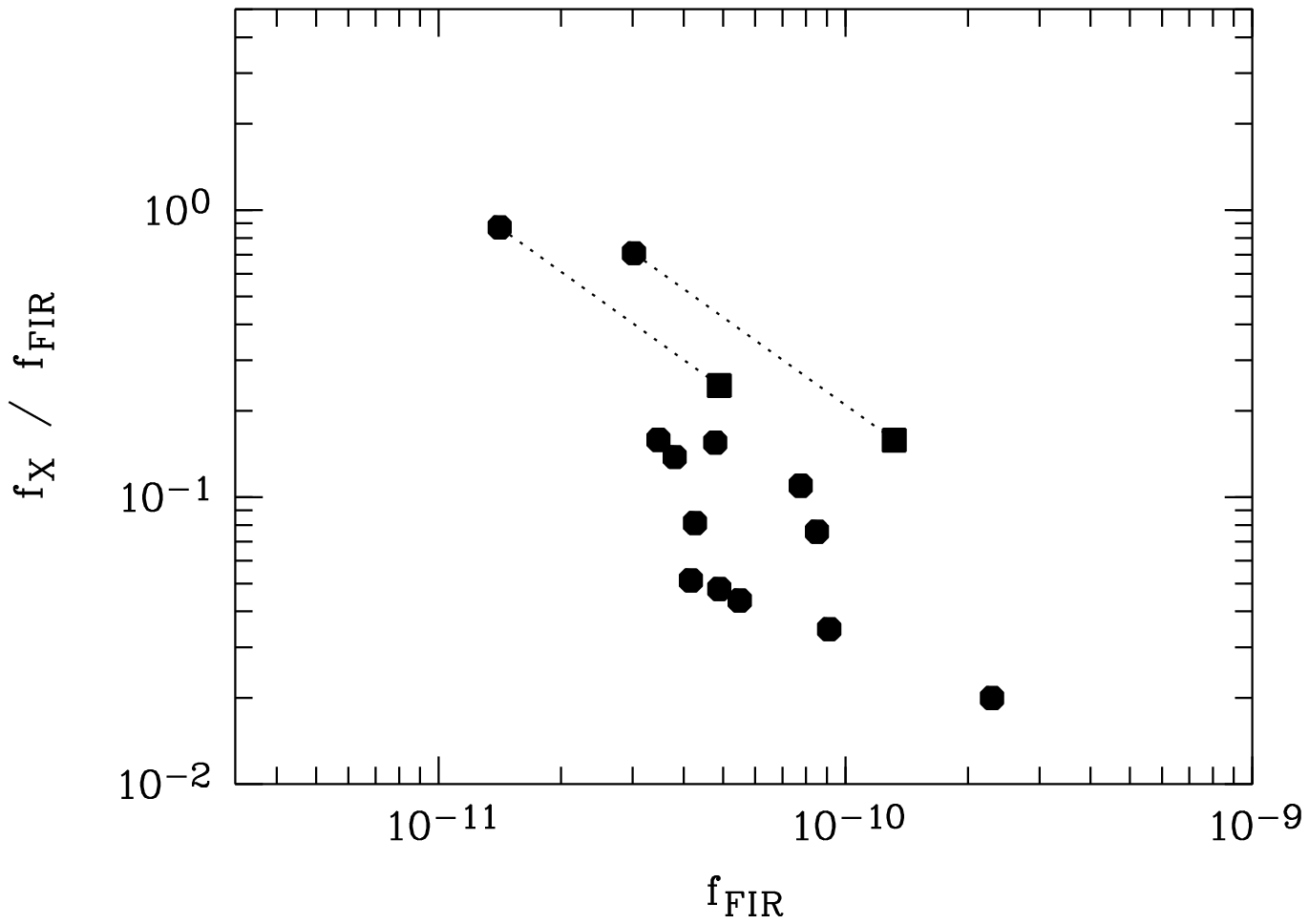,width=6cm,clip=}
  \end{minipage}
  \begin{minipage}{6cm}
     \psfig{figure=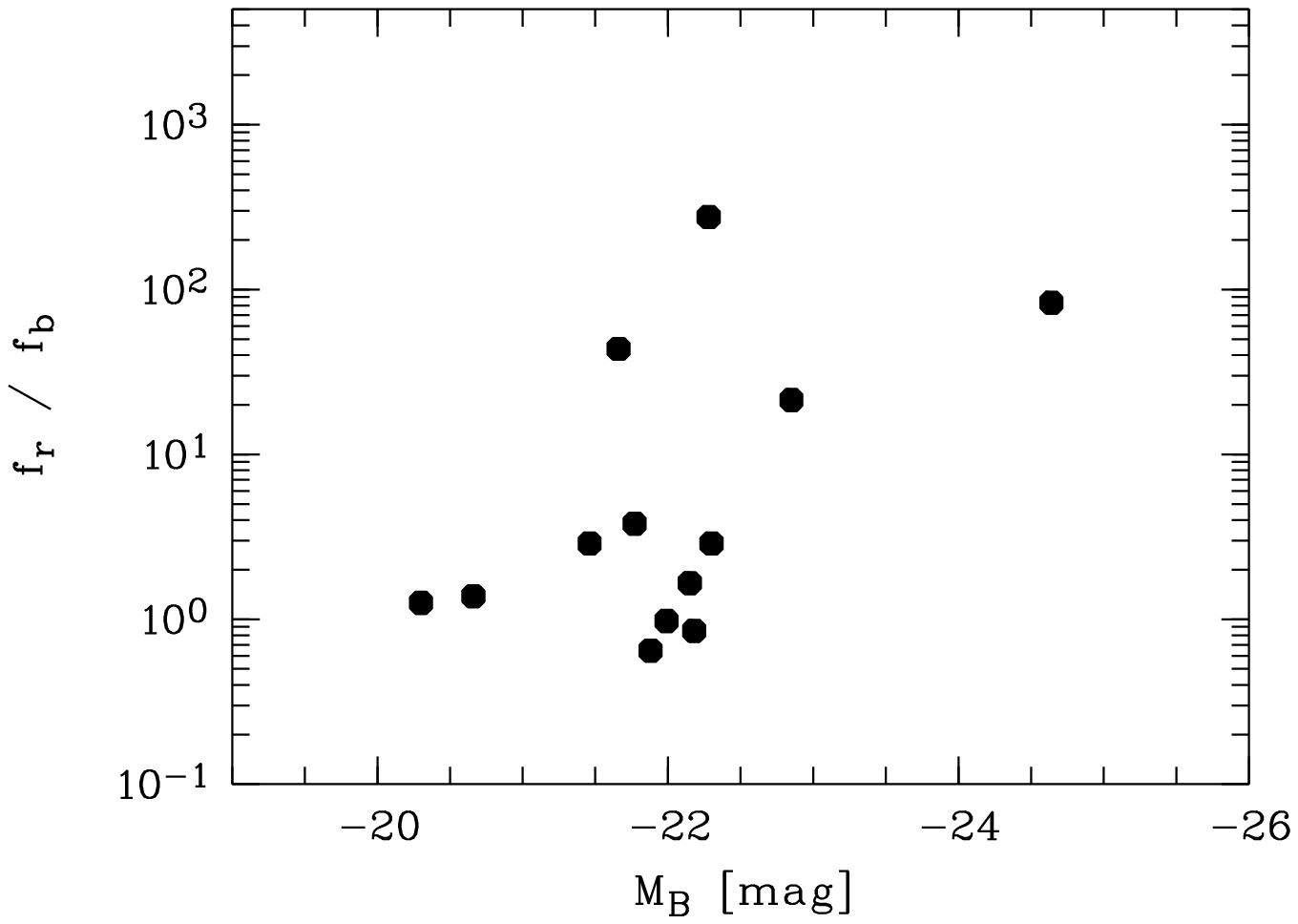,width=6cm,clip=}
  \end{minipage}
      \caption{Flux ratios as functions of absolute B magnitude or far infrared
               flux, respectively
               }
\end{figure*}
As shown in the last section the X-ray selected galaxies which were 
followed up optically cover redshifts from 0.014 to 0.13. Most of them
turn out to harvest an active nucleus or even
merging nuclei where at least one component is active. 
It is, however, not clear yet if these active nuclei, that we have classified 
using optical, radio and infrared data, really are the origin of the X-rays.

We have several ways to attack this problem and can propose
solutions for individual galaxies:

\begin{enumerate}
\item We can use the extent information from the RASS catalog. If a source
is extended the major part of the X-ray emission will not originate 
from the nucleus. This is the case for AM~0426-625~S, ESO~120-~G~017, and
ESO~122-~IG~016. From the optical spectra their nuclei have been classified 
as Sy2/LINER, non active, and LINER+non active, respectively. From this 
type of nucleus X-ray emission in the ROSAT band should be highly absorbed.
The extended X-rays may originate from hot gas in the early type galaxies
or in the groups they are member of.

\item An HRI detection of a source which is centered on the nucleus
and is unresolved strongly argues for a nuclear origin of the X-rays. This is
the case for 9 sources (see Table 2). In addition we searched 
the ROSAT public archive for additional coverage of the galaxies in our sample.
The only galaxy covered is UGC~716. It was observed serendipitously at 
10\arcmin\ offset from the center of an HRI pointing in January 1996. 
The X-ray source is clearly extended
and most likely represents emission from a galaxy cluster.

\item Time variability of the X-ray flux rules out emission from
extended gas clouds and strongly argues for an AGN. As described
above the comparison of the HRI and RASS fluxes for 9 sources 
indicates time variability supporting the AGN origin of the X-rays.
This further points at an AGN identification in addition to the HRI point
source detection argument in 2.

\item We can compare the measured X-ray luminosity with X-ray luminosities
expected from the AGN / morphology type of the galaxy or for the galaxy 
surroundings (group or cluster 
environment). While Seyfert 2 and LINER nuclei should not have luminosities 
in the ROSAT band above 10$^{41}$ \ergsec the hot gas in early
type galaxies or in groups or cluster may be more luminous by factors 
of 100 or 1000. This would argue for an origin of the X-ray emission from 
the surrounding group or cluster as listed in NED for
UGC 716, MCG~-01-05-031, AM~0426-625~S, 
ESO~120-~G~023, ESO~254-~G~017, and 
ESO~122-~IG~016. Hot gas in the early type host galaxy could explain 
the X-ray emission in IC~1867, NGC~1217, AM~0426-625~S, ESO~120-~G~023, and
UGC 12492. There are, however, galaxies for which these arguments to exclude
a Sy2/LINER nucleus as the origin of the X-ray emission do 
not hold. HRI observations of ESO~254-~G~017 show a X-ray variable nuclear
point source and strongly argue for an nuclear origin of the X-ray emission 
(see 2 and 3 above). The optical spectrum of the nucleus is heavily disturbed
by the host galaxy spectrum, but does not show any broad Balmer component
that would indicate a Sy1 type.   
For the Sy2 galaxy UGC~3134 neither the morphological
type (SBc) nor a known group or cluster surrounding can explain the
X-ray luminosity of $\sim 10^{43}$ erg s$^{-1}$. HRI observations could
clarify the interpretation if we see emission from a nuclear source.  

\item We have performed a line fitting analysis of the prominent
optical emission lines. 
The location of our objects in the diagnostic diagrams of Osterbrock
(1989)  clearly
demonstrates their AGN nature (cf. Fig. 4).
The flux ratios plotted in Fig. 5 are an additional indication for the
AGN character of the objects.
Using the nomograph shown in Fig. 1 of Maccacaro et al. (1988),
one sees that the combined X-ray flux and optical B-magnitude agree well
with those expected from AGN (cf. the left panel of Fig. 5).
The middle panel of Fig. 5 gives the ratio of the X-ray to far-infrared
(40-120 $\mu m$) flux. All of the objects show flux ratios above a
value of 0.01. As shown in Fig. 5 of  Boller et al. (1997) most AGN
exhibit flux ratios above that value. Theoretical models addressing the
X-ray and far-infrared emission of galaxies in different states of
nuclear activity suggest that flux ratios above about 0.01 require
AGN activity, whereas values below about 0.01 can be explained by
starburst activity (Bertoldi \& Boller 1998).
Using the formalism of Kellermann et al. (1989, see their Fig. 4)
one can quantify the radio-quiet or radio-loud nature of our objects.
The ratio of the radio to optical flux density for most objects is
below 10 (right panel of Fig. 5) and objects below such a value
are considered as radio-quiet (of course, there is no strict
dividing line between radio-quiet and radio-loud objects and there
is a continuous increase of the radio activity with increasing 
radio to optical flux ratio). We consider the six objects  
with flux ratios above 10 as radio-loud and the most intense
radio emitters in our sample (RX~J011232.8-320140 and NGC~1218) 
show a radio to optical flux ratio of about 275.
\end{enumerate}

The considerations above propose 25 AGN identifications  and 1 dubious case
for the 33 X-ray sources. This represents a success rate of about 75\% for
detecting active nuclei using our selection criteria. If we do not
consider the HRI pre-selection we still have a success rate of 
$\sim$ 50\%.

Of specific interest are the 3 BL Lac objects identified (one as the 
central source of the field and two nearby sources).
We tried in vain to identify possible optical absorption
lines to get a handle on the distance of the objects. Therefore no 
luminosity determination was possible. Their f$_x$/f$_r$ and 
f$_o$/f$_r$ values clearly put them in the regime of X-ray selected
BL Lac objects (see Fig. 11 of Brinkmann et al. (1995)).

The results of this first optical follow up observing run have demonstrated
that our selection strategy from the RASS bright source catalog / galaxy
correlations is rather efficient in detecting new active galactic nuclei. 
The results encourage further optical observations that will be reported 
in future papers of this series.

The detection of the more than 100 new members of the nearby AGN 
population expected from our program  
will allow to attack several important questions and will stimulate
further investigations:  
(i) sub-arcsec near-infrared imaging and spectroscopy to study the physical
conditions in the broad and narrow line regions; 
(ii) as already a few of our objects are strong Fe\,{\sc ii} emitters 
(MCG~-02-14-009, PMN~J0623-6436, MCG~-02-59-006) the
sample offers new possibilities to study the Fe\,{\sc ii} excitation
mechanisms, which are expected to be connected to the strength of
the X-ray emission;
(iii) high resolution optical spectroscopy of our objects with asymmetric
broad Balmer lines (VIII~Zw~36, ESO~416-~G~002, ESO~552-~G~039, PMN~J0623-6436,
ESO~419-~IG~026, ESO~209-~G~012) will allow to further constrain broad-line  
region models;
(iiii) Future X-ray missions will allow sub-arcsec imaging of the
nuclear regions (AXAF) and detailed spectral modeling (AXAF, XMM).

\acknowledgements
This research has made use of the SIMBAD database operated at CDS, Strasbourg,
France and of the NASA/IPAC Extragalactic Database (NED)
which is operated by the Jet Propulsion Laboratory, California Institute of
Technology, under contract with the National Aeronautics and Space
Administration.  

To overlay the X-ray data we used programs kindly provided by Andreas
Vogler and images
based on photographic data of the National Geographic Society -- Palomar
Observatory Sky Survey (NGS-POSS) obtained using the Oschin Telescope on
Palomar Mountain and the UK Schmidt Telescope.  The NGS-POSS was funded 
by a grant from the National
Geographic Society to the California Institute of Technology.  
The UK Schmidt Telescope was operated by the Royal Observatory
Edinburgh, with funding from the UK Science and Engineering Research
Council, until 1988 June, and thereafter by the Anglo-Australian
Observatory.  Original plate material is copyright (c) the Royal
Observatory Edinburgh and the Anglo-Australian Observatory.
The plates were processed into the present compressed digital form with
their permission.  The Digitized Sky Survey was produced at the Space
Telescope Science Institute under US Government grant NAG W-2166.

The ROSAT project is supported by the German Bundesministerium f\"ur
Bildung, Wissenschaft, Forschung
und Technologie (BMBF/DARA) and by the Max-Planck-Gesell\-schaft (MPG).

This work has been partially supported by Deut\-sche Agentur f\"ur
Raumfahrtangelegenheiten (DARA) grant 50\,OR\,9408\,9.

\onecolumn
\pagestyle{empty}
\landscape
\noindent {{\bf Table 2.} X-ray identification information
\scriptsize
  \begin{longtable}[l]{rrrlrrrrlll}
  \hline 
\multicolumn{1}{l}{{\it ROSAT} name} & \multicolumn{1}{l}{Gal.}       &\multicolumn{1}{l}{Count rate} 
&Name   &\multicolumn{2}{l}{~~~RA, Dec (J2000.0)}         &\multicolumn{1}{l}{$\Delta$}   
&\multicolumn{1}{l}{Redshift}&Morph. type&AGN type &Comment\\
\multicolumn{1}{l}{1RXS~J or RX~J}   &  \multicolumn{1}{l}{No.}       &\multicolumn{1}{l}{(cts s$^{-1}$)}&     
&\multicolumn{1}{l}{~~~(h m s)}& \multicolumn{1}{l}{(d m s)}& \multicolumn{1}{l}{(")}
&\multicolumn{1}{l}{(km s$^{-1}$)} &            & &\\
\multicolumn{1}{l}{(1)} & \multicolumn{1}{l}{(2)} & \multicolumn{1}{l}{(3)} & (4) 
& \multicolumn{1}{l}{~~~(5)} & \multicolumn{1}{l}{(6)} & \multicolumn{1}{l}{(7)} 
& \multicolumn{1}{l}{(8)} & (9) & (10) & (11) \\
\hline 
\endhead  
\hline 
\endfoot
000805.6$+$145027& 13&$0.179\pm0.018$&CGCG 433-025    & 0 08 05.6&  14 50 23&   4&$13520\pm~60$
&Sab          &Sy1  \\
003413.7$-$212619& 30&$0.330\pm0.035$&HCG 4a          & 0 34 13.5& -21 26 20&   3&$7900\pm~60$
&$^*$SBCbc pec: &$^*$Sy1.8\\
004236.9$-$104919& 38&$0.244\pm0.028$&VIII Zw 36      & 0 42 36.7& -10 49 22&   4&$12390\pm~60$
&E2           &Sy1.5  \\
010517.5$-$582618& 56&$0.182\pm0.039$&ESO 113- G 010  & 1 05 16.8& -58 26 13&   8&$7705\pm~30$
&$\Diamond$(R\_1)SB(rl)0/a&Sy1.8 &X-var\\
010517.3$-$582615&HRI&$0.123\pm0.005$&ESO 113- G 010  &          &          &   4\\
010918.0+131011& 62&$0.069\pm0.015$&UGC 716         & 1 09 18.4&  13 10 08&   7&$17710\pm150$
&S0           &non active&$^*$in cl, ID?\\
011219.5$-$320338& 68&$0.153\pm0.023$&NGC 427         & 1 12 19.1& -32 03 44&   8&$10035\pm~30$
&$\Diamond$(R\_1R'\_2?)SB(r)a&Sy1.2 &X-var\\
011219.3$-$320340&HRI&$0.025\pm0.002$&NGC 427         &          &          &   5\\
011232.8$-$320140&HRI&$0.033\pm0.002$&                & 1 12 32.6& -32 01 44&   5&
&star like&BL Lac &X-var\\
012018.8$-$440748& 72&$0.305\pm0.031$&ESO 244- G 017  & 1 20 19.6& -44 07 43&  10&$7045\pm~15$
&$\Diamond$(R:)SB(r)a   &$^*$Sy1.5 \\
014526.6$-$034945& 99&$0.175\pm0.029$&MCG -01-05-031  & 1 45 25.3& -03 49 39&  21&$5420\pm~60$
&Sbc pec &Sy2  &in cl, ID?\\
014739.5$-$660952&101&$0.091\pm0.020$&ESO 080- G 005  & 1 47 39.5& -66 09 49&   3&$8080\pm~60$
&$^*$Irr     &Sy1.8 &X-var\\
014739.2$-$660946&HRI&$0.010\pm0.002$&ESO 080- G 005  &          &          &   3\\
023513.9$-$293616&132&$0.356\pm0.034$&ESO 416- G 002  & 2 35 13.4& -29 36 18&   7&$17710\pm~15$
&$\Diamond$Sa:   &Sy1.9 &X-var\\
023513.5$-$293615&HRI&$0.041\pm0.004$&ESO 416- G 002  &          &          &   3\\
023454.8$-$293425&HRI&$0.006\pm0.002$&                & 2 34 54.5& -29 34 28&   5&$0.6785(5)$
&star like  &QSO \\
023536.8$-$293842&HRI&$0.043\pm0.004$&PHL 1389        & 2 35 36.6& -29 38 44&   3&
&star like &BL Lac \\
025552.4+091853&144&$0.083\pm0.019$&IC 1867         & 2 55 52.2&   9 18 42&  11&$7680\pm~30$
&E3           &Sy2/LINER&in group, ID?\\
030606.3$-$390212&151&$0.193\pm0.023$&NGC 1217        &$\Diamond$3 06 05.9& -39 02 10&  5&$6155\pm~30$
&$^*$S0/a &LINER &in cl, ID?\\
030825.9+040637&152&$0.175\pm0.021$&NGC 1218        & 3 08 26.3&   4 06 40&   6&$8590\pm~30$
&$\Diamond$S0/a         &Sy1 \\
034203.8$-$211428&174&$0.258\pm0.026$&ESO 548- G 081  & 3 42 02.8& -21 14 26&  14&$4110\pm~30$
&$^*$SBa pec &$^*$Sy1 \\
042710.2$-$624712&195&$0.097\pm0.015$&AM 0426-625 S   & 4 27 12.5& -62 47 09&  16&$5490\pm150$
&$\Diamond$Gpair, S0 pec &Sy2/LINER &$^*$in group, ID?\\ % pair ???, non pec
043520.2$-$780150&201&$0.274\pm0.025$&ESO 15- IG 011  & 4 35 16.2& -78 01 57&  14&$18280\pm~30 $
&$\Diamond$Gpair, S0: pec&$^*$Sy1.8 + LINER &merger, DN\\
043516.9$-$780157&HRI&$0.077\pm0.004$&ESO 15- IG 011  &          &          &   3\\
043813.8$-$104740&203&$0.066\pm0.016$&MCG -02-12-050  & 4 38 13.8& -10 47 45&   5&$10730\pm~30$
&SBb          &Sy1.2 \\
044148.1$-$011806&205&$0.083\pm0.016$&UGC 3134        & 4 41 48.3& -01 18 10&   5&$8665\pm~15$
&$^*$SBc      &Sy2 \\
045142.3$-$034834&210&$0.281\pm0.029$&MCG -01-13-025  & 4 51 41.4& -03 48 34&  15&$4765\pm~30$
&$^*$SB0&$^*$Sy1.2   \\
045840.4$-$215922&214&$0.106\pm0.018$&ESO 552- G 039  & 4 58 40.3& -21 59 32&   4&$11870\pm~30$
&$^*$SBa &Sy1.2 &X-var, in cl   \\ 
045840.2$-$215928&HRI&$0.062\pm0.006$&ESO 552- G 039  &          &          &   4\\
051621.5$-$103341&223&$0.307\pm0.029$&MCG -02-14-009  & 5 16 21.2& -10 33 40&   4&$8530\pm~30$
&SBb         &Sy1\\
055559.4$-$612438&238&$0.093\pm0.007$&ESO 120- G 023  & 5 55 52.6& -61 24 14&  55&$11280\pm150   $
&$\Diamond$SA0-        &non active&$*$pos? in group, ID?   \\
060635.4$-$472957&244&$0.213\pm0.016$&ESO 254- G 017  & 6 06 35.8& -47 29 56&   4&$8925\pm~30$
&$^*$E2 pec&Sy2/LINER &X-var, in cl \\ %lines  \\
060635.9$-$473001&HRI&$0.053\pm0.003$&ESO 254- G 017  &          &          &   5\\
062307.7$-$643618&254&$0.417\pm0.013$&PMN J0623-6436  & 6 23 07.7& -64 36 21&   3&$38640\pm~30$
&$^*$S0/a&$^*$Sy1    &X-var\\
062308.0$-$643619&HRI&$0.100\pm0.005$&PMN J0623-6436  &          &          &   3\\
063059.7$-$240636&258&$0.084\pm0.014$&PMN J0630-2406  & 6 30 59.4& -24 06 46&  11&
&star like   &BL Lac \\
063059.4$-$240643&HRI&$0.033\pm0.003$&PMN J0630-2406  &          &          &   3\\
064011.5$-$255337&261&$0.273\pm0.029$&ESO 490- IG 026 & 6 40 11.8& -25 53 43&   7&$7450\pm~30$
&$\Diamond$pec&Sy1.2  &$^*$X-var, merger \\
064011.9$-$255343&HRI&$0.056\pm0.004$&ESO 490- IG 026 &          &          &   1\\
071204.2$-$603005&276&$0.106\pm0.010$&ESO 122- IG 016 & 7 12 03.2& -60 30 30&  26&$9770\pm~80 $
&$^*$Irr &LINER + non active &$^*$DN, in group, ID?     \\
080157.7$-$494639&299&$0.163\pm0.016$&ESO 209- G 012  & 8 01 57.9& -49 46 42&   4&$12140\pm~30$
&$^*$Sa pec  &Sy1.5 \\
080157.8$-$494645&HRI&$0.117\pm0.009$&ESO 209- G 012  &          &          &   4\\
215656.8$-$113920&826&$0.285\pm0.031$&NGC 7158        &21 56 56.5& -11 39 32&  13&$8275\pm~30$
&SBa pec     &NLS1   \\
223656.5$-$221321&849&$0.422\pm0.043$&ESO 602- G 031  &22 36 55.9& -22 13 12&  13&$9895\pm~30$
&$^*$Sb&$^*$Sy1.8 &X-var\\
223655.8$-$221314&HRI&$0.021\pm0.002$&ESO 602- G 031  &          &          &   2\\
230921.5+004540&867&$0.091\pm0.018$&IC 5287         &23 09 20.3&   0 45 23&  25&$9750\pm~15$
&$\Diamond$(R')SB(r)b&Sy1.2   \\
231357.7$-$113027&870&$0.177\pm0.026$&MCG -02-59-006  &23 13 57.0& -11 30 19&  14&$22790\pm~30$
&SBc pec:   &Sy1.2  \\
231853.1$-$010338&876&$0.069\pm0.017$&UGC 12492       &23 18 53.7& -01 03 38&   9&$8710\pm~60$
&$\Diamond$S0&Sy2/LINER     \\ 
\end{longtable}
}
\noindent
$\Diamond$~Position or morphological type from NED\\
col 11: DN = double nucleus, ID? = identification questionable, in cl = in cluster, pos? = X-ray 
position questionable, X-var = indication for X-ray variability\\
{$^*$~\bf Notes to individual galaxies:}\\
{\bf HCG 4a} $\Diamond$(R'\_1)SB(rl)bc, member of HCG 4, 
candidate for AGN (Coziol et al. 1993) \\
{\bf UGC 716} in Abell 150, faint foreground star 3" north of nucleus\\
{\bf ESO 244- G 017} Sy1 candidate (Maia et al. 1989), Sy1 (Maza \& Ruiz 1992) \\
{\bf MCG -01-05-031} $\Diamond$SB(rs)bc pec:\\
{\bf ESO 080- G 005} $\Diamond$S...\\
{\bf NGC 1217} $\Diamond$SA:(s:)a pec\\
{\bf ESO 548- G 081} $\Diamond$SB(rs)a pec?, Sy1 (Stocke et al. 1991)\\
{\bf AM 0426-625} Pair of EGal, HD28667 nearby may contribute to X-ray emission \\
{\bf ESO 15- IG 011} interacting pair with plumes north, 
Sy1 single category object, only one nuclear component investigated (Sekiguchi \& Wolstencroft 1993).
We identify a Sy1 nucleus (center) cz=($18280\pm30$)km s$^{-1}$ 
and LINER (south) cz=($18430\pm30$)km s$^{-1}$ \\
{\bf UGC 3134} $\Diamond$SAB(s)c \\
{\bf MCG -01-13-025} $\Diamond$SAB(s)0+ pec:, Sy1 (Brissenden et al. 1987) \\
{\bf ESO 52- G 039} $\Diamond$S... \\
{\bf ESO 120- G 023} bright star 3' SW influences X-ray position and count rate, brightest in group \\
{\bf ESO 254- G 017} $\Diamond$E2? pec \\
{\bf PMN J0623-6436} $\Diamond$??Extended??, Sy1 with very strong Fe\,{\sc ii} emission (Keel et al. 1988) \\
{\bf ESO 490- IG 026} colliding pair \\
{\bf ESO 122- IG 016} $\Diamond$?GPair? \\
{\bf ESO 209- G 012} $\Diamond$Sa? \\
{\bf ESO 602- G 031} $\Diamond$(R')SAB(rs)b, Sy1 from poor spectrum (Coziol et al. 1994) \\

\vfill \eject

%\begin{table}
\normalsize
\noindent {{\bf Table 3.} Optical line properties. Line width 
W$_{\rm H_{\alpha} broad}$ (col. 11) is given as FWHM
\scriptsize
  \begin{longtable}[l]{lrrrrrrlrrrl}
  \hline
Name &Exp.&\multicolumn{1}{l}{cz(em)} 
&\multicolumn{1}{l}{cz(abs)}          &\multicolumn{1}{l}{cz(others)}
& Ref.& [O\,{\sc iii}]/H$_{\beta}$&[S\,{\sc ii}]/H$_{\alpha}$&[N\,{\sc ii}]/H$_{\alpha}$
&[O\,{\sc i}]/H$_{\alpha}$&\multicolumn{1}{l}{W$_{\rm H_{\alpha} broad}$}
& \multicolumn{1}{l}{comments$^*$} \\
&(s)& \multicolumn{1}{l}{(km s$^{-1}$)}&\multicolumn{1}{l}{(km s$^{-1}$)}
&\multicolumn{1}{l}{(km s$^{-1}$)}& & & & & &(km s$^{-1}$)&\\
\multicolumn{1}{l}{(1)} & \multicolumn{1}{l}{(2)} & \multicolumn{1}{l}{(3)} 
&\multicolumn{1}{l}{(4)}& \multicolumn{1}{l}{(5)} & \multicolumn{1}{l}{(6)} 
& \multicolumn{1}{l}{(7)}& \multicolumn{1}{l}{(8)} & \multicolumn{1}{l}{(9)} 
& \multicolumn{1}{l}{(10)} & \multicolumn{1}{l}{(11)} &\multicolumn{1}{l}{(12)}   \\
\hline 
\endhead
\hline 
\endfoot
CGCG 433-025          &1800&$13520\pm60$&$13750\pm150$&             & &38.0&  &  &  &8250
&H$\alpha$ self-absorbed\\
HCG 4a                &900&$ 7900\pm60$&$ 8000\pm150$&$ 8031\pm~30$&1& 8.2&0.7&1.7&0.2&2100&\\
VIII Zw 36            &900&$12390\pm60$&             &             & &13.4&0.5&0.9&0.2&1950
&H$\alpha$ broad asymmetric\\
ESO 113- G 010        &900&$ 7705\pm30$&$ 7480\pm150$&             & & 5.3&0.4&1.4&0.1&2000& \\
UGC 716               &900&            &$17710\pm150$&             & &    &   &   &   &    
&SGS \\
NGC 427               &2700&$10035\pm30$&$10060\pm150$&$ 9937\pm~38$&2&  &1.1&1.8&0.8&5500
&H$\beta$ narrow not detected\\
RX J011232.8$-$320140 &900&            &       &             & & \\
ESO 244- G 017        &900&$ 7045\pm15$&       &$ 7047\pm~19$&3& 7.1&0.3&1.5&0.2&3000&\\
MCG -01-05-031        &900&$ 5420\pm60$&$ 5520\pm150$&$ 5456\pm~22$&4& 3.7&0.4&0.7&0.2& & \\
ESO 080- G 005        &1800&$ 8080\pm60$&       &             & &13.0&0.3&0.8&0.1&3450&\\
ESO 416- G 002        &1500&$17710\pm15$&$17800\pm150$&             & &22.1&1.0&2.6&0.6&$\sim16000$
&FWHM of H$\alpha$ broad estimated \\
RX J023454.8$-$293425 &3000&$0.6785(5)$ &       &             & & \\
PHL 1389              &1200&            &       &             & & \\
IC 1867               &600&$ 7680\pm30$&$ 7540\pm150$&             & &    &   &6.7&0.8& 
&EL: H$\alpha$,[N\,{\sc ii}], [O\,{\sc i}], SGS\\
NGC 1217              &600&$ 6155\pm30$&$ 6350\pm150$&$ 6236\pm~34$&3&    &   &   &   &
&EL: H$\alpha$,[N\,{\sc ii}], [O\,{\sc i}], [O\,{\sc iii}], \\
& & & & & & & & & & &[S\,{\sc ii}], SGS\\
NGC 1218              &1500&$ 8590\pm30$&$ 8620\pm150$&$ 8644\pm~38$&3&    &9.5&21.5&  &2950
&SGS\\
ESO 548- G 081        &1500&$ 4110\pm30$&$ 4070\pm150$&$ 4341\pm~60$&3&    &1.7&7.9&0.5&4450&\\
AM 0426-625           &600&            &$ 5490\pm150$&             & &    &   &   &   &    &SGS\\
ESO 15- IG 011 N      &900&$18280\pm30$&       &$18332\pm~33$&1& 5.4&0.5&0.7&0.1 &2750& \\
ESO 15- IG 011 S      &1800&$18430\pm30$&$18550\pm150$&$18332\pm~33$&1& 1.4&0.6&0.9&0.2& & \\
MCG -02-12-050        &1800&$10730\pm30$&$10780\pm150$&             & &38.1&1.1&1.1&0.5&5400& \\
UGC 3134              &900&$ 8665\pm15$&$ 8720\pm150$&             & & 4.0&0.4&0.8&0.1&   & \\
MCG -01-13-025        &1200&$ 4765\pm30$&$ 4710\pm150$&  3897       &5& 2.9&0.3&0.9&0.4&4500&\\
ESO 552- G 039        &1500&$11870\pm30$&$11920\pm150$&             & &18.4&1.2&1.6&0.7&4000
&H$\alpha$ broad asymmetric \\
MCG -02-14-009        &900&$ 8530\pm30$&       &             & & 8.6&0.2&0.9&   &2850
&very strong Fe\,{\sc ii} multiplets  \\
ESO 120- G 023        &600&            &$11280\pm150$&             & &    &   &   &   &    
&SGS\\
ESO 254- G 017        &1500&$ 8925\pm30$&$ 9060\pm150$&             & &    &1.2&2.1&   &     
&EL: H$\alpha$,[N\,{\sc ii}], [O\,{\sc i}], [S\,{\sc ii}], \\
& & & & & & & & & & &SGS\\
PMN J0623-6436        &1200&$38640\pm30$&       & 38853       &6&    &   &   &   &2500
&H$\alpha$ + H$\beta$ broad asymmetric,\\ 
& & & & & & & & & & &very strong Fe\,{\sc ii} multiplets \\
PMN J0630-2406        &600&            &       &             & & \\
ESO 490- IG 026       &900&$ 7450\pm30$&       &  7751       &7&16.4&0.9&1.4&0.3&4100
&H$\alpha$ + H$\beta$ broad asymmetric \\
ESO 122- IG 016 NE    &1200&            &$ 9830\pm150$&             & &    &   &   &   &    
&EL: [N\,{\sc ii}]?, SGS\\
ESO 122- IG 016 SW    &1200&            &$ 9770\pm~80$&             & &    &   &   &   &
&SGS \\
ESO 209- G 012        &1200&$12140\pm30$&             &$11743\pm120$&8& 8.6&0.8&1.6&0.2&3400
&H$\alpha$ broad asymmetric \\
NGC 7158              &900&$ 8275\pm30$&              &             & &10.9&0.1&0.9&0.1&2100& \\
ESO 602- G 031        &1800&$ 9895\pm30$&             &$ 9828\pm~30$&1&10.1&0.4&1.3&0.1&5900& \\
IC 5287               &1800&$ 9750\pm15$&$ 9730\pm150$&             & &    &1.8&4.8&   &4200& \\
MCG -02-59-006        &900&$22790\pm30$&              &             & &    &0.1&1.3&   &3700
&redshifted H$\alpha$ and H$\beta$ narrow \\
& & & & & & & & & & &strong Fe\,{\sc ii} multiplets \\
UGC 12492             &600&$ 8710\pm60$&$ 8900\pm150$& 9019        &3&    &   &   &   &
&EL: [N\,{\sc ii}], SGS \\
\end{longtable}
}
\noindent
$^*$ SGS = strong galaxy spectrum, EL = emission lines\\
\noindent
References:
(1)~Da Costa et al. 1991; 
(2)~Ramella et al. 1996;
(3)~De Vaucouleurs et al. 1991;
(4)~Zabludoff et al. 1993;
(5)~Brissenden et al. 1987;
(6)~Keel et al. 1988;
(7)~Takata et al. 1994;
(8)~Fisher et al. 1995

\vfill \eject

\normalsize
\noindent {{\bf Table 4.} Multi-wavelength properties. Col. (1): Galactic  N$_{\rm H}$ 
in units of 10$^{21}$cm$^{-2}$  (Dickey \& Lockman 1990), (5) and (8): in units of 
erg cm$^{-2}$ s$^{-1}$ Hz$^{-1}$, (13): in units of erg s$^{-1}$ Hz$^{-1}$\\
\scriptsize
  \begin{longtable}[l]{lllllllllllllrrr}
  \hline
name&$\rm N_{Hgal}$&R&$\rm m_B$&$\rm log$&$\rm log$&$\rm log$&$\rm log$&$\rm M_B$&$\rm log $&$\rm log$&
$\rm log $&$\rm log$&$\rm log $&$\rm log $&$\rm log $\\
& & & &$\rm f_{\nu B}$&$\rm f_X$&$\rm f_{FIR}$&$\rm f_r$& &$\rm L_B$ &$\rm L_X$ &$\rm L_{FIR}$ &
$\rm l_r$ &$\rm {\frac{f_X}{f_B}}^\dagger$ &$\rm \frac{f_X}{f_{FIR}}$ & $\rm \frac{f_r}{f_{\nu B}}$ \\
& &(Mpc) &(mag) & &\multicolumn{2}{l}{$\rm (erg\ cm^{-2} s^{-1})$}&&(mag) 
&\multicolumn{3}{c}{$\rm (erg\ s^{-1}$)}& &  & &  \\
(1)& (2)& (3)& (4)& (5)& (6)& (7)& (8)& (9)& (10)& (11)& (12)& (13)& (14)& (15)& (16) \\
\hline 
\endhead
\hline 
\endfoot
CGCG 433-025         & 0.44&   180.3& 15.7&-25.62&-11.31&  $-$&  $-$&-21.46& 44.02& 43.64&  $-$&  $-$&  0.34&  $-$&  $-$\\
HCG 4a               & 0.16&   105.3& 13.7&-24.82&-11.34& -9.64&-24.36&-22.30& 44.35& 43.14& 44.84& 30.10& -0.49& -1.70&  0.46\\
VIII Zw 36           & 0.26&   165.2& 14.6&-25.18&-11.33&  $-$&  $-$&-22.37& 44.38& 43.55&  $-$&  $-$& -0.12&  $-$&  $-$\\
ESO 113- G 010       & 0.30&   102.7& 14.6&-25.18&-11.07&-10.11&  $-$&-21.34& 43.97& 43.40& 44.34&  $-$&  0.14& -0.96&  $-$\\
UGC 716              & 0.41&   236.1& 15.6&-25.58&-11.75&  $-$&  $-$&-22.15& 44.29& 43.45&  $-$&  $-$& -0.14&  $-$&  $-$\\
NGC 427              & 0.21&   133.8& 15.1&-25.38&-11.82&  $-$&  $-$&-21.42& 44.00& 42.88&  $-$&  $-$& -0.41&  $-$&  $-$\\
RX J011232.8$-$320140& 0.21&     $-$& 21.1$^1$&-27.78&-11.70&  $-$&-25.34&  $-$&  $-$&  $-$&  $-$&  $-$&  2.10&  $-$&  2.45\\
ESO 244- G 017       & 0.24&    93.9& 14.6&-25.18&-11.26&-10.46&  $-$&-21.15& 43.89& 43.12& 43.92&  $-$& -0.05& -0.80&  $-$\\
MCG -01-05-031       & 0.28&    72.3& 13.3&-24.66&-11.45&-10.37&-24.86&-21.88& 44.19& 42.70& 43.78& 29.28& -0.76& -1.09& -0.19\\
ESO 080- G 005       & 0.25&   107.7& 16.0&-25.74&-12.19&  $-$&  $-$&-20.05& 43.45& 42.32&  $-$&  $-$& -0.42&  $-$&  $-$\\
ESO 416- G 002       & 0.17&   236.1& 14.9&-25.30&-11.64&  $-$&-23.97$^3$&-22.85& 44.57& 43.56&  $-$& 31.18$^3$& -0.31&  $-$&  1.33$^3$\\
RX J023454.8$-$293425& 0.17&  2712.1& 17.9$^1$&-26.50&-12.47&  $-$&  $-$&-25.13& 45.49& 45.00&  $-$&  $-$&  0.06&  $-$&  $-$\\
PHL 1389             & 0.17&     $-$& 16.1$^2$&-25.78&-11.62&$-$&  -25.26&  $-$&  $-$&  $-$&  $-$&  $-$&  0.19& $-$&  $-0.28$\\
IC 1867              & 1.18&   102.4& 14.6&-25.18&-11.44&  $-$&  $-$&-21.33& 43.97& 43.02&  $-$&  $-$& -0.23&  $-$&  $-$\\
NGC 1217             & 0.21&    82.1& 13.3&-24.66&-11.50&-10.04&-24.44&-22.15& 44.29& 42.77& 44.22& 29.80& -0.81& -1.46&  0.22\\
NGC 1218             & 0.73&   114.5& 13.9&-24.90&-11.20&  $-$&-22.46$^3$&-22.28& 44.35& 43.35&  $-$& 32.07$^3$& -0.27&  $-$&  2.44$^3$\\
ESO 548- G 081       & 0.28&    54.8& 12.8&-24.46&-11.28&-10.42&  $-$&-21.77& 44.14& 42.63& 43.49&  $-$& -0.79& -0.86&  $-$\\
AM 0426-625          & 0.24&    73.2& 13.0&-24.54&-11.75&  $-$&  $-$&-22.20& 44.32& 42.41&  $-$&  $-$& -1.18&  $-$&  $-$\\
ESO 15- IG 011       & 0.70&   243.7& 14.2&-25.02&-11.13&-10.32&  $-$&-23.62& 44.88& 44.10& 44.88&  $-$& -0.08& -0.81&  $-$\\
MCG -02-12-050       & 0.60&   143.1& 15.2&-25.42&-11.67&-10.38&-24.97&-21.46& 44.02& 43.08& 44.36& 29.76& -0.22& -1.29&  0.46\\
UGC 3134             & 0.47&   115.5& 14.2&-25.02&-11.63&-10.31&-25.03&-21.99& 44.23& 42.93& 44.24& 29.51& -0.58& -1.32& -0.01\\
MCG -01-13-025       & 0.41&    63.5& 14.6&-25.18&-11.14&  $-$&-25.09&-20.30& 43.55& 42.90&  $-$& 28.94&  0.07&  $-$&  0.10\\
ESO 552- G 039       & 0.28&   158.3& 16.3&-25.86&-11.37&  $-$&  $-$&-20.58& 43.67& 43.47&  $-$&  $-$&  0.52&  $-$&  $-$\\
MCG -02-14-009       & 0.91&   113.7& 15.5&-25.54&-10.92&-10.85$^4$&-25.41&-20.66& 43.70& 43.64& 43.69$^4$& 29.12&  0.65& -0.06$^4$&  0.14\\
ESO 120- G 023       & 0.49&   150.4& 15.3&-25.46&-11.57&  $-$&  $-$&-21.47& 44.02& 43.23&  $-$&  $-$& -0.08&  $-$&  $-$\\
ESO 254- G 017       & 0.51&   119.0& 14.6&-25.18&-11.34&  $-$&-23.54$^3$&-21.66& 44.10& 43.25&  $-$& 31.03$^3$& -0.13&  $-$&  1.64$^3$\\
PMN J0623-6436       & 0.56&   515.2& 14.8$^1$&-25.26&-11.05&  $-$&-23.34$^3$&-24.64& 45.29& 44.84&  $-$& 32.46$^3$&  0.24&  $-$&  1.92$^3$\\
PMN J0630-2406       & 0.97&     $-$& 15.4$^1$&-25.50&-11.44&  $-$&-23.89$^3$&  $-$&  $-$&  $-$&  $-$&  $-$&  0.09&  $-$&  1.62$^3$\\
ESO 490- IG 026      & 1.09&    99.3& 14.1&-24.98&-11.19&-10.07&-24.41&-21.77& 44.14& 43.24& 44.35&30.01
& -0.18& -1.12&  0.58\\
ESO 122- IG 016      & 0.97&   130.3& 14.6&-25.18&-11.37&  $-$&  $-$&-21.86& 44.18& 43.30&  $-$&  $-$& -0.16&  $-$&  $-$\\
ESO 209- G 012       & 2.68&   161.9& 15.3&-25.46&-10.68&-10.52$^5$&  $-$&-21.63& 44.09& 44.19& 44.32$^5$
&  $-$&  0.81& -0.15$^5$&  $-$\\
NGC 7158             & 0.39&   110.3& 17.3$^2$&-26.26&-11.15&  $-$&  $-$&-18.79& 42.95& 43.38&  $-$&  $-$&  -1.14&  $-$&  $-$\\
ESO 602- G 031       & 0.22&   131.9& 14.3&-25.06&-11.89&  $-$&-25.14&-22.18& 44.31& 42.79&  $-$& 29.52& -0.80&  $-$& -0.07\\
IC 5287              & 0.42&   130.0& 14.8&-25.26&-11.62&  $-$&  $-$&-21.65& 44.09& 43.05&  $-$&  $-$& -0.33&  $-$&  $-$\\
MCG -02-59-006       & 0.27&   303.9& 15.7&-25.62&-11.46&  $-$&  $-$&-22.59& 44.47& 43.96&  $-$&  $-$&  0.19&  $-$&  $-$\\
UGC 12492            & 0.42&   116.1& 14.6&-25.18&-11.74&  $-$&  $-$&-21.61& 44.08& 42.83&  $-$&  $-$& -0.53&  $-$&  $-$\\
\end{longtable}
}
\noindent
$\dagger$ $\rm log\ \frac{f_X}{f_B} = 5.37 + log(f_X) + \frac{m_B}{2.5}$ 
(following Maccacaro et al. 1988)\\
$^1$ $\rm m_B$ from ROE/NRL digitized sky survey (Yentis et al. 1992)\\
$^2$ $\rm m_B$ from SIMBAD \\
$^3$ $\rm 4.85$ GHz radio flux from NED\\
$^4$ upper limit in 100 $\mu m$ flux,  log $\rm f_{FIR}$ is below -10.31 (\ergcm), log $\rm L_{FIR}$ below 44.22 
(\ergsec), $log \rm \frac{f_X}{f_{FIR}}$ above -0.61 \\
$^5$ upper limit in 100 $\mu m$ flux,  log $\rm f_{FIR}$ is below -9.89 (\ergcm), log $\rm L_{FIR}$ below 44.95
(\ergsec), $log \rm \frac{f_X}{f_{FIR}}$ above -0.79 \\

\normalsize
\vfill \eject
\end{document}